\documentclass[useAMS,usenatbib]{mn2e}
\usepackage{amssymb}
\usepackage{graphicx}
\usepackage{amsmath}
\usepackage{natbib}
\usepackage{epsfig}
\usepackage{epstopdf}
\newcommand{\asec} {\mbox{$^{\prime \prime}$} }

\usepackage{mathrsfs}
\usepackage{amsfonts}

\title[Mixing Diagram]{A Quasar-Galaxy Mixing Diagram: Quasar Spectral Energy Distribution
Shapes in the Optical to Near-Infrared}

\author[Heng Hao et al.]{Heng Hao$^{1,2}$\thanks{E-mail:henghao@post.harvard.edu}, Martin Elvis$^{2}$,
Angela Bongiorno$^{3, 4}$, Gianni Zamorani$^{5}$, Andrea
\newauthor
Merloni$^{3}$, Brandon C. Kelly$^{6}$, Francesca Civano$^{2,7}$,
Annalisa Celotti$^{1,8}$, Luis C. Ho$^{9}$,
\newauthor
 Knud Jahnke$^{10}$, Andrea Comastri$^{5}$, Jonathan R.
Trump$^{11}$, Vincenzo Mainieri$^{12}$,
\newauthor
Mara Salvato$^{13, 14}$, Marcella Brusa$^{3, 15}$, Chris D.
Impey$^{16}$, Anton M. Koekemoer$^{17}$,
\newauthor
Giorgio Lanzuisi$^{3}$, Cristian Vignali$^{5,15}$, John D.
Silverman$^{18}$, C. Megan Urry$^{19}$,
\newauthor
Kevin Schawinski$^{20}$\\
$^{1}$SISSA, Via Bonomea 265, I-34136 Trieste, Italy\\
$^{2}$Harvard-Smithsonian Center for Astrophysics, 60 Garden Street,
Cambridge, MA 02138, USA\\
$^{3}$Max Planck Institute f\"ur Extraterrestrische Physik, Postfach
1312, 85741, Garching bei M\"{u}nchen, Germany\\
$^{4}$INAF-Osservatorio Astronomico di Roma, Via di Frascati 33,
00040, Monteporzio Catone, Rome, Italy\\
$^{5}$INAF - Osservatorio Astronomico di Bologna, via Ranzani 1,
I-40127 Bologna, Italy\\
$^{6}$Department of Physics, Broida Hall, University of California,
Santa Barbara, CA 93106, USA\\
$^{7}$Dartmouth College, Department of Physics and Astronomy, 6127
Wilder Lab, Hanover, NH 03755\\
$^{8}$INAF - Osservatorio Astronomico di Brera, via E. Bianchi 46,
I-23807 Merate, Italy\\
$^{9}$The Observatories of the Carnegie Institute for Science, Santa
Barbara Street, Pasadena, CA 91101, USA\\
$^{10}$Max-Planck-Institut f\"ur Astronomie, K\"onigstuhl 17,
Heidelberg, D-69117, Germany\\
$^{11}$UCO/Lick Observatory, University of California, Santa Cruz,
CA 95064, USA\\
$^{12}$European Southern Observatory, Karl-Schwarzschild-Strasse 2,
D-85748 Garching bei M\"{u}nchen, Germany\\
$^{13}$IPP - Max-Planck-Institute for Plasma Physics, Boltzmann
Strasse 2, D-85748, Garching bei M\"{u}nchen, Germany\\
$^{14}$Excellence Cluster, Boltzmann Strasse 2, D-85748,
Garching bei M\"{u}nchen, Germany\\
$^{15}$Dipartimento di Fisica e Astronomia, Universit\`{a} degli
studi di Bologna, viale Berti Pichat 6/2 40127 Bologna, Italy\\
$^{16}$Steward Observatory, University of Arizona, 933 North Cherry
Avenue, Tucson, AZ 85721, USA\\
$^{17}$Space Telescope Science Institute, 3700 San Martin Drive,
Baltimore, MD 21218, USA\\
$^{18}$Kavli Institute for the Physics and Mathematics of the
Universe, Todai Institutes for Advanced Study, the University of
Tokyo,\\
Kashiwa, Japan 277-8583 (Kavli IPMU, WPI)\\
$^{19}$Physics Department and Yale Center for Astronomy and
Astrophysics, Yale University, New Haven, CT 06511, USA\\
$^{20}$Institute for Astronomy, Department of Physics, ETH
Zurich, Wolfgang-Pauli-Strasse 16, CH-8093 Zurich, Switzerland}

\begin{document}
%%%%%%%%%%%%%%%%%%

\date{Version May 8th, 2013.}

\pagerange{\pageref{firstpage}--\pageref{lastpage}}\pubyear{2013}

\maketitle

\label{firstpage}

%%%%%%%%%%%%%%%%%%%%%%%%%%%%%%%%%%%%%%%%%%%%%%%%%%%%%%
\begin{abstract}
We define a quasar-galaxy mixing diagram using the slopes of their
spectral energy distributions (SEDs) from $1~\mu$m to 3000~\AA\ and
from $1~\mu$m to 3~$\mu$m in the rest frame. The mixing diagram can
easily distinguish among quasar-dominated, galaxy-dominated and
reddening-dominated SED shapes. By studying the position of the 413
XMM selected Type 1 AGN in the wide-field ``Cosmic Evolution Survey"
(COSMOS) in the mixing diagram, we find that a combination of the
Elvis et al. (1994, hereafter E94) quasar SED with various
contributions from galaxy emission and some dust reddening is
remarkably effective in describing the SED shape from $0.3-3~\mu$m
for large ranges of redshift, luminosity, black hole mass and
Eddington ratio of type 1 AGN. In particular, the location in the
mixing diagram of the highest luminosity AGN is very close (within
1$\sigma$) to that of the E94 SED. The mixing diagram can also be
used to estimate the host galaxy fraction and reddening in quasar.
We also show examples of some outliers which might be AGN in
different evolutionary stages compared to the majority of AGN in the
quasar-host galaxy co-evolution cycle.
\end{abstract}


%%%%%%%%%%%%%%%%%%%%%%%%%%%%%%%%%%%%%%%%%%%%%%%%%%%%%%%%%%%%%%%%%%%%%%%%
\section{Introduction}

The masses of the super massive black holes (SMBHs) that exist in
most, if not all, galaxy nuclei (e.g. Kormendy \& Richstone 1995),
are proportional to their host galaxy bulge stellar mass, as
measured by either luminosity (Kormendy \& Richstone 1995; Marconi
\& Hunt 2003) or velocity dispersion (Ferrarese \& Merritt 2000;
Gebhardt et al.  2000). As most SMBH growth occurs during their
active phases (the `Soltan argument', Soltan 1982), most bulges must
have gone through an active phase, being seen as a quasar or active
galactic nucleus (AGN). It is observed that both galaxies and AGN
exhibit coordinated ``downsizing'': massive galaxy star formation
peaks at $z\sim2$, while high luminosity quasars have their peak
space density at $z=2-3$ (Silverman et al. 2005; Brusa et al. 2010;
Civano et al. 2011); lower mass galaxies star formation peaks at
$z=1-1.5$, as do lower luminosity AGN (Franceschini et al. 1999;
Ueda et al. 2003; Brandt \& Hasinger 2005; Bongiorno et al. 2007). A
close co-evolutionary link between SMBH activity and host galaxy
evolution seems to be required.

In principle, we could study whatever feedback process controls this
co-evolution, by separately analyzing the emission associated to the
SMBH and the host galaxy in the same objects. Observationally,
however, it is difficult to disentangle the emission from quasar and
host galaxy in the optical-IR range, especially for high redshift
($z>1$) objects. Spatially decomposing a point-source AGN and an
extended host requires expensive high-resolution Hubble Space
Telescope imaging. Even the 0.1\asec angular resolution of the
Hubble cannot easily resolve the extended host emission from the
point like AGN emission at $z>1$ (e.g. Cisternas et al. 2011). SED
fitting techniques can do so, but have to assume one or several
quasar and galaxy SED models (e.g. Merloni et al. 2010), which might
lead to systematic errors that are difficult to quantify.

As an alternative approach, we have made use of the fact that the
spectral energy distributions (SEDs) of a quasar and of a galaxy
near $1~\mu$m are completely different.  Quasar SEDs show a
pronounced dip near $1~\mu$m (e.g. Elvis et al. 1994, E94
hereinafter; Richards et al. 2006, R06 hereinafter), while, in
contrast, a galaxy SED peaks at around $1-2~\mu$m. This dichotomy
allows us to define a diagram of near-infrared (NIR) versus optical
(OPT) slopes on either side of $1~\mu$m (rest frame) that cleanly
separates the two SED forms.

In this diagram (Figure~\ref{slp}), galaxies lie in a well-defined
region ($\alpha_{\rm{OPT}}<0$, $\alpha_{\rm{NIR}}>0.8$), that is
clearly distinct from the location of the standard AGN SED
($\alpha_{\rm{OPT}}>0$, $\alpha_{\rm{NIR}}<0$, E94). Reddening moves
objects almost perpendicularly to a line joining the galaxy locus to
the AGN locus in the diagram. Thus this diagram can distinguish the
quasar-dominated, host-dominated or reddening-dominated SEDs easily,
without strong model assumptions, and can pick out AGN with mixtures
of these three components. Hence we call this the Quasar-Galaxy
mixing diagram (hereinafter ``mixing diagram'' for short).

With this convenient tool, we can more easily study the evolution of
quasar SEDs with physical parameters, identify outliers, and
estimate host/reddening contributions.  This mixing diagram is a
generalization of the quasar-galaxy mixing curves in the
$(U-B)(B-V)$ color-color plane defined for ``N galaxies'' by Sandage
(1971) and Weedman (1973). The plot is equivalent to a color-color
plot, but utilizes more photometric bands and is defined in the rest
frame. As a result, the mixing diagram can be used for sources at
any redshift.

In this paper, we use the mixing diagram to study the the SED shape
in the optical to near-infrared decade ($3~\mu$m to 3000~\AA) for
three type 1 AGN samples: the large XMM-COSMOS type 1 AGN sample
(Elvis et al. 2012, Paper I hereinafter), the SDSS-Spitzer quasar
sample (R06) and the bright quasar sample (E94). Detailed
description of the three samples are in \S~\ref{s:sample}. We
primarily focus on the XMM-COSMOS type 1 AGN sample to demonstrate
the major applications of the mixing diagram.

All the wavelengths considered in this paper are in the rest frame.
We adopt the WMAP 5-year cosmology (Komatsu et al. 2009), with
$\rm{H_0 =71~km~s^{-1}~Mpc^{-1}}$, $\Omega_{\rm{M}}=0.26$ and
$\Omega_{\rm{\Lambda}}=0.74$.

%%%%%%%%%%%%%%%%%%%%%%%%%%%%%%%%%%%%%%%%%%%%%%%%%%%%%%%%%%%%%%%%%%%%%%%%%%%%%
\section{Quasar-Galaxy Mixing Diagram}
\label{s:md}

%%%%%%%%%%%%%%%%%%%%%%%
\begin{figure*}
\includegraphics[angle=0,width=0.45\textwidth]{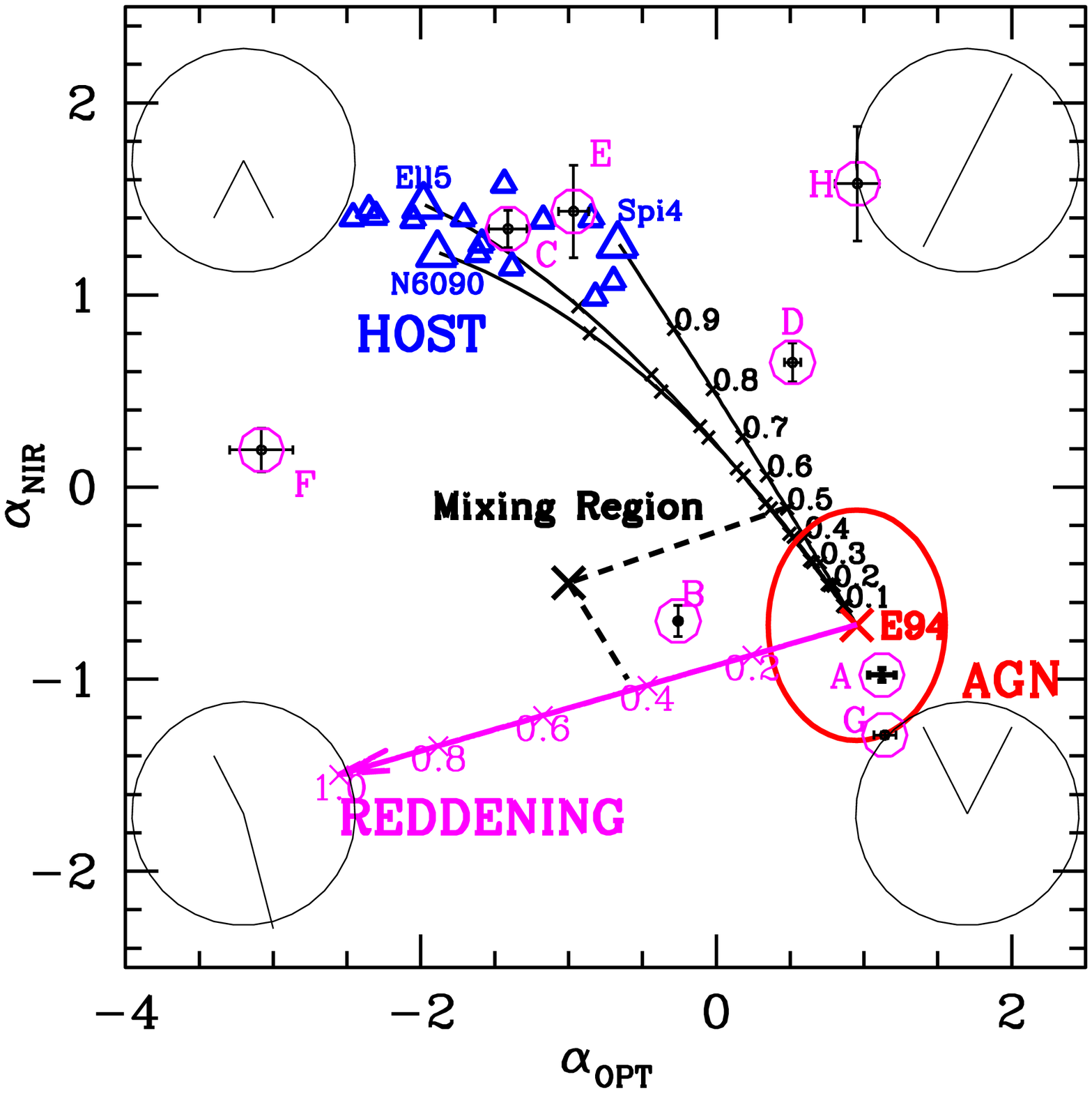}
\includegraphics[angle=0,width=0.45\textwidth]{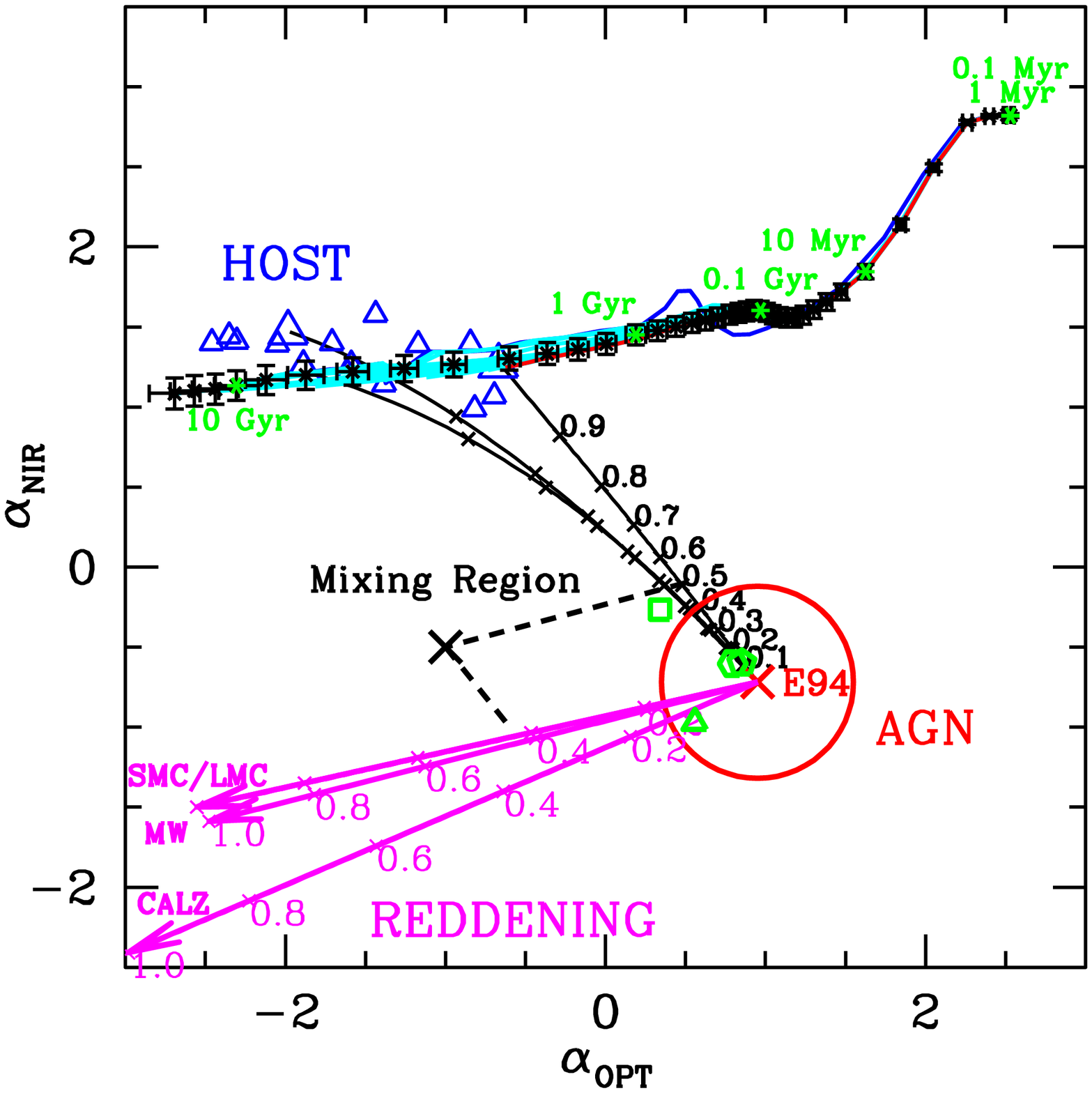}
\caption{Two representations of the quasar-galaxy mixing diagram,
$\alpha_{\rm{NIR}}$ ($3~\mu$m to $1~\mu$m) versus
$\alpha_{\rm{OPT}}$ ($1~\mu$m to $0.3~\mu$m). Note that these slopes
are defined in the log$~\nu L_{\nu}$ versus log$~\nu$ plane. The E94
radio-quiet mean SED is shown as a red cross
($\alpha_{\rm{OPT}}=0.95\pm0.04$ and
$\alpha_{\rm{NIR}}=-0.72\pm0.05$). The red circle shows the
dispersion of the quasar samples (\S~\ref{s:intsdisp}). The blue
triangles indicate the 16 galaxy templates from ``SWIRE Template
Library'' (Polletta et al. 2007). The black lines connecting the
SWIRE galaxy templates and the E94 mean SED are mixing curves
(\S~\ref{s:fgal}), showing where mixed quasar-galaxy SEDs would
locate. The numbers beside the mixing curves are the galaxy fraction
at $1~\mu$m. The magenta arrow shows how reddening affects the E94
radio-quiet mean SED. The numbers under the reddening vector show
the E(B-V) values. \emph{Left:} Different regions of the plot
correspond to different SED shapes, as shown in the black circles at
the four corners. In these four black circles, the SEDs are in the
log$~\nu L_{\nu}$ versus log$~\nu$ plane, with wavelength increasing
to the left. The reddening vector here is calculated using SMC
reddening law. The points circled in magenta show the position of
the outliers in the sample discussed in Elvis et al. (2012) (A, B,
C, D) and in \S~\ref{s:outliers} of this paper (E, F, G, H).
\emph{Right:} The green square represents mean SED of the 203
XMM-COSMOS quasars (Paper I). The green pentagon represents the R06
mean SED ($\alpha_{\rm{OPT}}=0.85\pm0.03$ and
$\alpha_{\rm{NIR}}=-0.60\pm0.05$). The green hexagon represents the
Hopkins et al. (2007) quasar SED template
($\alpha_{\rm{OPT}}=0.79\pm0.14$ and
$\alpha_{\rm{NIR}}=-0.60\pm0.05$). The green triangle represents the
Shang et al. (2011) quasar SED template
($\alpha_{\rm{OPT}}=0.55\pm0.06$ and
$\alpha_{\rm{NIR}}=-0.97\pm0.32$). The solid lines in the upper
region show 16 different Bruzual \& Charlot (2003, BC03 hereinafter)
galaxy models. These 16 models using exponentially declining star
formation history with e-folding timescale $\tau=(0.01, 0.05, 0.1,
0.3, 0.5, 0.6, 1, 2, 3, 5, 10, 15, 30, 50, 80, 100)$ Gyr. The blue
line is for $\tau=0.01$ Gyr, while the red line is for $\tau=100$
Gyr, and the other lines are in cyan. The black tick marks on the
lines are for the galaxy model with $\tau=1$ Gyr with age of the
galaxy ($t_{\rm{age}}$) running from 0.1~Myr to 20~Gyr, in steps of
$\log_{10} t_{\rm{age}}=0.1$. The magenta arrows in the bottom show
tracks for four different reddening curves (Small Magellan
Cloud-`SMC', Large Magellan Cloud-`LMC', Milky Way-`MW', Calzetti et
al. 2000-`Calz') applied to E94 radio quiet mean SED. \label{slp}}
\end{figure*}
%%%%%%%%%%%%%%%

The mixing diagram axes are the $1-3~\mu$m (rest-frame) SED
power-law slope ($\alpha_{\rm{NIR}}$) versus the $0.3-1~\mu$m
(rest-frame) power-law slope ($\alpha_{\rm{OPT}}$), where $\nu
F_{\nu}\propto \nu ^{\alpha}$. These ranges lie on either side of
the $1~\mu$m dip, or inflection point, of the rest frame SED.

The $1~\mu$m wavelength point is not chosen as the central point
arbitrarily. This is where the Wien tail of the black body thermal
emission of the hottest dust (at the maximum sublimation temperature
of $\sim1500~K$, Barvainis et al. 1987) begins to outshine the
optical band power-law ($\alpha\sim$ -0.3) of the SMBH accretion
disk (Malkan \& Sargent 1982; Sanders et al. 1989; E94; Glikman et
al. 2006).

We tried several different wavelength ranges to calculate the slopes
and found that the adopted ranges best represent the dip around
$1~\mu$m. If a smaller wavelength range is chosen, the number of
photometric points in the range will be greatly reduced, due to the
relatively limited photometry coverage (only J H K band) in the NIR
range. If a longer wavelength range is chosen, a variety of problems
would make the estimates of the slope more difficult. For example,
shorter wavelengths, into the UV, are more affected by variability
and by the FeII `small bump' (Wills, Netzer \& Wills, 1985); longer
wavelengths in the NIR encounter a range of cooler dust emission
which adds noise to the NIR slope. In the chosen wavelength range,
the XMM-COSMOS type 1 AGN SED dispersion is invariant in a large
range of $z$ and $L_{bol}$ (Hao et al. 2013a, Paper II hereinafter),
which implies an invariant intrinsic dispersion of SED shape in this
wavelength range.

To ensure reliable slopes, we require at least 3 photometric points
to define each slope. The robustness of the slope measurement using
3 or more photometric points was tested in Hao et al. (2011) and
found to be good. For the XMM-COSMOS quasar sample, the optical data
set is so rich that the mean number of photometry points used in
calculating $\alpha_{\rm{OPT}}$ is 11.4$\pm$6.1, while the infrared
data is less rich and the mean number of photometry points used in
calculating $\alpha_{\rm{NIR}}$ is 4.3$\pm$0.7. The errors on the
slopes ($\alpha_{\rm{OPT}}$ and $\alpha_{\rm{NIR}}$) are the
standard errors of the linear fit. The measurement error on the
photometry is used in the fitting.

The major characteristics of the mixing diagram are shown in Figure
\ref{slp}. The E94 radio-quiet (RQ) mean SED template is shown by a
red cross. This template is bluer than almost all COSMOS XMM quasars
(Paper I), probably due to the $(U-B)$ selection criterion used to
select it (Schmidt \& Green 1983). The 16 galaxy
templates\footnotemark ~from the ``SWIRE Template Library" (Polletta
et al. 2007) are shown as blue triangles in the left panel of
Figure~\ref{slp}. Lines joining the E94 mean SED to three
representative galaxy templates are drawn. These mixing curves are
marked at 10\% intervals of host galaxy contribution (see
\S~\ref{s:fgal} for details).

\footnotetext{The 16 galaxy templates in the ``SWIRE Template
Library'' (Polletta et al. 2007) include: 3 elliptical galaxy
templates ``Ell2'', ``Ell5'', ``Ell13'' representing elliptical
galaxy of age 2~Gyr, 5~Gyr and 13~Gyr respectively; 7 spiral galaxy
templates ``S0'', ``Sa'', ``Sb'', ``Sc'', ``Sd'', ``Sdm'', ``Spi4'';
and 6 starburst galaxy templates ``NGC6090'', ``M82'', ``Arp220'',
``IRAS20551-4250'', ``IRAS22491-1808'', ``NGC6240''.}

Note that the slopes are defined in $\log\nu L_{\nu}$ versus
$\log\nu$ plane. Different SED shapes lie in different regions of
the mixing diagram, as sketched inside the circles in the four
corners of the left panel (wavelength increases to the left in these
circles): the {\em bottom right} corner shows the $1~\mu$m
inflection of an AGN dominated SED; the {\em upper left} corner
shows the cool starlight peak of a galaxy dominated SED; the {\em
bottom left} corner shows the rapid drop in the optical
characteristics of a dust reddening dominated SED.  The {\em top
right} corner shows an SED falling throughout the entire optical-NIR
range. This was not a known SED shape until the recent discovery of
``hot dust poor'' AGN (HDP hereinafter, Jiang et al. 2010; Hao et
al. 2010, 2011), which make up 10\% of the quasar population (Hao et
al. 2011).

We will discuss in detail the major characteristics of the mixing
diagram as shown in the right panel of Figure~\ref{slp} in the
following sub-sections.

\subsection{Quasar Templates}

Besides the E94 quasar SED template, there are several recent
updates (R06; Hopkins et al. 2007; Shang et al. 2011; Paper I). The
comparison of these SED templates were discussed in Paper I.

The R06 SED template was compiled from the Spitzer-SDSS sample,
containing 259 AGN and used a ``gap repair'' technique that replaces
the missing photometry with the normalized E94 mean SED to the
adjacent available photometry bands. Due to the limited coverage in
near-infrared, the R06 mean SED is therefore, by construction, very
similar to the E94 mean SED. Hopkins et al. (2007) simply combined
the R06 mean SED with the composite quasar SED (Vanden Berk et al.
2001), thus it has a shape similar to both R06 and E94. As we can
see in the right panel of Figure~\ref{slp}, the R06
($\alpha_{\rm{OPT}}=0.85\pm0.03$ and
$\alpha_{\rm{NIR}}=-0.60\pm0.05$) represented with a pentagon and
the Hopkins et al. (2007) template ($\alpha_{\rm{OPT}}=0.79\pm0.14$
and $\alpha_{\rm{NIR}}=-0.60\pm0.05$) represented with a hexagon are
very close to the E94 template ($\alpha_{\rm{OPT}}=0.95\pm0.04$ and
$\alpha_{\rm{NIR}}=-0.72\pm0.05$) represented with a cross.

The Shang et al. (2011) mean SED ($\alpha_{\rm{OPT}}=0.55\pm0.06$
and $\alpha_{\rm{NIR}}=-0.97\pm0.32$) was calculated using 27 nearby
bright radio-quiet quasars. As there is limited coverage in infrared
(only 3 points in the near-infrared range from the template), there
is a large error bar in the near-infrared slope calculation. The
Shang et al. (2011) template is represented with a triangle in the
right panel of Figure~\ref{slp}.

Paper I studied 413 XMM selected COSMOS type 1 AGN. Due to the X-ray
selection, there are more quasars in this sample having a large host
contribution (see also \S~\ref{s:mdsample}). 203 quasars in the
sample can be corrected for host galaxy contribution from the
Marconi \& Hunt (2003) scaling relationship adding an evolutionary
term (Bennert et al. 2010, 2011). The mean host-corrected SED of the
203 XMM-COSMOS quasars is represented with a square in the right
panel of Figure~\ref{slp}. We can see that there is still an
indication of a excess of host contribution, that remains
un-corrected. This is likely due to the dispersion in the scaling
relationship.

Given the similar location of these templates in the mixing diagram,
the results derived from the mixing diagram (e.g., host galaxy
fraction, reddening etc) would not be significantly affected if R06
or Hopkins et al. (2007) templates are chosen instead of E94. Shang
et al. (2011) and Paper I have other contamination factors in the
templates themselves that render them not proper to be chosen as the
pure quasar template. So we will use E94 template to represent pure
quasar SED template for future discussion in this paper.

\subsection{Galaxy Templates}

The 16 SWIRE galaxy SED templates are all from the observations of
various types of galaxies (Polletta et al. 2007). Theoretically,
models of the galaxy SEDs have been developed based on the stellar
population synthesis technique (Bruzual \& Charlot 2003, BC03
hereinafter). These models have been successfully used in SED
fitting especially in the optical range (e.g., Ilbert et al. 2009;
Ilbert et al. 2010; Bongiorno et al. 2012). However, these models do
not include the dust attenuation and re-radiation, and we are still
not sure if all the SEDs produced from these models exist in real
universe. Here we plot (Figure~\ref{slp}, right) the BC03 SED models
on the mixing diagram in comparison with the Polletta et al. (2007)
observed galaxy templates.

In the right panel of Figure~\ref{slp}, the galaxy SED model is
computed using the preferred Padova 1994 evolutionary tracks (Alongi
et al. 1993; Bressan et al. 1993; Fagotto et al. 1994 a, b; Girardi
et al. 1996) assuming a universal initial mass function (IMF) from
Chabrier (2003) and an exponentially declining star formation
history. The star formation rate $\psi(t)$ is expressed as
$\psi(t)=1M_{\bigodot}\tau^{-1}exp(-t/\tau)$, where $\tau$ is the
e-folding timescale. We show models for 16 different e-folding
timescales $\tau=(0.01, 0.05, 0.1, 0.3, 0.5, 0.6, 1, 2, 3, 5, 10,
15, 30, 50, 80, 100)$ Gyr, ranging a variety of star-formation
history.

The lines shown are for galaxies with ages ($t_{\rm{age}}$) running
from 0.1~Myr to 20~Gyr for each e-folding timescale model. The black
tick points on the lines are for a galaxy model with $\tau=1$~Gyr in
steps of $\log_{10} t_{\rm{age}}=0.1$. For different e-folding
timescales, the young galaxies ($t_{\rm{age}}<0.01$~Gyr) are quite
similar to each other. However, for older galaxies
($t_{\rm{age}}>0.1$~Gyr), the positions in this plot are quite
different for different e-folding timescales. For example, for
$\tau=0.01$~Gyr model (blue solid line), the oldest galaxy
($t_{\rm{age}}=20$~Gyr) reaches the leftmost region of the diagram
and for $\tau=100$~Gyr model (red solid line), the oldest galaxy
($t_{\rm{age}}=20$~Gyr) only reaches the Spi4 position. All the
galaxies in the $\tau=100$~Gyr model fail to overlap with the
observed Polletta et al. (2007) galaxy region. When we increase the
e-folding timescale, the position of the galaxies with the same age
at $t_{\rm{age}}>0.1$~Gyr move from the left to the right on the
diagram. The distances among the lines on the mixing diagram with
different e-folding timescales lie within the error bar of the
slopes except for the $\tau=0.01$~Gyr model, which show a wave at
$0.01~\rm{Gyr}<t_{\rm{age}}<1~\rm{Gyr}$. This wave also exists in
other lines with smaller size and at $1<\alpha_{\rm{OPT}}<1.5$. The
wave may caused by the molecular feature in the atmosphere of
cool/old stars.

From $t_{\rm{age}}>0.1$~Gyr, the $\alpha_{\rm{NIR}}$ values are
almost constant compared to the huge change in $\alpha_{\rm{OPT}}$.
This is reasonable because the near-infrared SED mainly comes from
the emission of old stars whereas the optical SED mainly comes from
the emission of young stars. As an exponentially declining star
formation history is assumed, when the galaxy gets old enough, the
star-formation rate is low, which means the young star population
becomes very small.  So the optical SED changes a lot, but the
near-infrared SED almost stays constant.

%%%%%%%%%%%%%%%%%%%%%%%%%%%%%%%%%%%%%%%
\begin{table*}
\begin{minipage}{\textwidth}\centering
\caption{Spectral slopes for different $f_g$ values (Mixing Curve)
assuming E94 mean SED as the pure quasar SED. $^1$ \label{t:mc}}
\begin{tabular}{c|c@{\hspace{1mm}}c|c@{\hspace{1mm}}c|c@{\hspace{1mm}}c|c@{\hspace{1mm}}c|c@{\hspace{1mm}}c}
\hline $f_g$ & \multicolumn{2}{c}{Spi4} & \multicolumn{2}{c}{Ell5} &
\multicolumn{2}{c}{Sb} & \multicolumn{2}{c}{S0} &
\multicolumn{2}{c}{NGC6090}
\\
 & $\alpha_{\rm{OPT}}$ & $\alpha_{\rm{NIR}}$ & $\alpha_{\rm{OPT}}$ &
 $\alpha_{\rm{NIR}}$ & $\alpha_{\rm{OPT}}$ & $\alpha_{\rm{NIR}}$ &
 $\alpha_{\rm{OPT}}$ & $\alpha_{\rm{NIR}}$ & $\alpha_{\rm{OPT}}$ & $\alpha_{\rm{NIR}}$
 \\ \hline
 0.0 &   0.950 &  -0.719 &   0.950 &  -0.719 &   0.950 &  -0.719 &   0.950 &  -0.719 &   0.950 &  -0.719 \\
 0.1 &   0.876 &  -0.621 &   0.857 &  -0.618 &   0.862 &  -0.617 &   0.860 &  -0.617 &   0.864 &  -0.620 \\
 0.2 &   0.794 &  -0.513 &   0.752 &  -0.506 &   0.762 &  -0.506 &   0.759 &  -0.506 &   0.766 &  -0.511 \\
 0.3 &   0.702 &  -0.394 &   0.634 &  -0.382 &   0.648 &  -0.383 &   0.641 &  -0.383 &   0.655 &  -0.392 \\
 0.4 &   0.598 &  -0.262 &   0.496 &  -0.244 &   0.515 &  -0.246 &   0.505 &  -0.245 &   0.525 &  -0.260 \\
 0.5 &   0.480 &  -0.112 &   0.334 &  -0.086 &   0.357 &  -0.090 &   0.341 &  -0.089 &   0.372 &  -0.111 \\
 0.6 &   0.342 &   0.060 &   0.138 &   0.097 &   0.164 &   0.089 &   0.140 &   0.091 &   0.184 &   0.059 \\
 0.7 &   0.178 &   0.263 &  -0.109 &   0.316 &  -0.082 &   0.302 &  -0.119 &   0.304 &  -0.053 &   0.257 \\
 0.8 &  -0.024 &   0.509 &  -0.439 &   0.586 &  -0.415 &   0.563 &  -0.475 &   0.566 &  -0.373 &   0.497 \\
 0.9 &  -0.286 &   0.823 &  -0.934 &   0.941 &  -0.925 &   0.902 &  -1.037 &   0.906 &  -0.857 &   0.800 \\
 1.0 &  -0.656 &   1.264 &  -1.972 &   1.469 &  -2.044 &   1.396 &  -2.447 &   1.403 &  -1.876 &   1.218 \\
\hline
\end{tabular}\\
$^1$ A portion of the table is shown here for guidance. The complete
table for 16 SWIRE galaxy templates will be available online.
\end{minipage}
\end{table*}
%%%%%%%%%%%%%%%%%%%%%%%%%%%%%%%%%%%%%%%%%%%%%

Compared to SWIRE galaxy templates, the BC03 models show very blue
galaxies located on the upper right corner of the mixing diagram and
show less spread in the $\alpha_{\rm{NIR}}$ direction. The blue
galaxies are typically very young and they would be expected to
contain large amount of gas. These galaxies are very rare in the
redshift range of current major surveys and tend to be more common
for high reshifts ($z>6$, Bouwens et al. 2012). These very blue
galaxies are so extremely short-lived that they are expected to be a
very small fraction in any sample of galaxies. For example, in
Ilbert et al. (2010), SED fitting analysis is performed to $\sim
200,000$ IRAC selected galaxies with $0.2<z<2$ in the COSMOS field.
They find that only a few percent have $t_{age}<0.5$~Gyr and most of
them are fitted with a significant extinction.

One would suspect that the presence of a young population on top of
an older population could mimic a blue non-thermal quasar-like
optical spectrum. However, in practice, the chance is low, because,
to reach the slope, the star-formation rate of the young population
would be too extreme. Also, if this were commonly true, optically
selected type 1 AGN would be severely diluted by starbursts, which
is not seen.

The possible presence of very young galaxy models also provides an
alternative explanation to the HDP AGN SED that is the normal quasar
SED with large fraction of young host galaxy. But this explanation
would require the quasar to be active simultaneously with a strong
starburst, which is not seen in large samples (e.g. Kewley et al.
2006, Schawinski et al. 2009, Wild et al. 2010).

In this paper, to be consistent with the quasar template we use
(E94, which has been derived from observed SEDs), we only use the
observed Polletta et al. (2007) SWIRE galaxy templates for further
discussion.

\subsection{Galaxy Fraction Mixing Curves}
\label{s:fgal}

We can quantify the host galaxy contribution fraction $f_g$ at
$1~\mu$m for any quasar, assuming that the E94 RQ template
represents a pure AGN SED. A definition similar to $f_g$ is widely
used in SED fitting with different normalization wavelengths (e.g.
Salvato et al. 2009, Merloni et al. 2010). The parameter $f_g$ is
defined as the galaxy fraction at $1~\mu$m, and describes how close
the observed SED is to the galaxy templates. First, we normalize
both the galaxy and AGN template at $1~\mu$m. Then the mixture of
some fraction of galaxy ($f_g$) and some fraction of AGN (1-$f_g$)
emission can be calculated accordingly. Suppose that at frequency
$\nu$ the galaxy template SED luminosity is $\nu L^{G}_{\nu}$ and
the AGN template (E94) SED luminosity is $\nu L^{A}_{\nu}$, then the
mixing of the two SEDs luminosity is $$\nu L^{mix}_{\nu}=f_g\nu
L^{G}_{\nu}+(1-f_g)\nu L^{A}_{\nu}$$

The black curves in Figure~\ref{slp} show the slopes of SED
templates obtained by mixing the AGN and galaxy templates with
values of $f_g= 0 - 1$. The mixing curves of the starburst galaxy
``NGC6090'' and the spiral galaxy ``Spi4'' define the red and blue
boundaries of the possible slopes obtained by mixing the E94 SED
with all 16 galaxy templates in the SWIRE library. The spectral
slopes for mixtures of E94 with Spi4, Ell5, Sb, S0, and NGC6090 for
11 values of $f_g$ are listed in Table \ref{t:mc}. The complete
table for all the 16 templates is available on line.

%%%%%%%%%%%%%%%%%%%%%%%%%%%%%%%%%%%%%
\subsection{Reddening Vectors}
\label{s:redvector}

%%%%%%%%%%%%%%%%%%%%%%%%%%%%%%%%%%%%%%%%%%%%%
\begin{table*}
\begin{minipage}{\textwidth}
\centering \caption{Reddening Vector for E94\label{t:redv}}
\begin{tabular}{c|c@{\hspace{1mm}}c|c@{\hspace{1mm}}c|c@{\hspace{1mm}}c|c@{\hspace{1mm}}c}
\hline Ext. Law & \multicolumn{2}{c}{SMC} & \multicolumn{2}{c}{LMC}
& \multicolumn{2}{c}{MW} &
\multicolumn{2}{c}{Calz}\\
$E(B-V)$ & $\alpha_{\rm{OPT}}$ & $\alpha_{\rm{NIR}}$ &
$\alpha_{\rm{OPT}}$ & $\alpha_{\rm{NIR}}$ & $\alpha_{\rm{OPT}}$ &
$\alpha_{\rm{NIR}}$ & $\alpha_{\rm{OPT}}$ &
$\alpha_{\rm{NIR}}$ \\
\hline

 0.0 &  0.950 & -0.719 &  0.950 & -0.719 &  0.950 & -0.719 &  0.950 & -0.719 \\
 0.1 &  0.596 & -0.798 &  0.596 & -0.798 &  0.604 & -0.807 &  0.553 & -0.890 \\
 0.2 &  0.242 & -0.877 &  0.242 & -0.877 &  0.257 & -0.895 &  0.156 & -1.060 \\
 0.3 & -0.112 & -0.956 & -0.112 & -0.956 & -0.089 & -0.982 & -0.241 & -1.231 \\
 0.4 & -0.465 & -1.034 & -0.465 & -1.034 & -0.435 & -1.070 & -0.639 & -1.402 \\
 0.5 & -0.819 & -1.113 & -0.819 & -1.113 & -0.781 & -1.158 & -1.036 & -1.572 \\
 0.6 & -1.173 & -1.192 & -1.173 & -1.192 & -1.128 & -1.246 & -1.433 & -1.743 \\
 0.7 & -1.527 & -1.271 & -1.527 & -1.271 & -1.474 & -1.333 & -1.830 & -1.914 \\
 0.8 & -1.881 & -1.350 & -1.881 & -1.350 & -1.820 & -1.421 & -2.227 & -2.084 \\
 0.9 & -2.235 & -1.428 & -2.235 & -1.428 & -2.166 & -1.509 & -2.624 & -2.255 \\
 1.0 & -2.588 & -1.507 & -2.588 & -1.507 & -2.513 & -1.597 & -3.021 & -2.426 \\

\hline
\end{tabular}
\end{minipage}
\end{table*}
%%%%%%%%%%%%%%%%%%%%%%%%%%%%%%%%%%%%%%%%%%%%%%

Intrinsic reddening in AGN is often important in defining their SEDs
(e.g., Ward et al. 1987; O'Brien et al. 1988; Young et al. 2008;
Shang et al. 2011). The magenta arrows in Figure~\ref{slp} show
$\alpha_{\rm{OPT}}$ and $\alpha_{\rm{NIR}}$ for the E94 SED when
reddened by $E(B-V) = 0 - 1~\rm{mag}$.

We consider four different reddening laws: Small Magellanic Cloud
(SMC), Large MC (LMC), Milky Way (MW), and Calzetti et al. (2000,
Calz). For the SMC, LMC, and MW reddening laws, the reddening of the
E94 SED is derived with the IDL de-reddening routines
\texttt{`FM\_UNRED.PRO'}(for SMC and LMC) and
\texttt{`CCM\_UNRED.PRO'} (for MW), which all use the Fitzpatrick
(1999) parameterizations of the SMC (Gordon et al. 2003), LMC
(Misselt et al. 1999), and the MW (Cardelli et al. 1989; O'Donnell
1994) extinction curves. For the Calz reddening law, the reddening
is derived with IDL de-reddening routine \texttt{`CALZ\_UNRED.PRO'},
which uses the Calzetti et al. (2000) recipe developed for galaxies
where massive stars dominate the radiation output. The SMC reddening
law (Gordon et al. 2003) is typically used for quasars, and is shown
to fit reddening in quasars more effectively than a LMC or MW
reddening law (Hopkins et al. 2004; Richards et al. 2003). Reddening
primarily affects $\alpha_{\rm{OPT}}$. The effect of reddening is
reported in Table \ref{t:redv}.

Using the reddening vector, we can estimate $f_g$ and $E(B-V)$ from
the mixing diagram for sources lying off the E94-host mixing curves
toward the lower left. For each source we can draw a line parallel
to the reddening curve (black dashed line in Figure~\ref{slp}). The
crossing point of this line and the mixing curve shows approximately
the value of $f_g$. The length of the parallel line gives an
estimate of $E(B-V)$. We use this technique in \S~\ref{s:fg} and
\S~\ref{s:febv}.

Different reddening laws could cause different $f_g$ and $E(B-V)$
derived from the mixing diagram. As we can see from the right panel
of Figure~\ref{slp}, the SMC, LMC, and MW reddening vectors are
closely similar to each other. The Calz reddening vector is
significantly different from the other three reddening laws. This
reddening law is generally used for star-forming galaxies (Calzetti
et al. 2000) not AGN. Gordon et al. (2003) performed a comparison
between SMC, LMC, and MW reddening laws and found that the
extinction curves only begin to diverge shortward of $\sim 2000$\AA\
and at rest frame near-UV ($\sim 2500$\AA) through near infrared
($\sim 1~\mu$m), the three laws are extremely similar. So for the
rest frame wavelength range in which the mixing diagram is defined
(3000\AA\ to $1~\mu$m and $1~\mu$m to $3~\mu$m), the results will
not be significantly different if we choose either SMC, LMC or MW
reddening law. We will only consider the SMC reddening law in the
following discussion.

%%%%%%%%%%%%%%%%%%%%%%%%%%%%%%%%%%%%%%%%%%%%%%%%%%%%%%%%%%%%%%%%%%%%%%%%%%
\section{Mixing Diagram for Type 1 AGN Samples}

\subsection{Type 1 AGN Samples}
\label{s:sample}
%%%%%%%%%%%%%%%
\begin{figure}
\epsfig{file=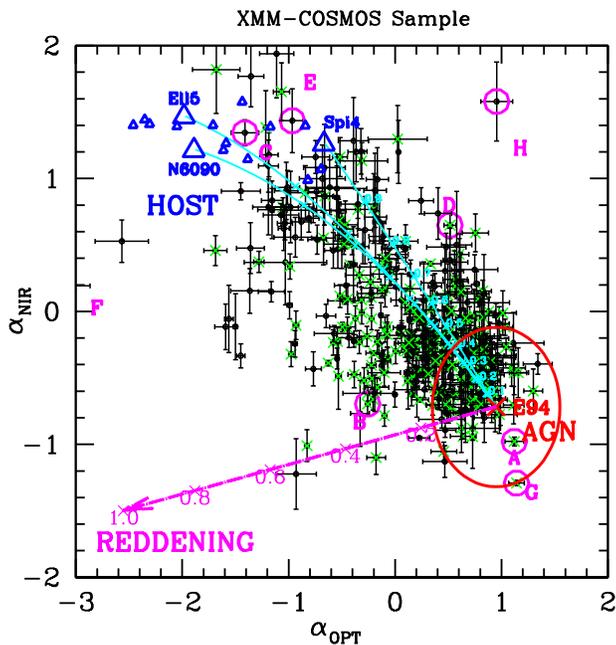, angle=0,width=\linewidth} \caption{The
mixing diagram of the XMM-COSMOS type 1 AGN sample (XC413).  The
other points and lines are color-coded as in Figure~\ref{slp}. The
green crosses show the 206 quasars with black hole mass estimates
(the SS206 sub-sample, Hao et al. 2013a). \label{slpxmm}}
\end{figure}
%%%%%%%%%%%%%%%

%%%%%%%%%%%%%%%%%%%%%%%
\begin{figure*}
\includegraphics[angle=0,width=0.45\textwidth]{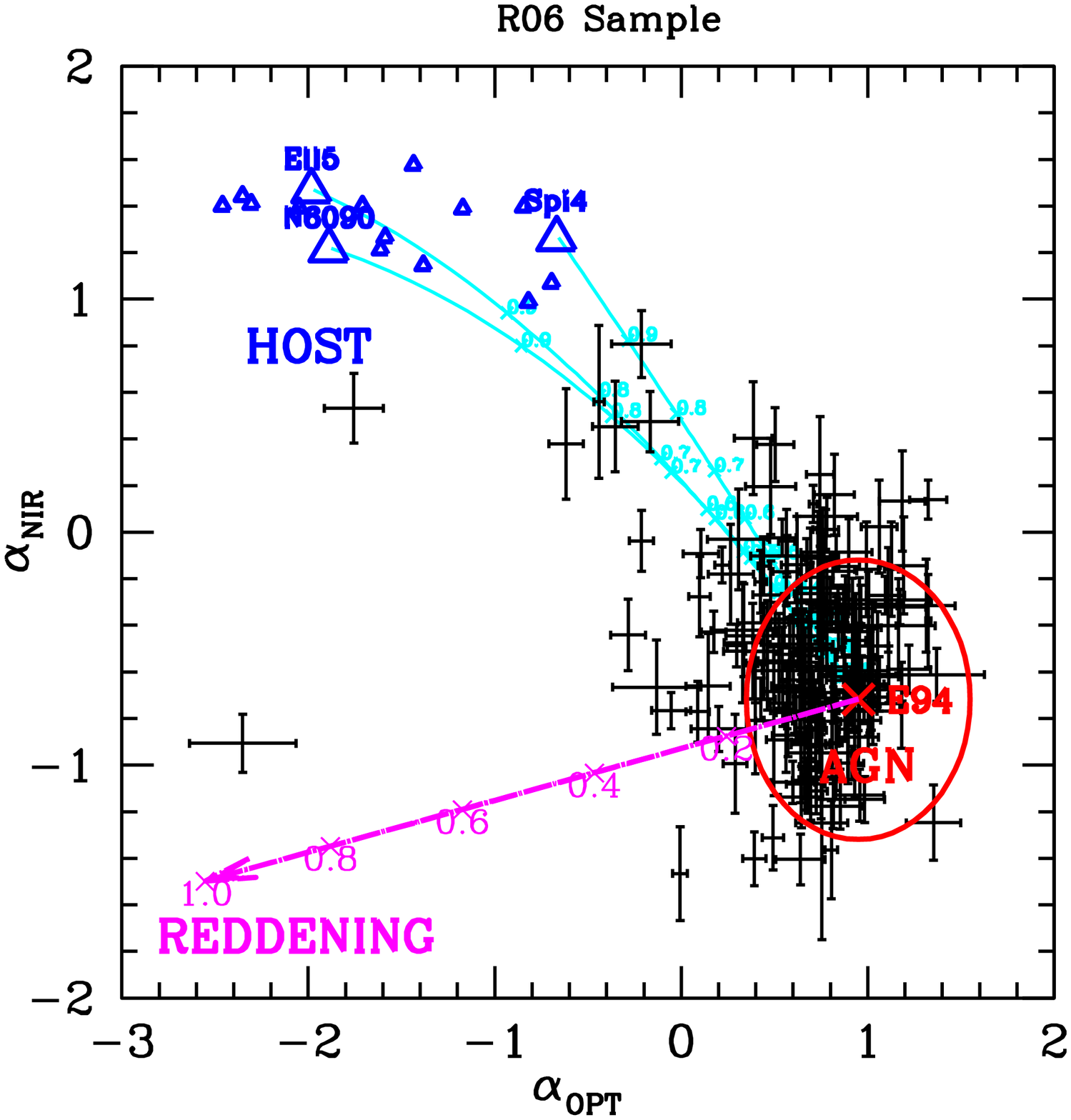}
\includegraphics[angle=0,width=0.45\textwidth]{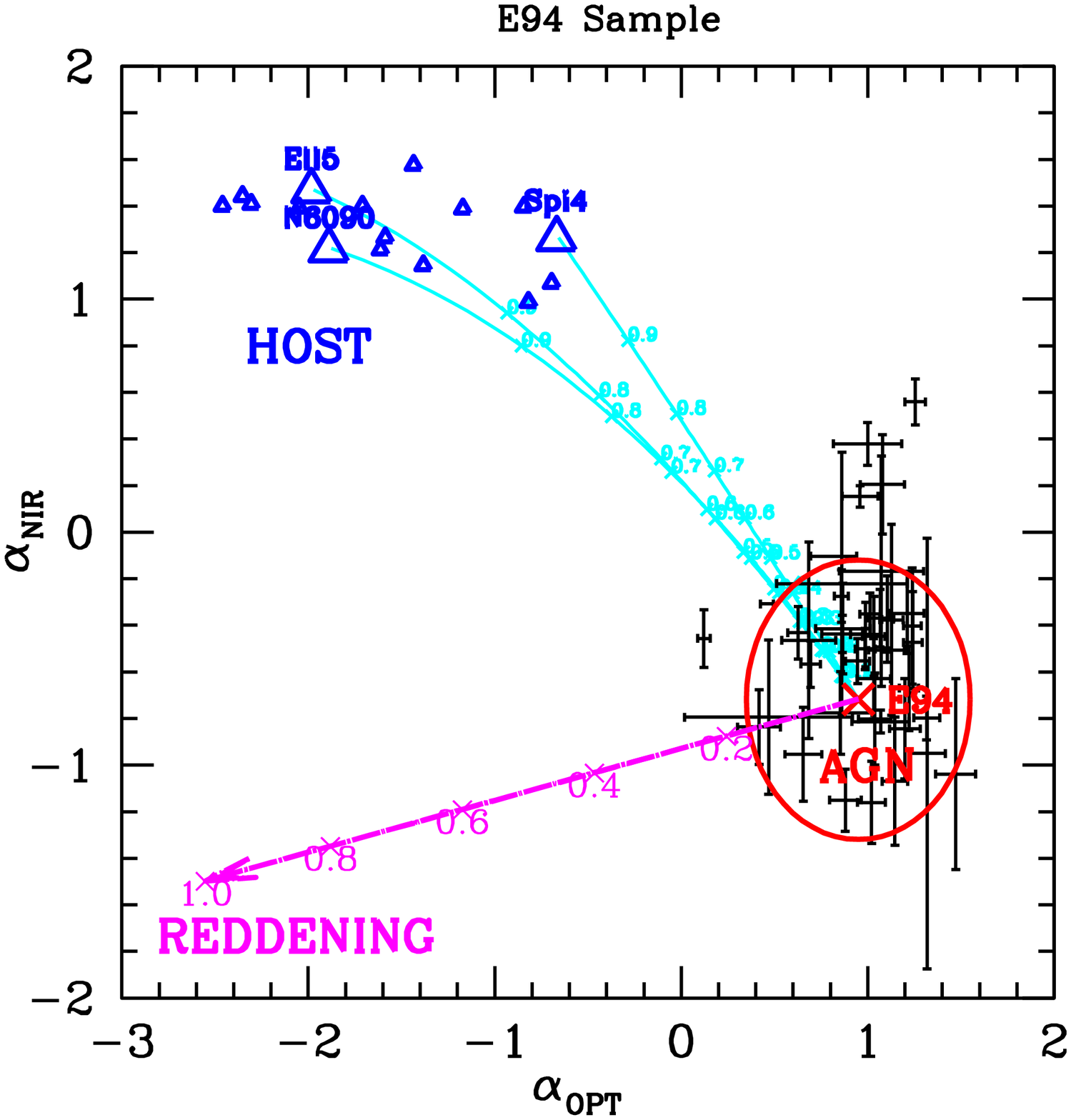}
\caption{The mixing diagram of the SDSS-Spitzer quasar sample (R06,
left) and the bright quasar sample (E94, right). The other points
and lines are color-coded as in Figure~\ref{slpxmm}.
\label{slpR06E94}}
\end{figure*}
%%%%%%%%%%%%%%%

The three type 1 AGN samples we used in this paper are:

(1) The XMM-COSMOS type 1 AGN Sample (XC413, Paper I). The COSMOS
field (Scoville et al. 2007) was imaged in X-rays with
\emph{XMM-Newton} for a total of $\sim1.5$~Ms (Hasinger et al. 2007;
Cappelluti et al. 2007, 2009). Optical identifications were made by
Brusa et al. (2007, 2010) for the entire XMM-COSMOS sample, who gave
photometric properties and redshifts for each X-ray point source.
From this complete sample, we extracted a sample of 413 type 1 AGN,
defined by having broad line FWHM$>$2000 km s$^{-1}$. The XC413
catalog was described in detail in Paper I.

This sample has full wavelength coverage from radio to X-ray (for a
total of 43 photometric bands, Paper I) and high confidence level
spectroscopic redshifts (Trump et al. 2009; Schneider et al. 2007;
Lilly et al.  2007, 2009). In this paper, we also add the recently
released H band photometry from CFHT/WIRCAM (McCracken et al. 2010).
Now 405 out of the 413 XMM-COSMOS quasars have H band photometry,
compared to 252 out of 413 in Paper I. As described in Paper I, the
photometric data obtained from different telescopes and with
different seeing were matched and the aperture fluxes were all
transformed to total flux according to the point spread function
simulation for each telescope (e.g., Capak et al. 2007, Brusa et al.
2007). As in Paper I, in order to reduce the extra error in the SED
slope measurement that can be caused by variability of quasars, we
used only the optical photometric data obtained in a shorter time
period (2004-2008) close to the time of the infrared Spitzer-IRAC
data. The COSMOS type 1 AGN sample has an extremely rich coverage
(36 bands) in the optical to near-infrared range. The objects have
redshifts $0.1\leq z \leq4.3$ and magnitudes $16.9\leq
i_{AB}\leq24.8$, with 94\% - 98\% being radio-quiet (Hao et al.
2013b).

In this sample, 206 quasars have published black hole mass
measurements (Trump et al. 2009b; Merloni et al. 2010), which are
based on the scaling relationship between broad emission line (BEL)
FWHM and black hole mass (Vestergaard 2004). For the quasars with
only zCOSMOS spectrum, the black hole mass was estimated for only
those with MgII lines in the spectrum (Merloni et al. 2010), using
the calibration of McLure \& Jarvis (2002). For the rest of the
sample, the BELs are located close to the ends of the spectra, so
reliable black hole mass estimates cannot be made. We call the
sub-sample with black hole mass estimates SS206 (`SS' stands for
sub-sample) hereinafter.

(2) The SDSS-Spitzer Sample (R06). The R06 sample consists of 259
{\it Spitzer} sources identified with Sloan Digital Sky Survey
(SDSS) quasars in four different degree-scale fields, and is,
therefore, mid-IR identified and optically selected. The redshift
range covered is $z=0.14 - 5.2$ with 93\% being at $z<3$. Most
(215/259) of the R06 sources did not have 2MASS J H K photometry.
Details about how we measured the slopes with this sample were
described in Hao et al. (2011).

(3) The bright quasar sample  (E94). This sample consists of 42
quasars in the redshift range $z= 0.025 - 0.94$, with 80\% of them
being at $z<0.3$. The optical photometry was obtained at the FLWO
(F. L. Whipple Observatory) 24 inch telescope within one week of the
MMT FOGS (Faint Object Grism Spectrograph) spectroscopic
observations. The NIR data were obtained with MMT and IRTF. More
details on the observation can be found in E94. The E94 SEDs have
been corrected for host galaxy contamination by subtracting the host
galaxy template SED based on the Sbc galaxy model of Coleman et al.
(1980). The E94 sample has bolometric luminosities (log$L_{bol}$) in
the range of $44.6-47.2$~erg/s with mean of 45.75~erg/s. We
recalculated the E94 bolometric luminosities with the same
cosmological parameters used for XC413. Compared to XC413, E94
sample is on average more luminous and contains less low luminosity
quasars than the XC413. The Eddington ratio of the PG quasars
(including E94 sample) is comparable to that of the XC413 (Sikora et
al. 2007, Paper II).

\subsection{Mixing Diagram for the Quasar Samples}
\label{s:mdsample}

We plot the XC413 sample on the mixing diagram in
Figure~\ref{slpxmm}. The distribution is continuous and largely lies
between the E94 mean SED and the galaxy templates, along the mixing
curves, with some spread in the reddening direction to values as
large as $E(B-V)\sim0.6$, but mostly with $E(B-V)<0.3$. The green
crosses represent objects with black hole mass estimates (Paper II),
which span the range of the entire sample in the mixing diagram.

The diagram shows that about 90\% of the sources lie in the left
hand triangular `mixing wedge' between the mixing curves and the
reddening vector.  The SEDs of these AGN can be accounted for with a
simple combination of an E94 quasar SED, plus a galaxy contribution
and reddening. This suggests that the AGN sample is consistent with
a single intrinsic SED shape, closely resembling the E94 mean quasar
SED (see also in \S~\ref{s:evolution}, where we compare in detail
the XC413 and the E94 quasar sample).

There are several sources outside the wedge, which are outliers with
respect to the bulk of the type 1 AGN population (see
\S~\ref{s:outliers} for details). As the galaxy SED dispersion is
expected to be broader than the 16 Polletta templates, it is not
surprising to see three sources (XID=4, 1559, 5617) beyond the SWIRE
galaxy template region (still within 1$\sigma$) that would formally
require $f_g>1$. We excluded these sources when using the mixing
diagram to calculate $f_g$ (see \S~\ref{s:fg} for details).

We also plot the mixing diagram for the SDSS-spitzer quasar sample
(R06) and the bright quasar sample (E94), shown in
Figure~\ref{slpR06E94} (see also Hao et al. 2011). For the
optically-selected R06 sample, quasars by selection are more
clustered in the quasar dominated region unlike the X-ray-selected
XC413, which includes more sources with low quasar to host galaxy
contrast. The E94 quasars have been corrected for host galaxy
contribution. Thus they are, by construction, clustered around the
E94 mean (red cross) in the quasar dominated region. Compared to the
E94 sample, the R06 sample is not as blue in the optical.

%%%%%%%%%%%%%%%%%%%%%%%%%%%%%%%%%%%%%%%%%%%%%%%%%%%%
\subsection{Intrinsic Slope Dispersion}
\label{s:intsdisp}
%%%%%%%%%%%%%%%%%%%%%%%
\begin{figure*}
\includegraphics[angle=0,width=0.32\textwidth]{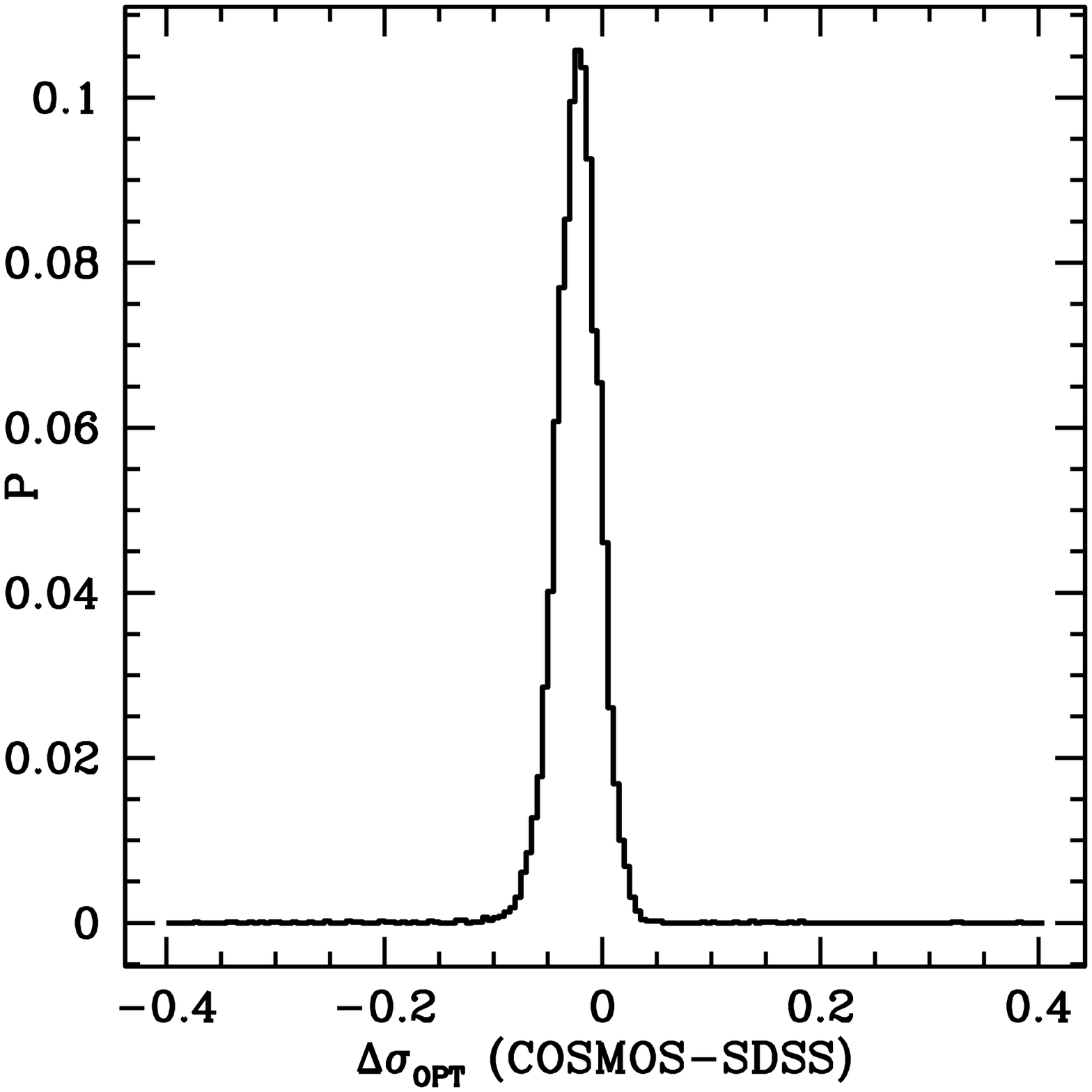}
\includegraphics[angle=0,width=0.32\textwidth]{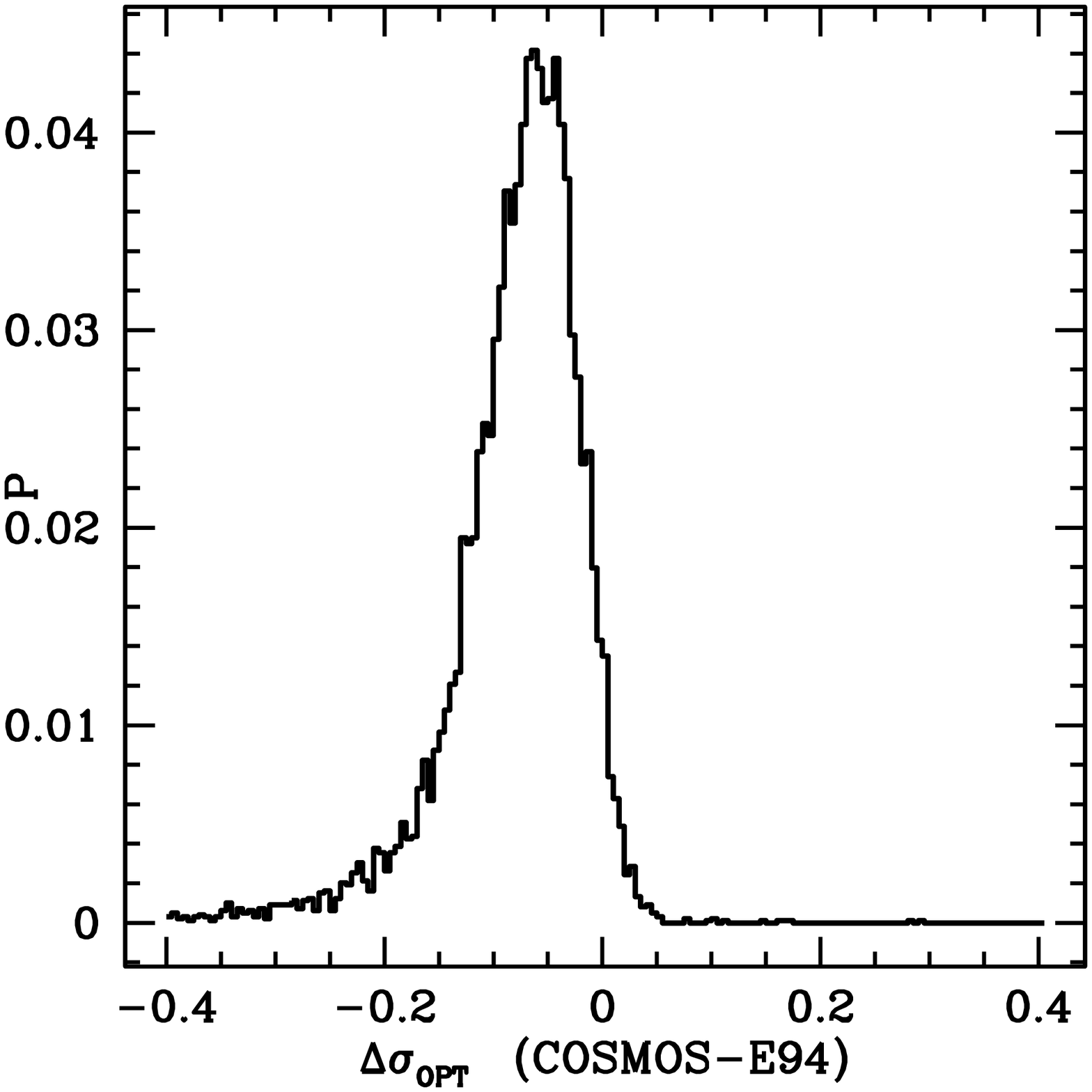}
\includegraphics[angle=0,width=0.32\textwidth]{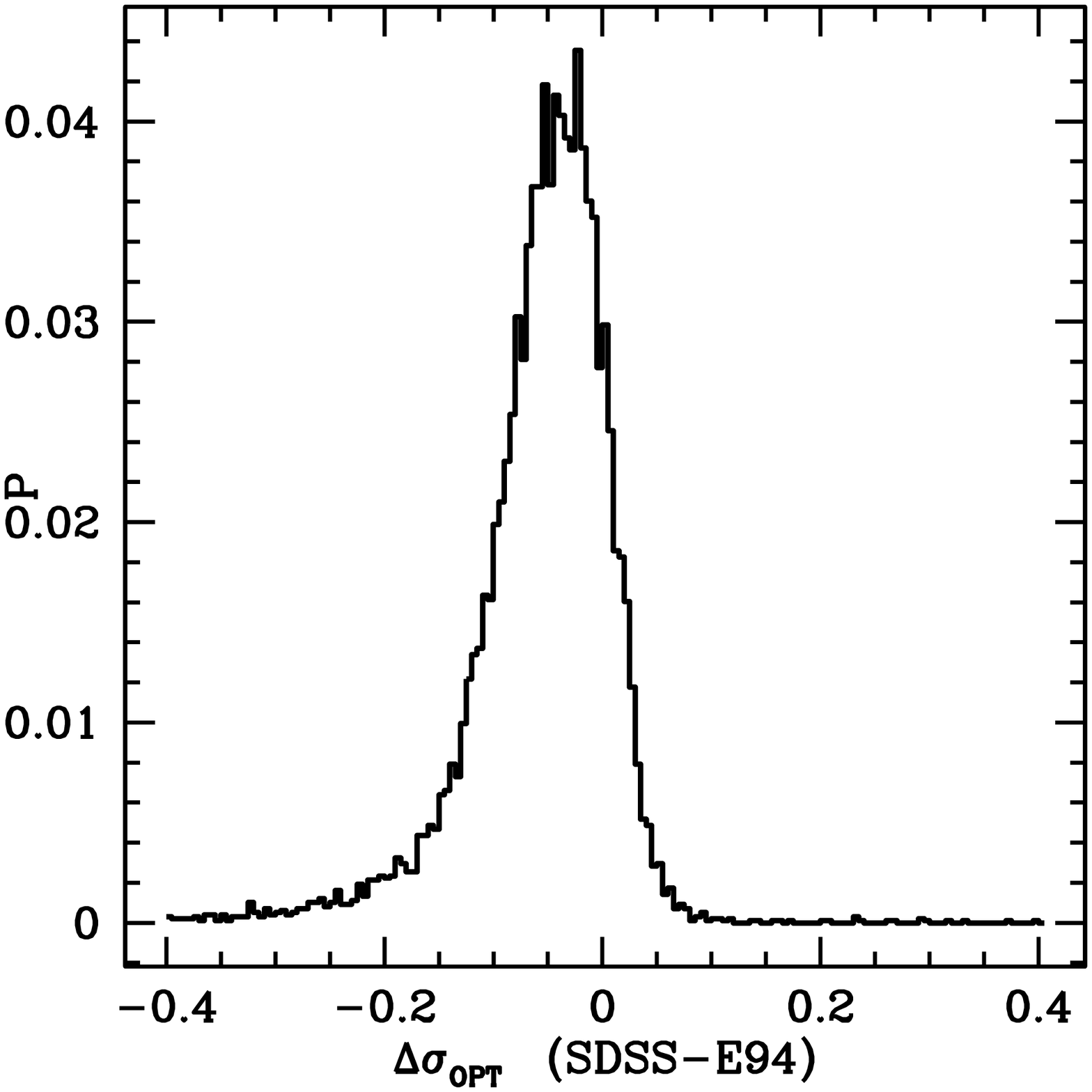}
\includegraphics[angle=0,width=0.32\textwidth]{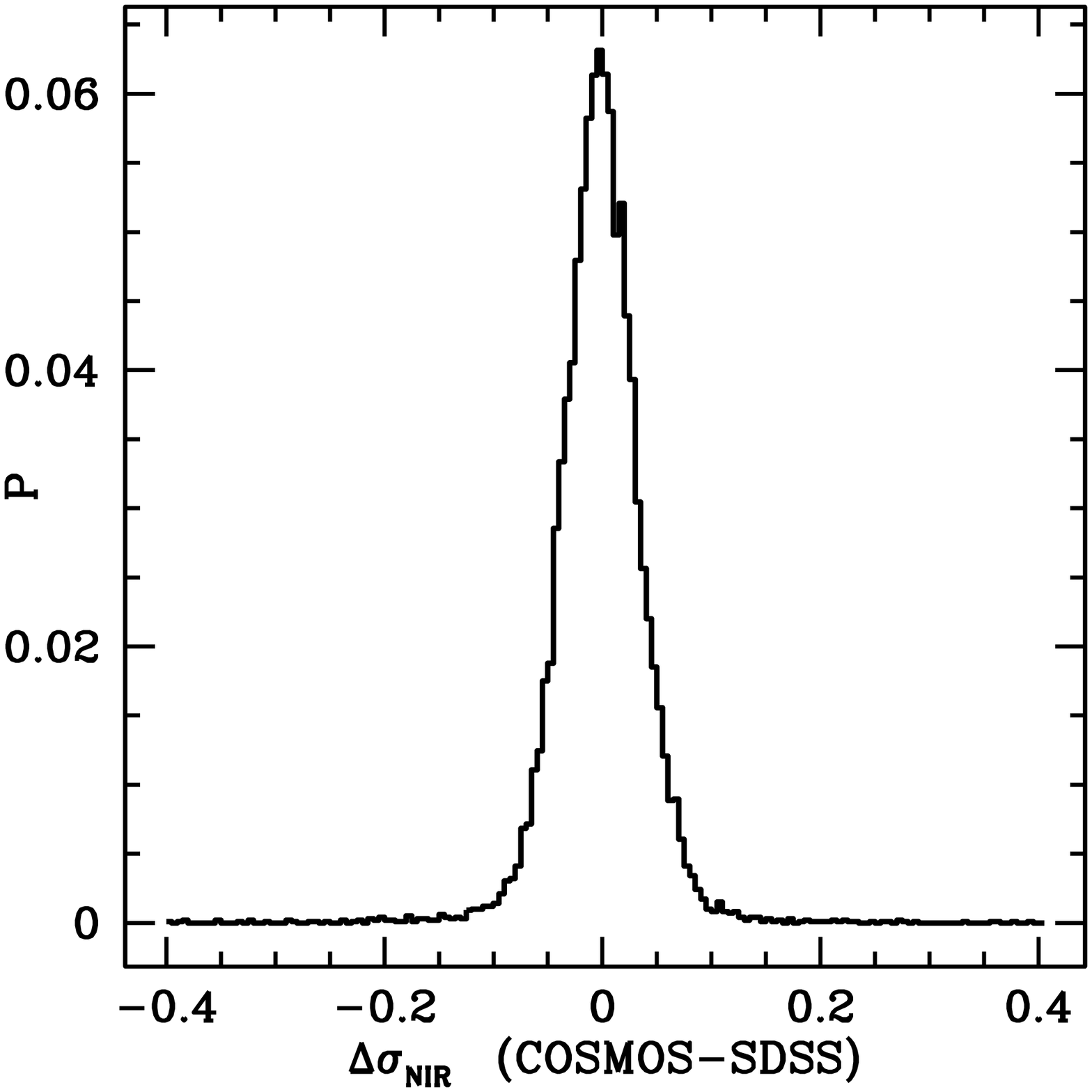}
\includegraphics[angle=0,width=0.32\textwidth]{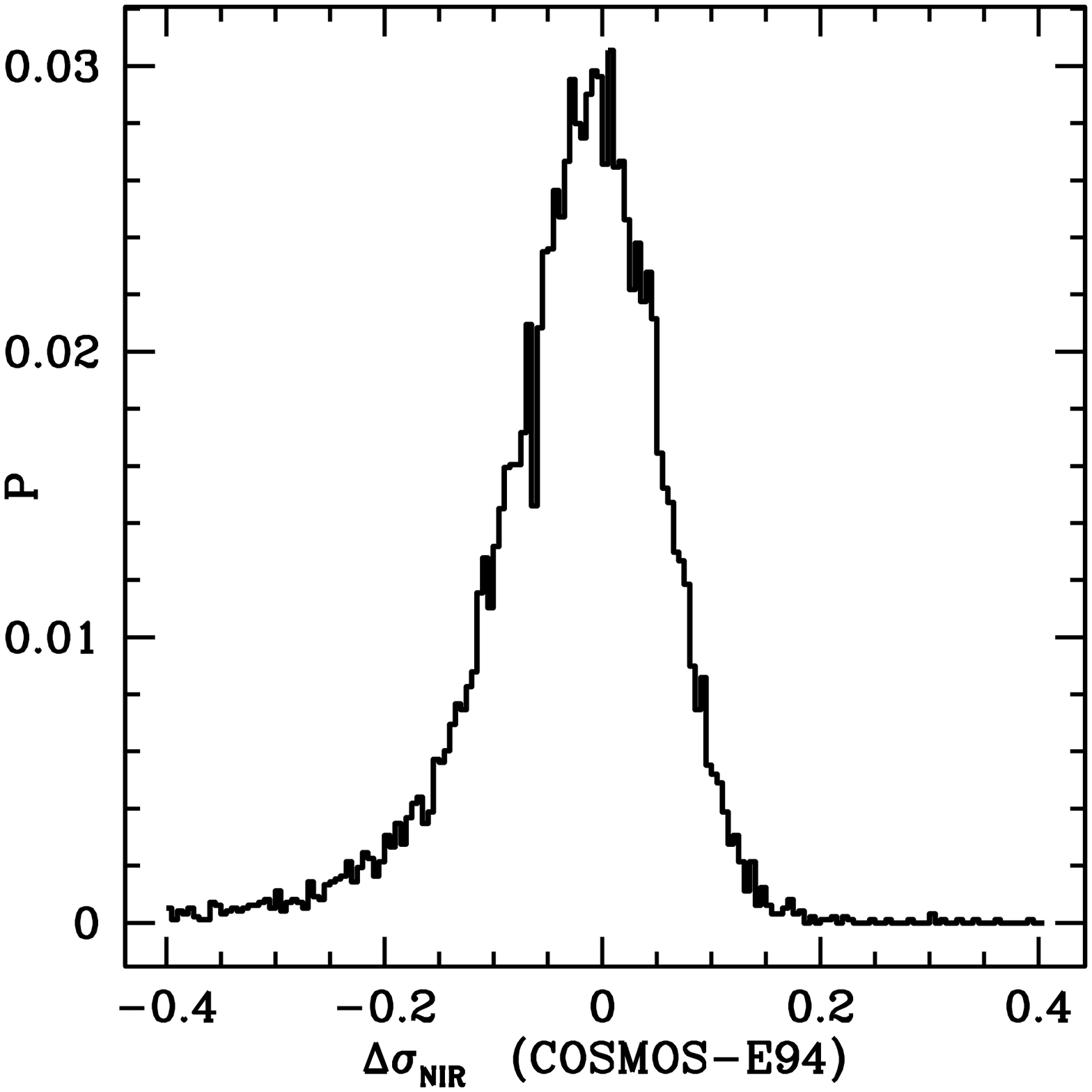}
\includegraphics[angle=0,width=0.32\textwidth]{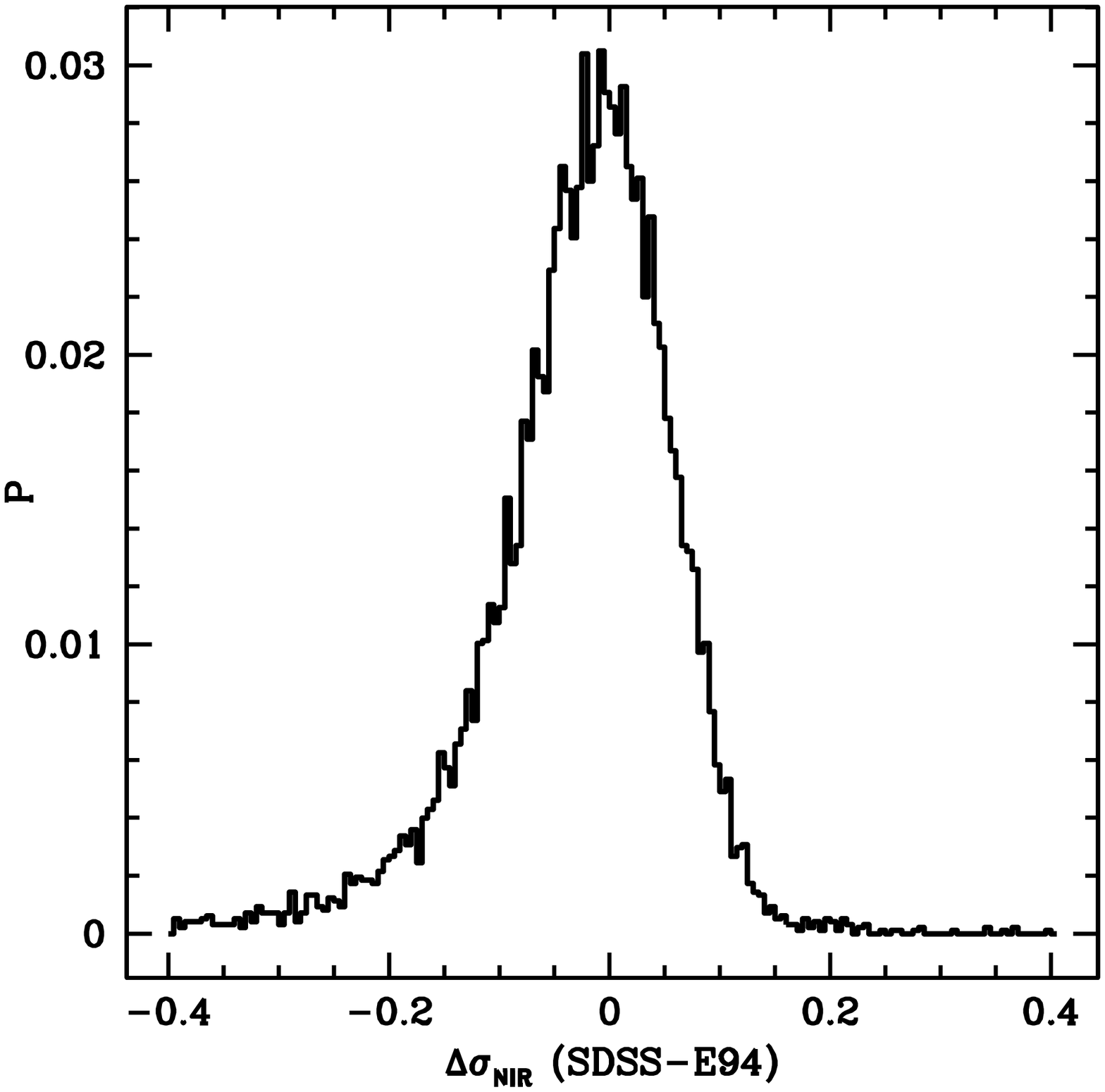}
\caption{The probability distribution of the difference in slope
dispersion in the three samples.\label{sigcmp}}
\end{figure*}
%%%%%%%%%%%%%%%%%%%%%%

We selected a sub-sample of AGN-dominated XC413 SEDs with
$\alpha_{\rm{OPT}}>0.2$ in order to exclude galaxy- or
reddening-dominated sources. This sample has a mean slope
$\bar{\alpha}_{\rm{OPT}}$=0.63 (standard deviation
$\sigma_{\rm{OPT}}$=0.24), and $\bar{\alpha}_{\rm{NIR}}=-0.31$
(standard deviation $\sigma_{\rm{NIR}}$=0.36). The E94 RQ mean
(${\alpha}_{\rm{OPT}}$(E94) = 0.95,
${\alpha}_{\rm{NIR}}$(E94)$=-0.72$, see values in Table~\ref{t:mc}
for $f_g$=0) lies at the extreme blue end of the distribution.

To estimate the intrinsic dispersion within the AGN-dominated XC413
sub-sample we removed the effect of measurement error, namely:
$\sigma_{\rm{INT}} = \sqrt{{\sigma}^2-Err^2}$. The mean of the
measurement error for $\alpha_{\rm{OPT}}$ is $Err_{\rm{OPT}}$ =
0.09, and for $\alpha_{\rm{NIR}}$ is $Err_{\rm{NIR}}$ = 0.12. The
intrinsic dispersion thus is $\sigma_{\rm{INT,OPT}}$ = 0.22 and
$\sigma_{\rm{INT,NIR}}$ = 0.34 respectively. Therefore, the
intrinsic dispersion of the SED shape is 2 - 3 times the measurement
error and seems to be significant.

The equivalent intrinsic dispersions in the E94 and R06 sample were
estimated by Hao et al. (2011), who found: $\sigma_{\rm{E94, INT,
OPT}}$ = 0.25, $\sigma_{\rm{E94, INT, NIR}}$ = 0.32,
$\sigma_{\rm{R06, INT, OPT}}$ = 0.23, and $\sigma_{\rm{R06, INT,
NIR}}$ = 0.36, respectively. The intrinsic dispersions are thus
similar for all the three samples.

To compare the intrinsic dispersion of these three samples more
rigorously, we applied the Bayesian method of Kelly et al. (2007).
This assumes that the intrinsic distribution of the slopes is a
mixture of Gaussians. The probability distributions of the
differences in slope dispersion between the samples are shown in
Figure~\ref{sigcmp}. For the dispersion in $\alpha_{\rm{OPT}}$, the
significance of the difference between the XMM-COSMOS and R06 sample
is 0.16$\sigma$; between XMM-COSMOS and E94 sample is 0.11$\sigma$
and between R06 and E94 sample is 0.08$\sigma$. Therefore, the
intrinsic dispersions of the $\alpha_{\rm{OPT}}$ are consistent with
being the same for all the three samples. For $\alpha_{\rm{NIR}}$,
the significance of the difference between the XMM-COSMOS and R06
sample is 0.02$\sigma$; between the XMM-COSMOS and E94 sample is
0.08$\sigma$ and between the R06 and E94 sample is 0.06$\sigma$. As
in the simpler analysis, the intrinsic dispersions of the
$\alpha_{\rm{OPT}}$ and $\alpha_{\rm{NIR}}$ are consistent with
being the same for all the three samples.

Using this result we can create a more rigorous AGN-dominated sample
using the intrinsic dispersion to define a radius in the
($\alpha_{\rm{OPT}}$, $\alpha_{\rm{NIR}}$) plane within which such
AGN must lie. As the distribution of the quasars is continuous,
different radii define different populations of quasars. We define a
circle centered on the E94 RQ mean SED template with a radius of 0.6
on the mixing diagram to define AGN-dominated sources. This is
approximately 3$\sigma_{\rm{OPT}}$ and 1.5$\sigma_{\rm{NIR}}$ of the
intrinsic dispersion. Note that the AGN dominated circle chosen here
is somewhat arbitrary and is used just for illustration. Different
radii or even shapes of the AGN-dominated region can be chosen for
different purposes. For the XC413 sample, the sources within the
dispersion circle populate mainly the left upper quadrant, similar
to the R06 sample, but unlike the host-corrected E94 sample. We will
discuss this more in section \S~\ref{s:evolution}.

%%%%%%%%%%%%%%%%%%%%%%%%%%%%%%%%%%%%%
\section{Application of the Mixing Diagram to the XC413 Sample}
\subsection{SED Evolution on the Mixing Diagram} \label{s:evolution}
%%%%%%%%%%%%%%%%%%
\begin{figure*}
\includegraphics[angle=0,width=0.245\textwidth]{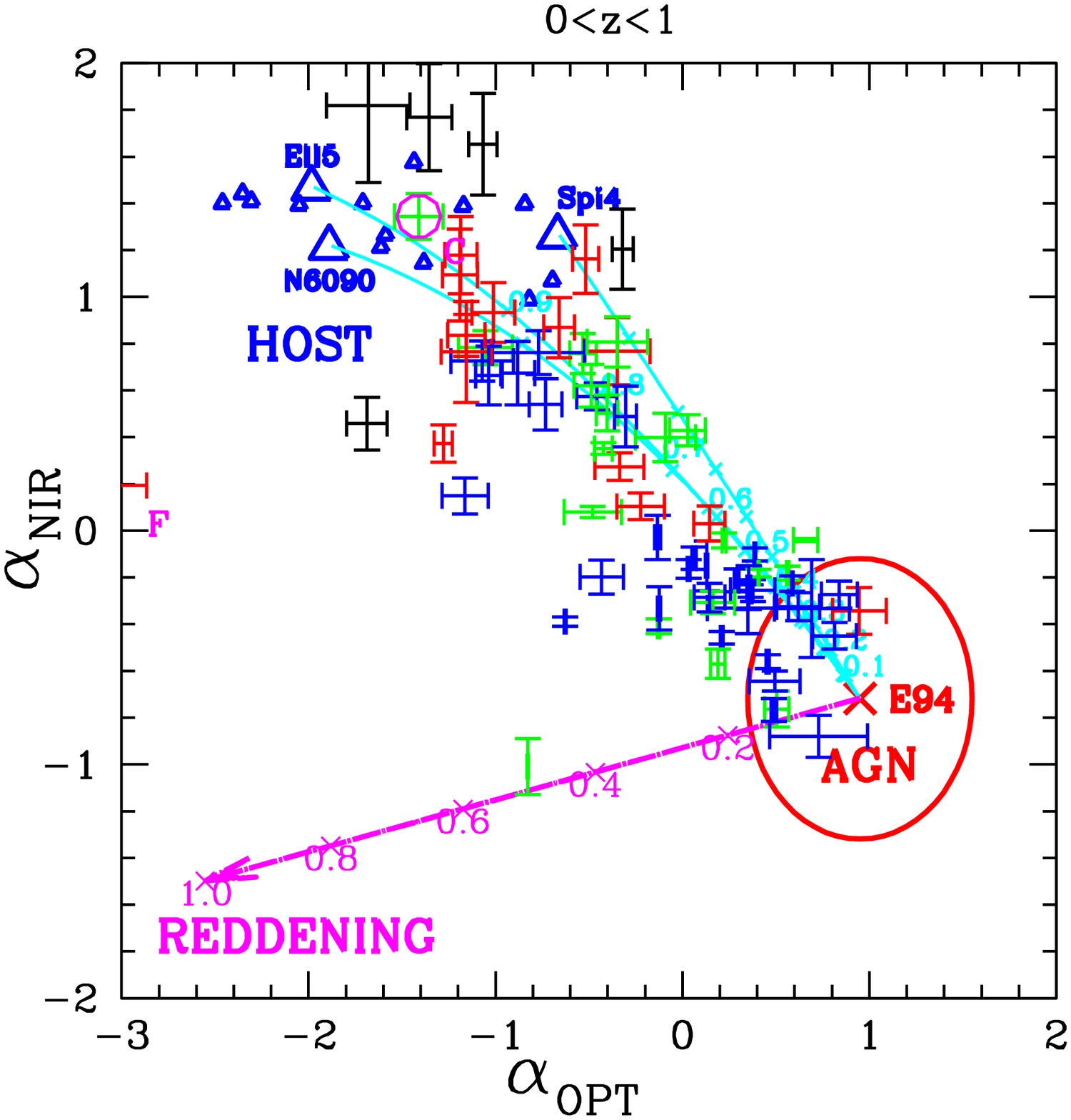}
\includegraphics[angle=0,width=0.245\textwidth]{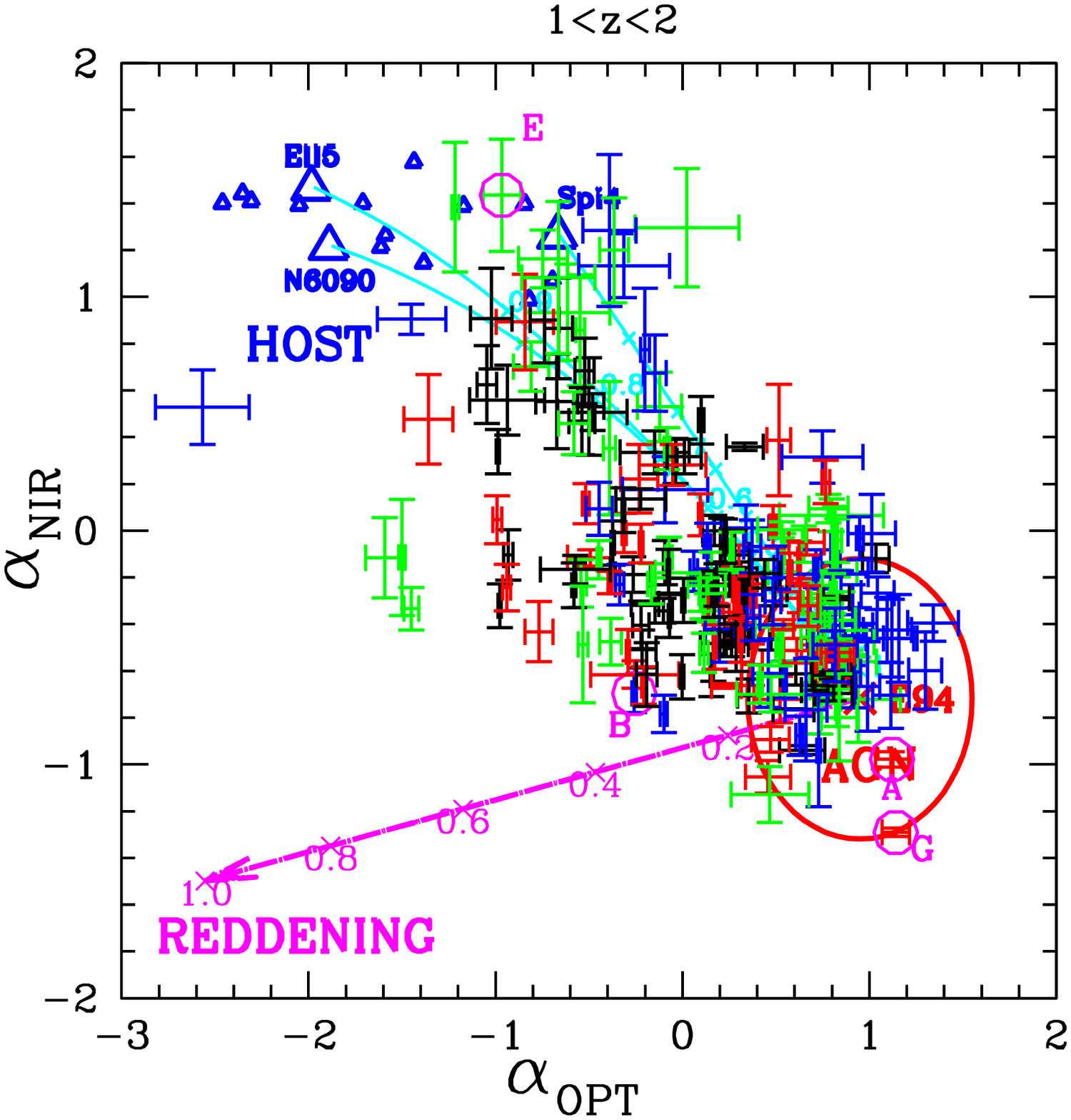}
\includegraphics[angle=0,width=0.245\textwidth]{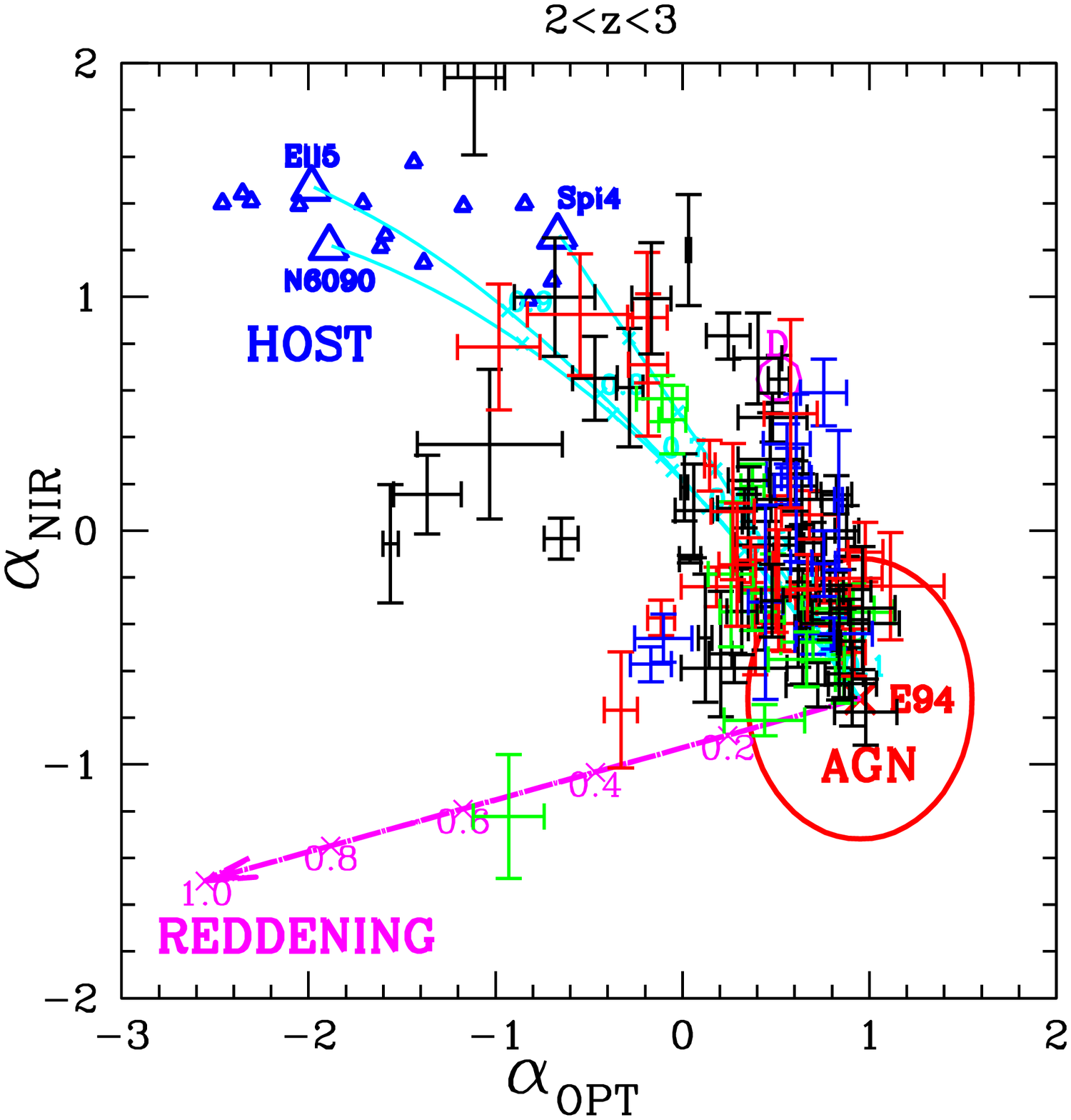}
\includegraphics[angle=0,width=0.245\textwidth]{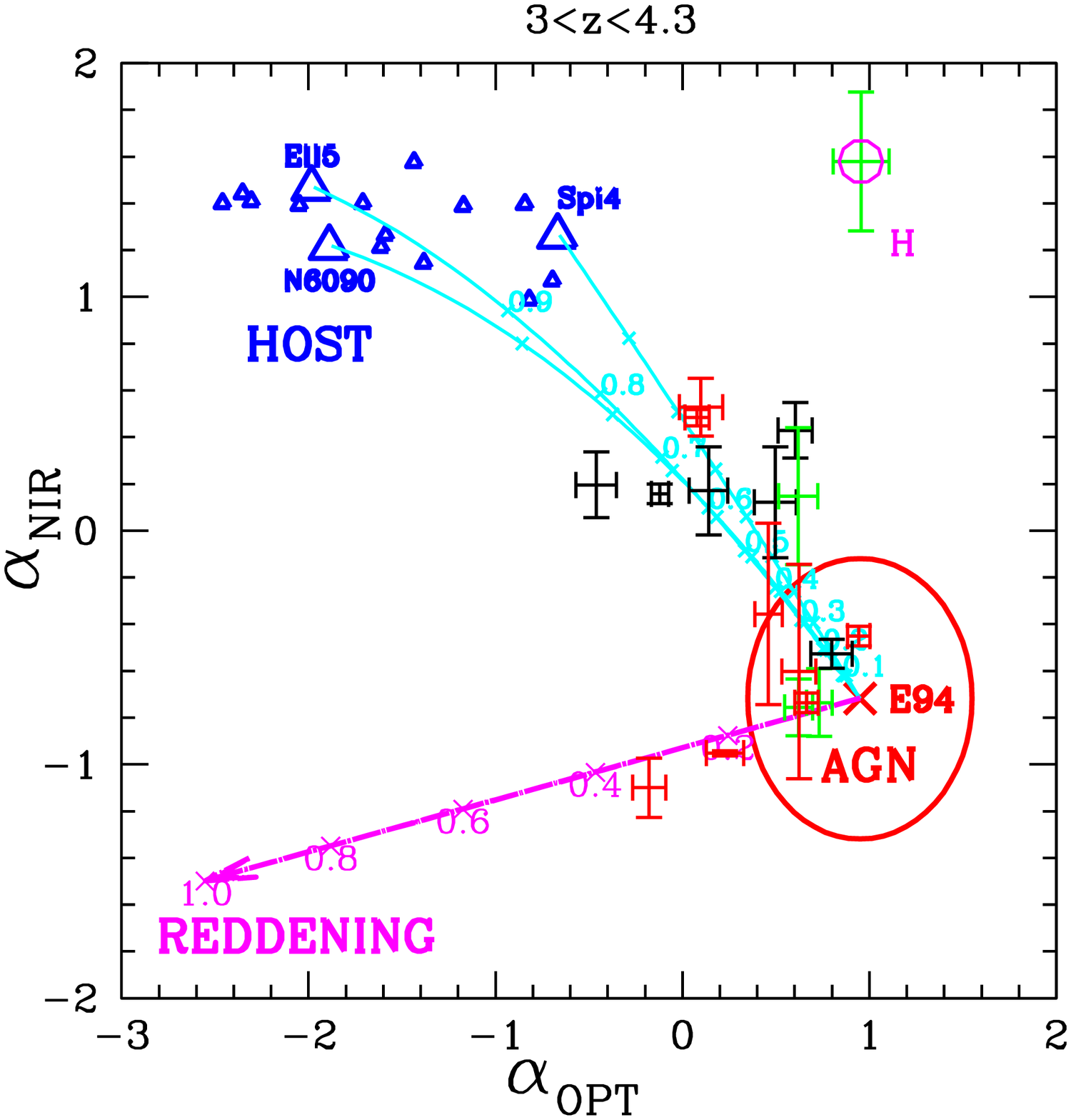}
\includegraphics[angle=0,width=0.245\textwidth]{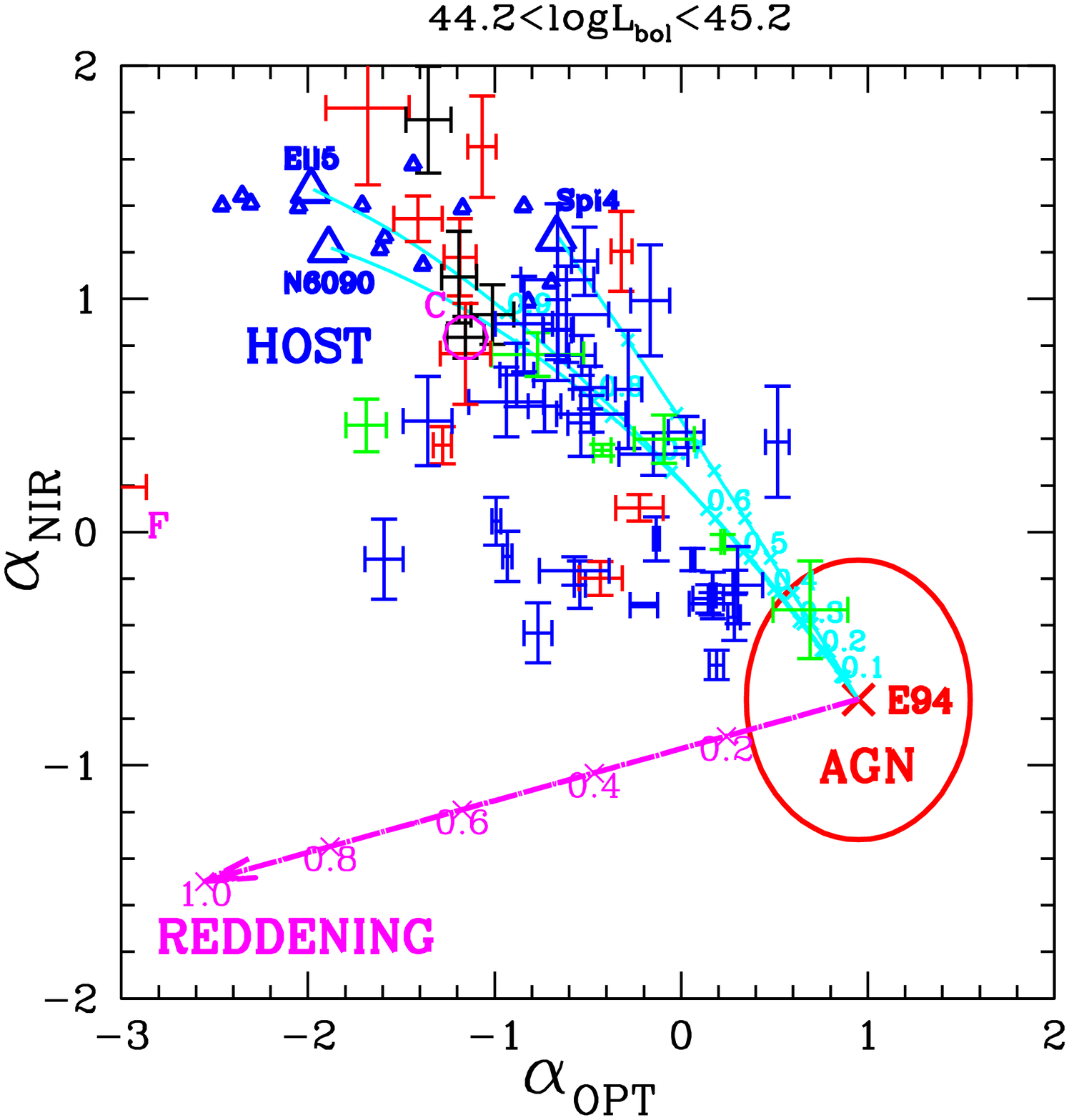}
\includegraphics[angle=0,width=0.245\textwidth]{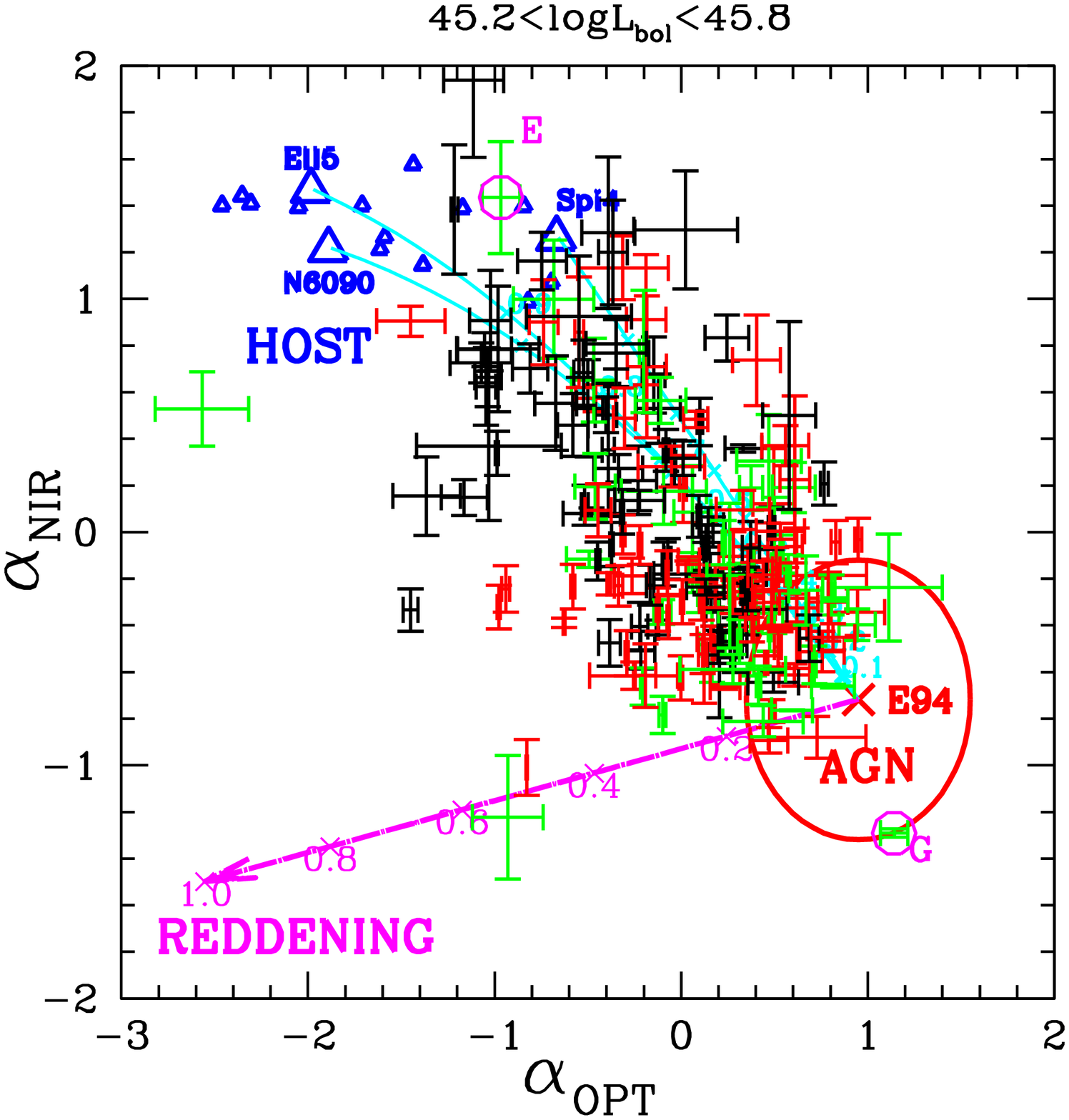}
\includegraphics[angle=0,width=0.245\textwidth]{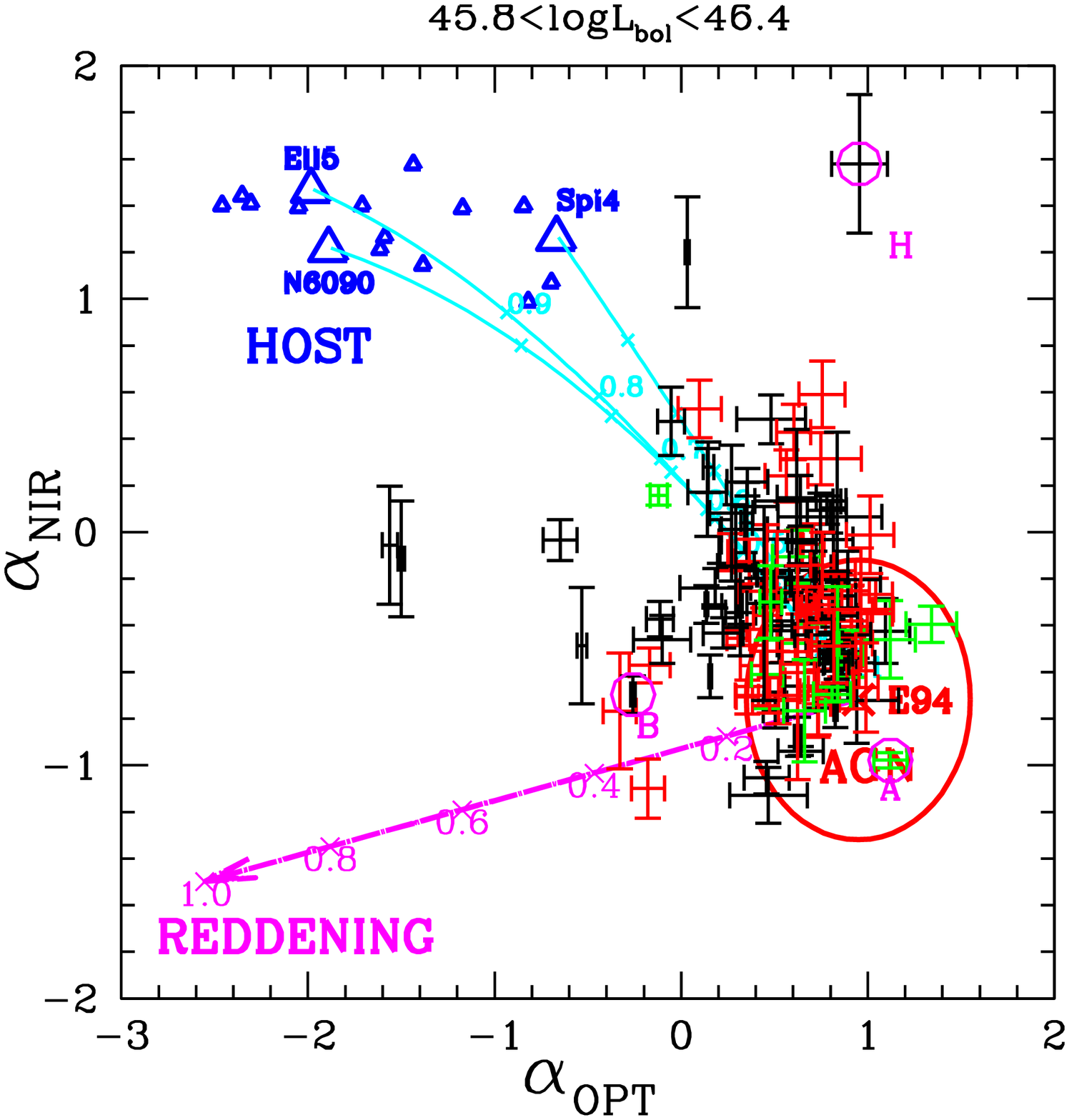}
\includegraphics[angle=0,width=0.245\textwidth]{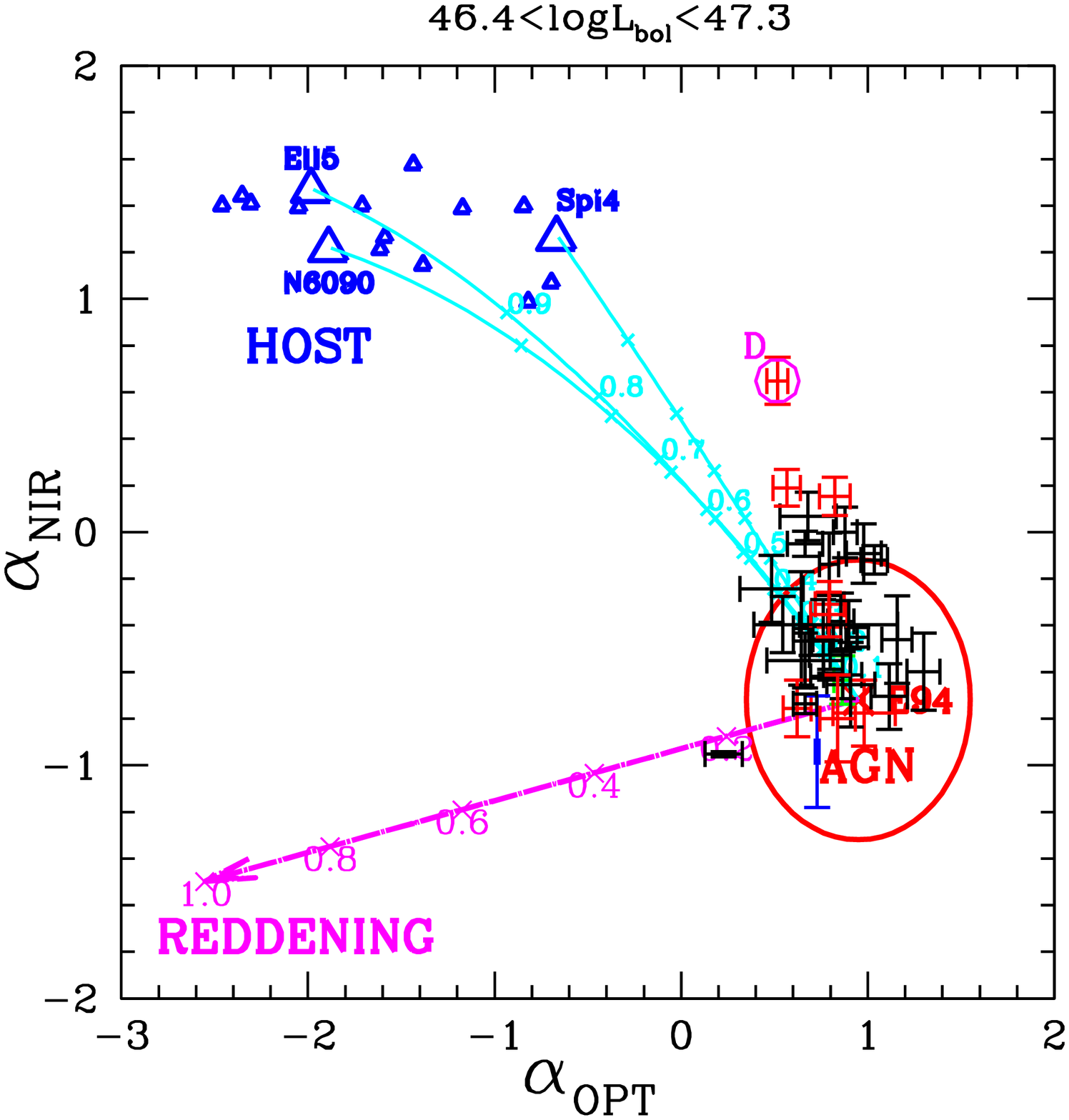}
\caption{$\alpha_{\rm{NIR}}$ v.s. $\alpha_{\rm{OPT}}$ plot for the
XC413 sample in $z$ bins [0 -- 1 -- 2 -- 3 -- 4.3] (top row) and
log$L_{bol}$ bins [44.2 -- 45.2 -- 45.8 -- 46.4 -- 47.3] (bottom
row). Different colors of the points in each plot represent quasars
in different sub-bins, with bin width 0.25, from low to high: black,
red, green and blue. The E94 mean SED is shown as the red cross,
with the galaxy templates from the SWIRE (Polletta et al. 2007, blue
triangles). The cyan lines are the quasar-host mixing curves. The
purple line is the reddening vector. The red circle shows the
dispersion circle. \label{slopezLpage}}
\end{figure*}
%%%%%%%%%%%%%%%%%%

%%%%%%%%%%%%%%%%%%
\begin{figure*}
\includegraphics[angle=0,width=0.245\textwidth]{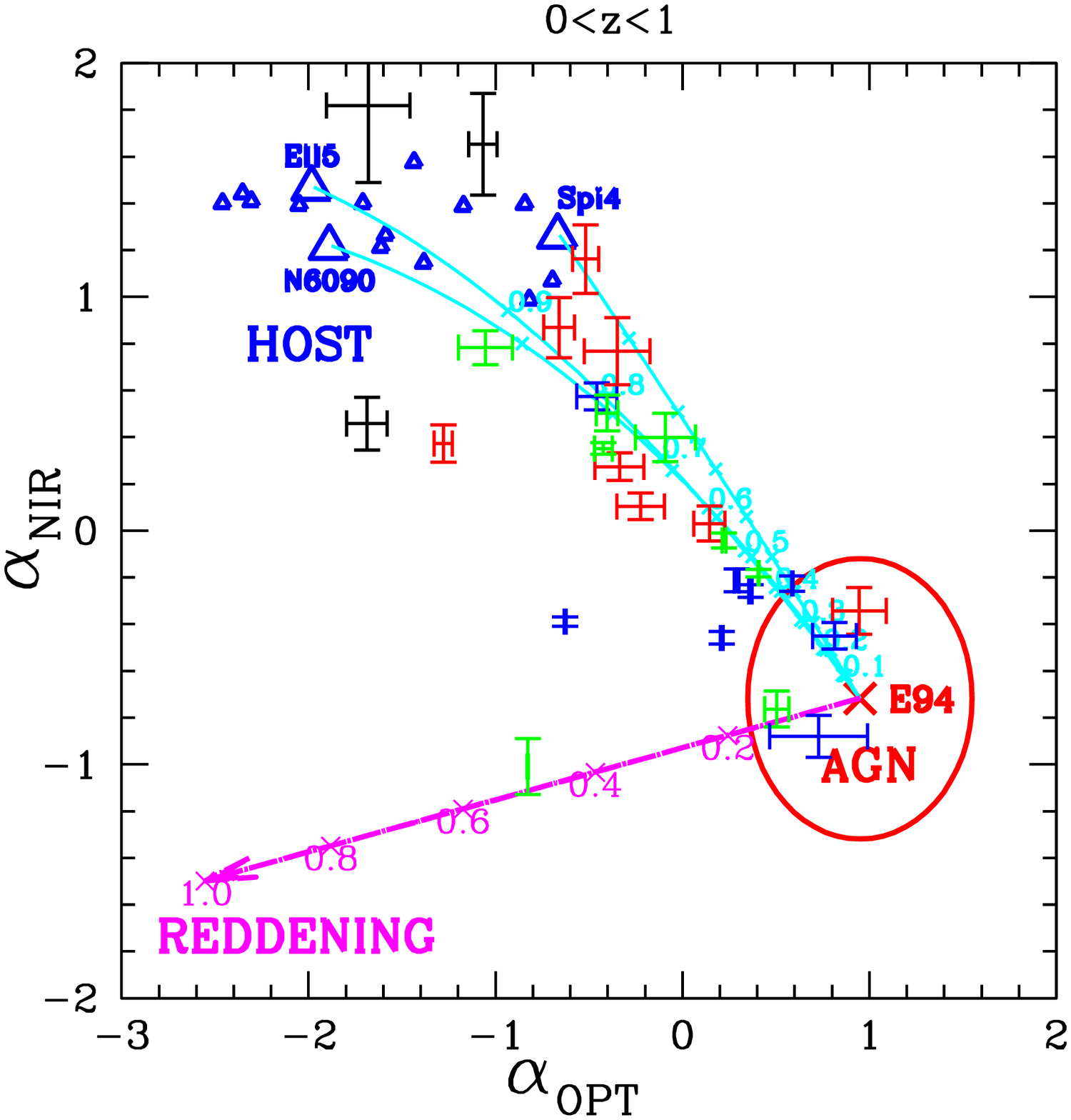}
\includegraphics[angle=0,width=0.245\textwidth]{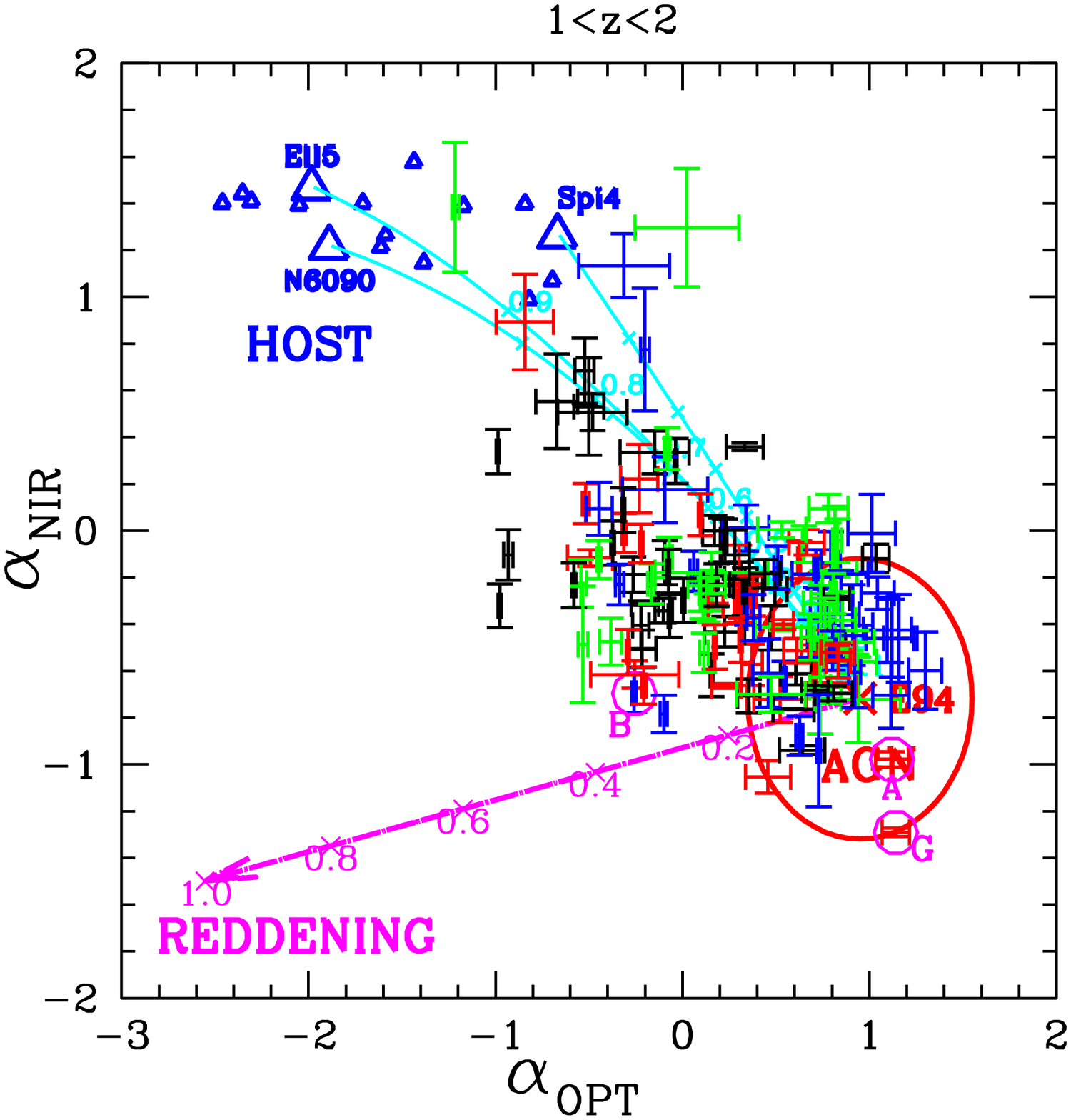}
\includegraphics[angle=0,width=0.245\textwidth]{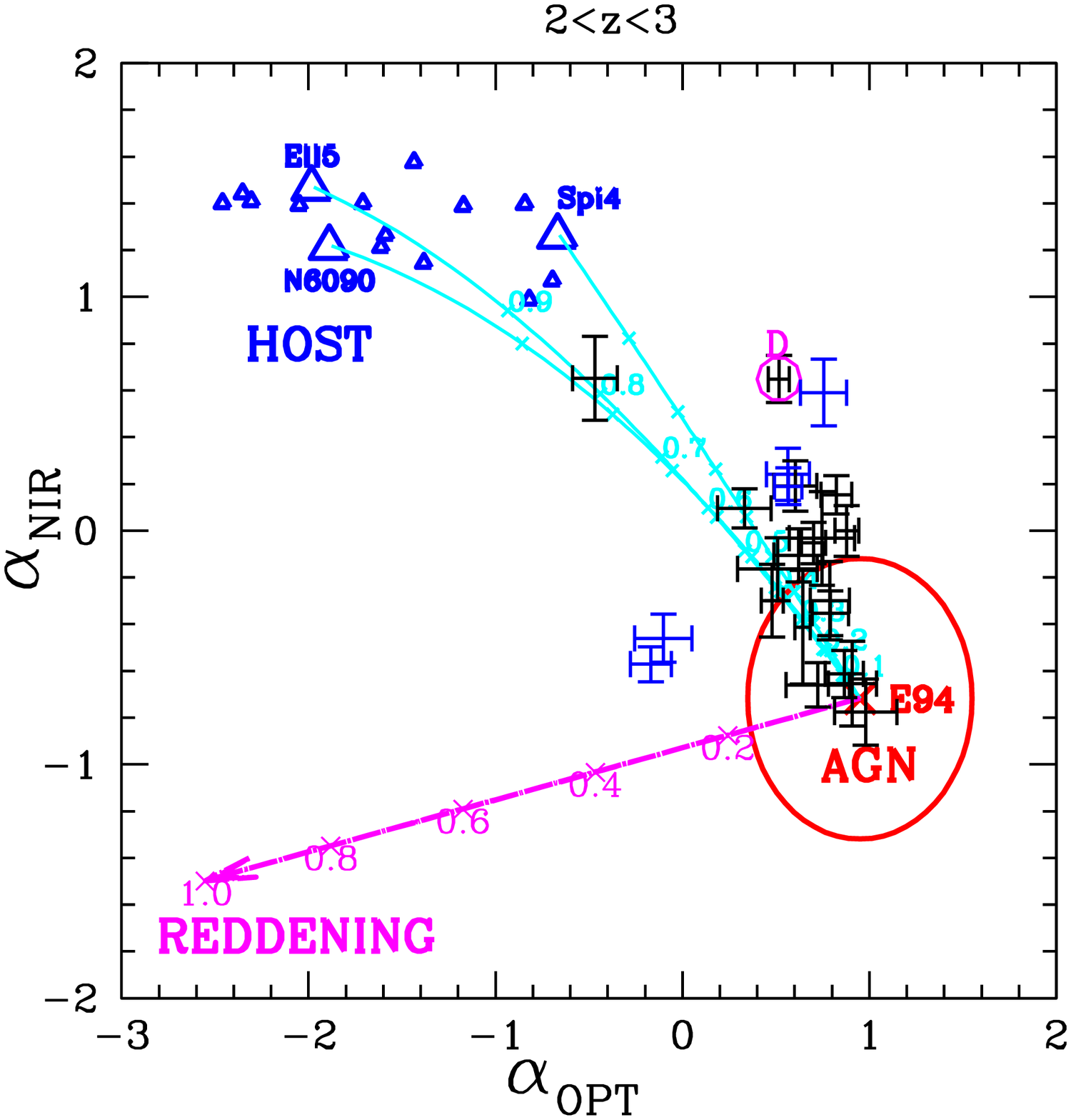}
\includegraphics[angle=0,width=0.245\textwidth]{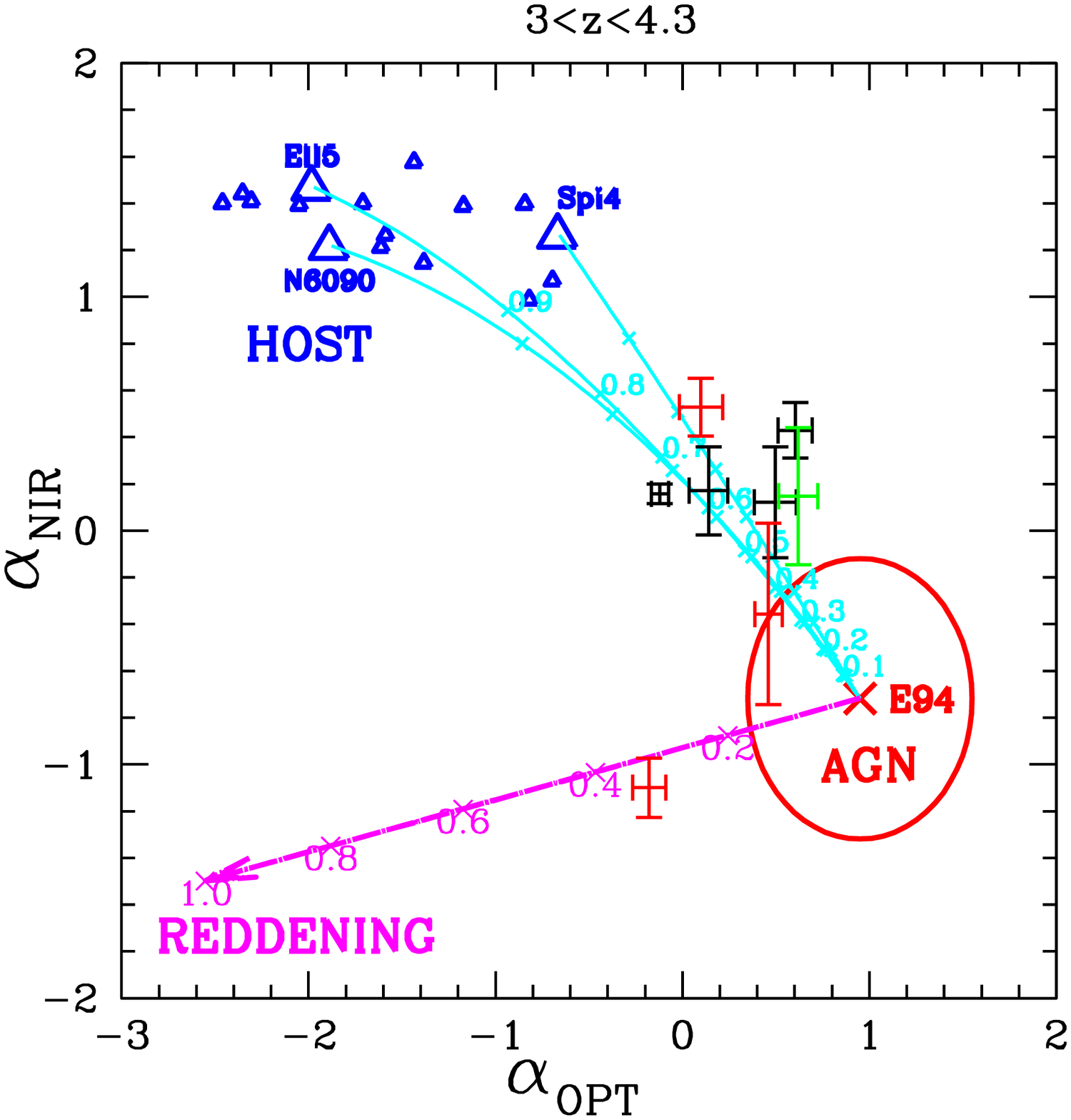}
\includegraphics[angle=0,width=0.245\textwidth]{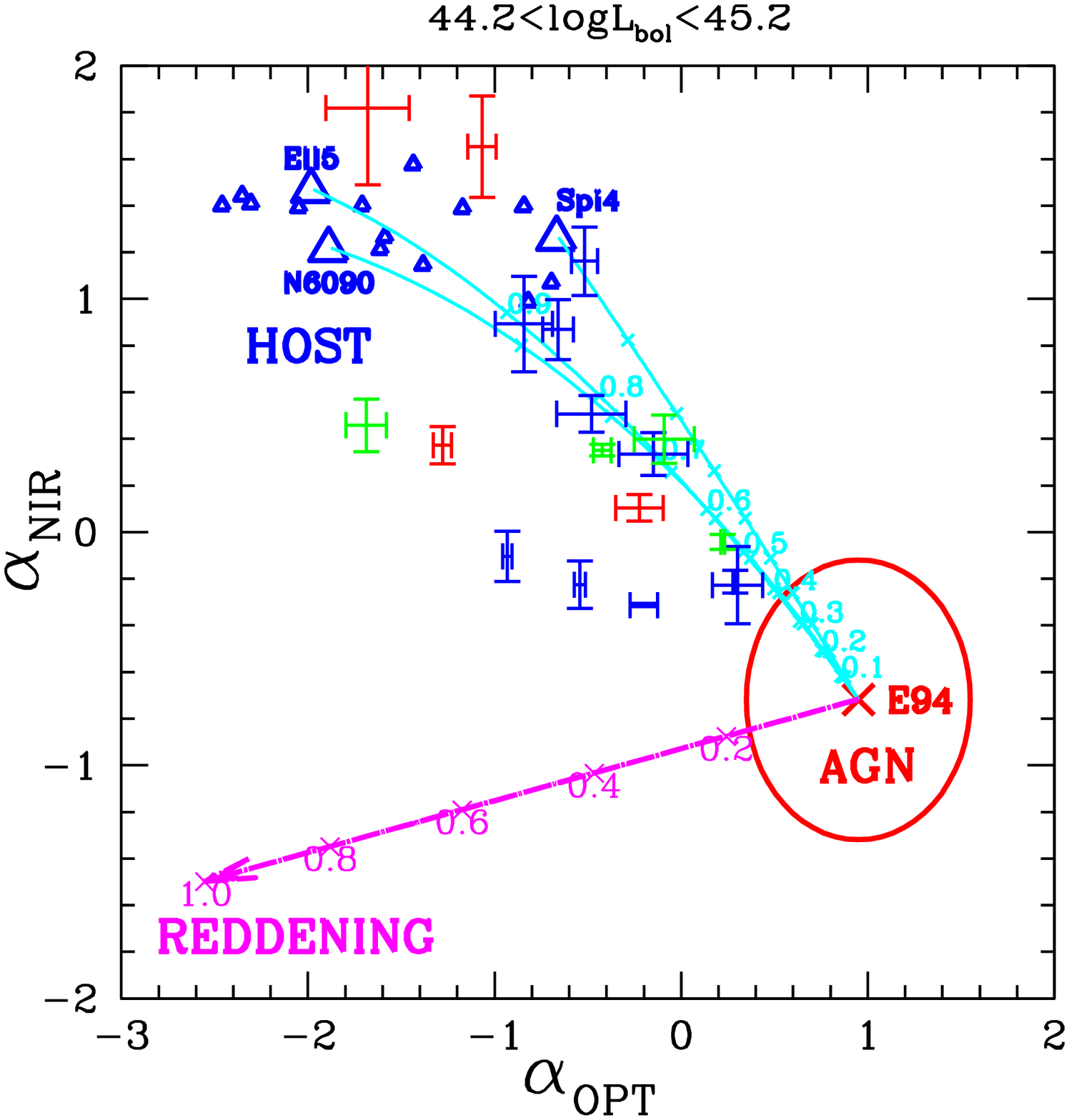}
\includegraphics[angle=0,width=0.245\textwidth]{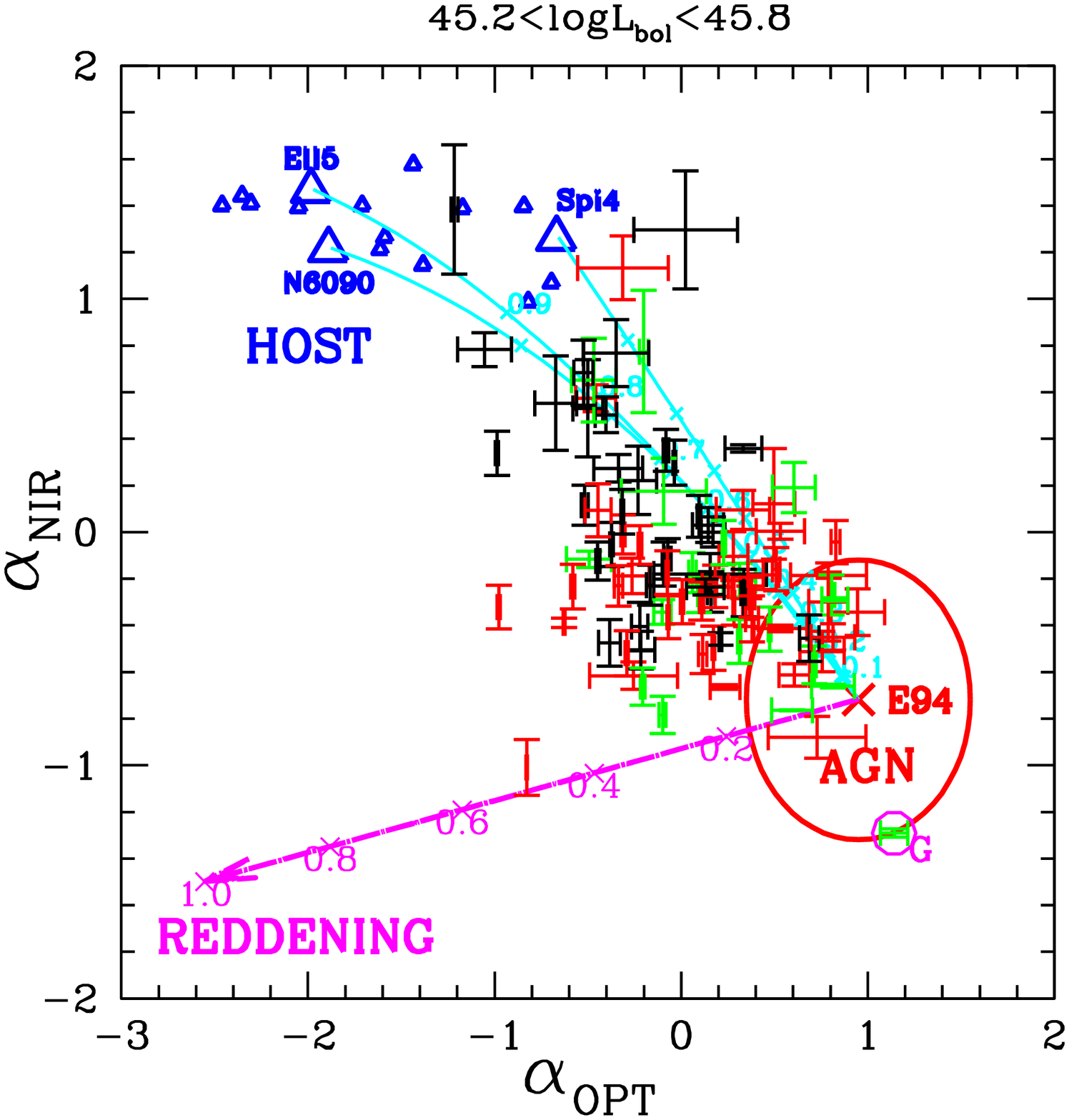}
\includegraphics[angle=0,width=0.245\textwidth]{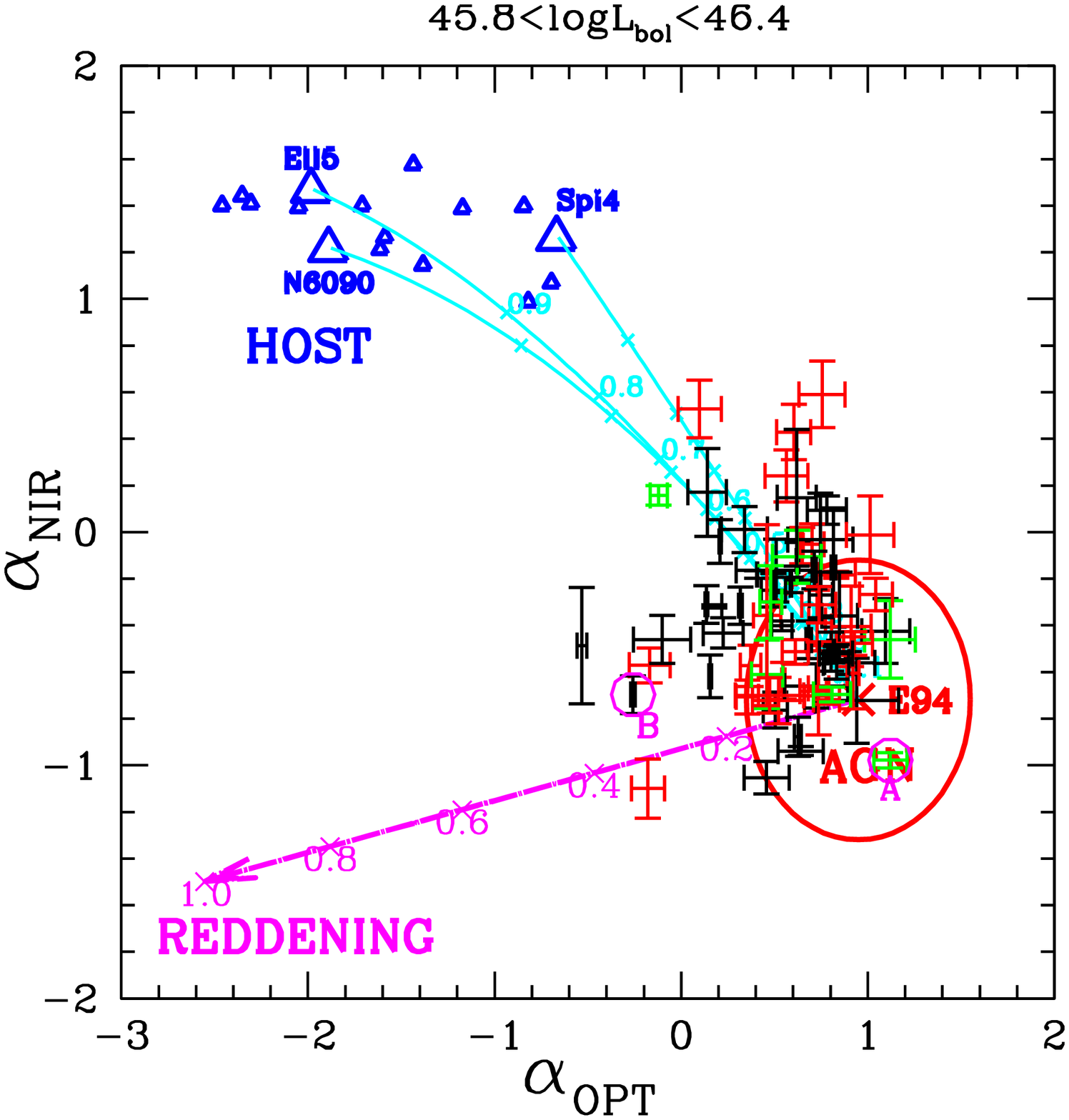}
\includegraphics[angle=0,width=0.245\textwidth]{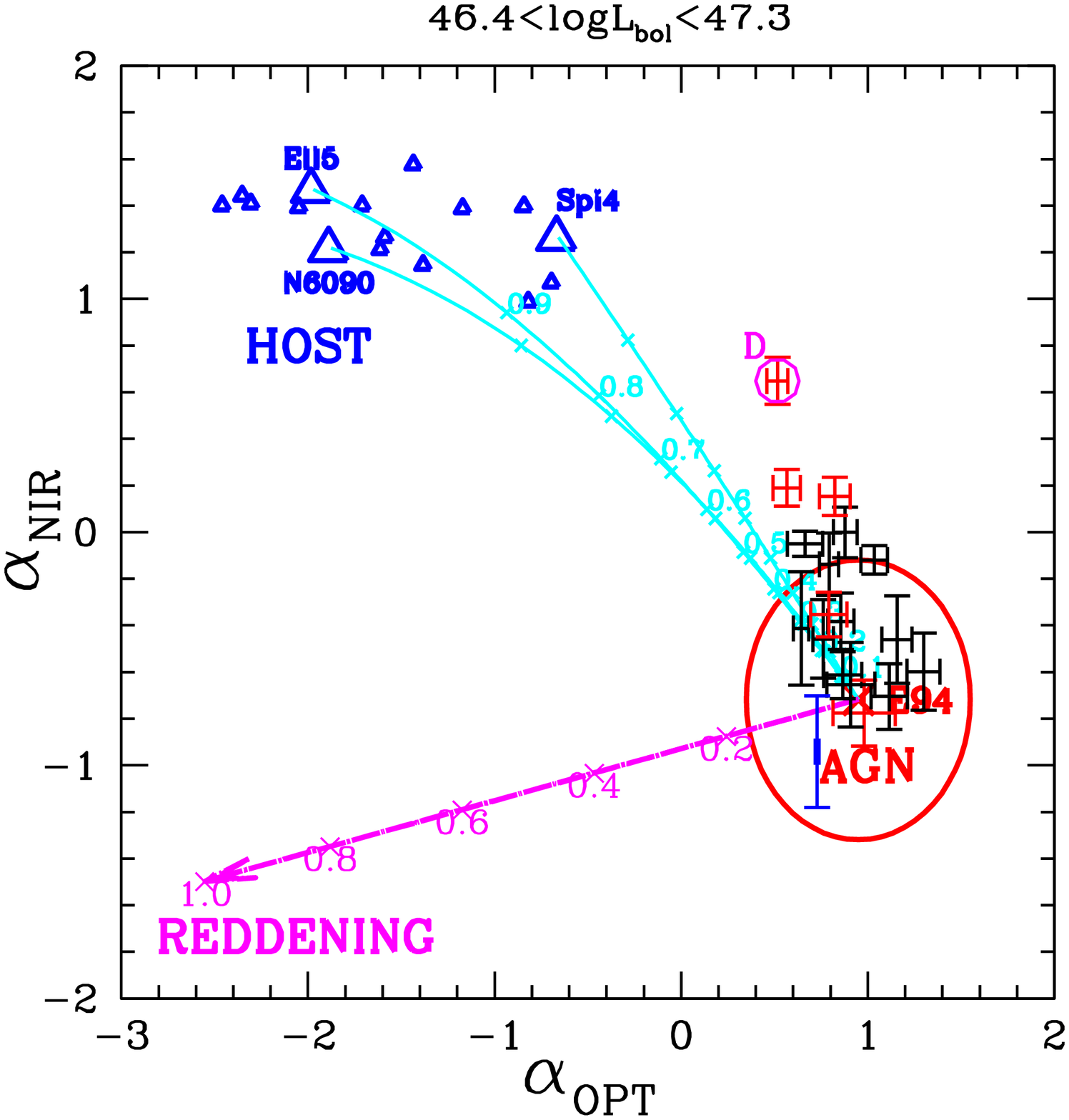}
\includegraphics[angle=0,width=0.245\textwidth]{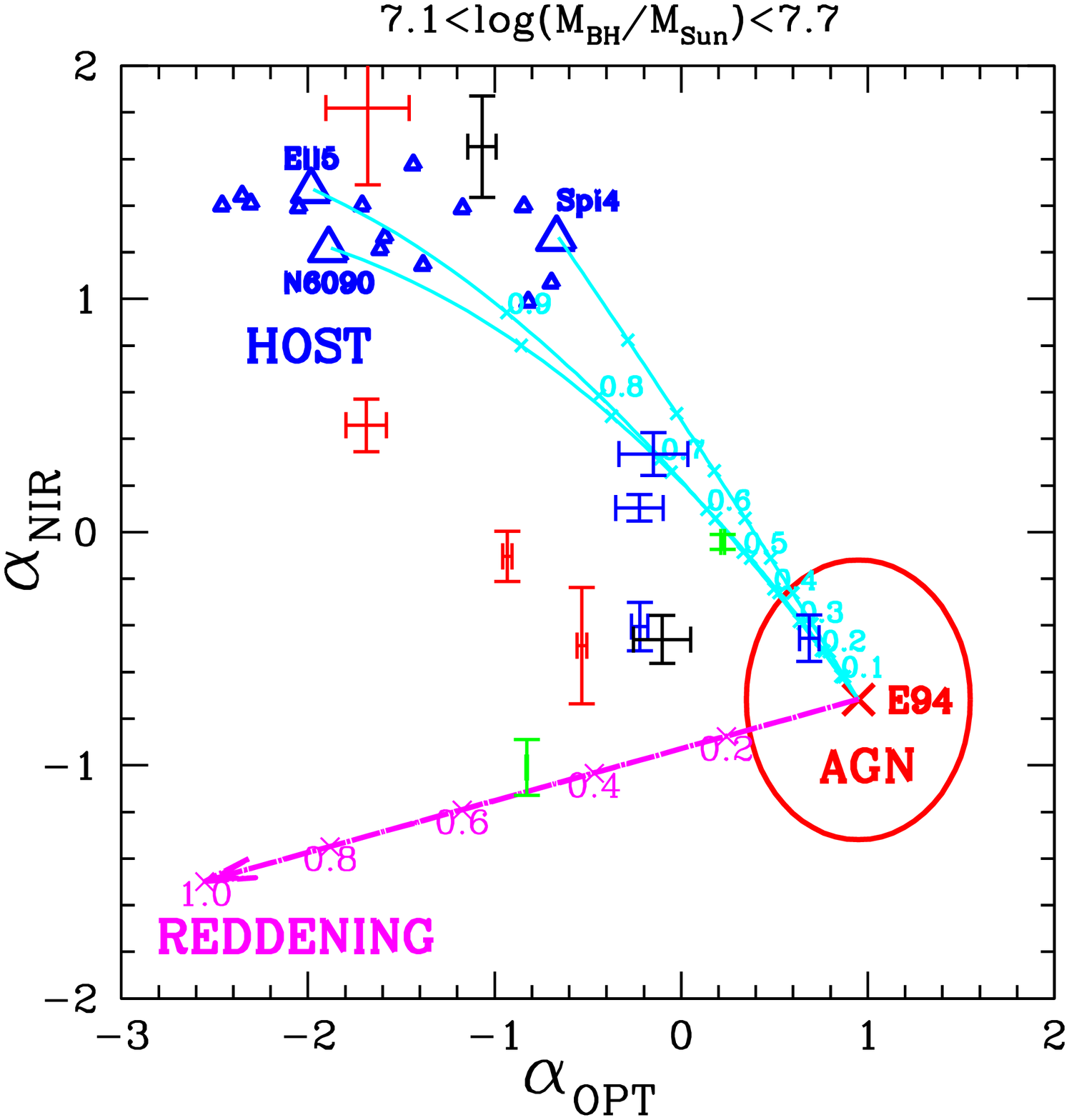}
\includegraphics[angle=0,width=0.245\textwidth]{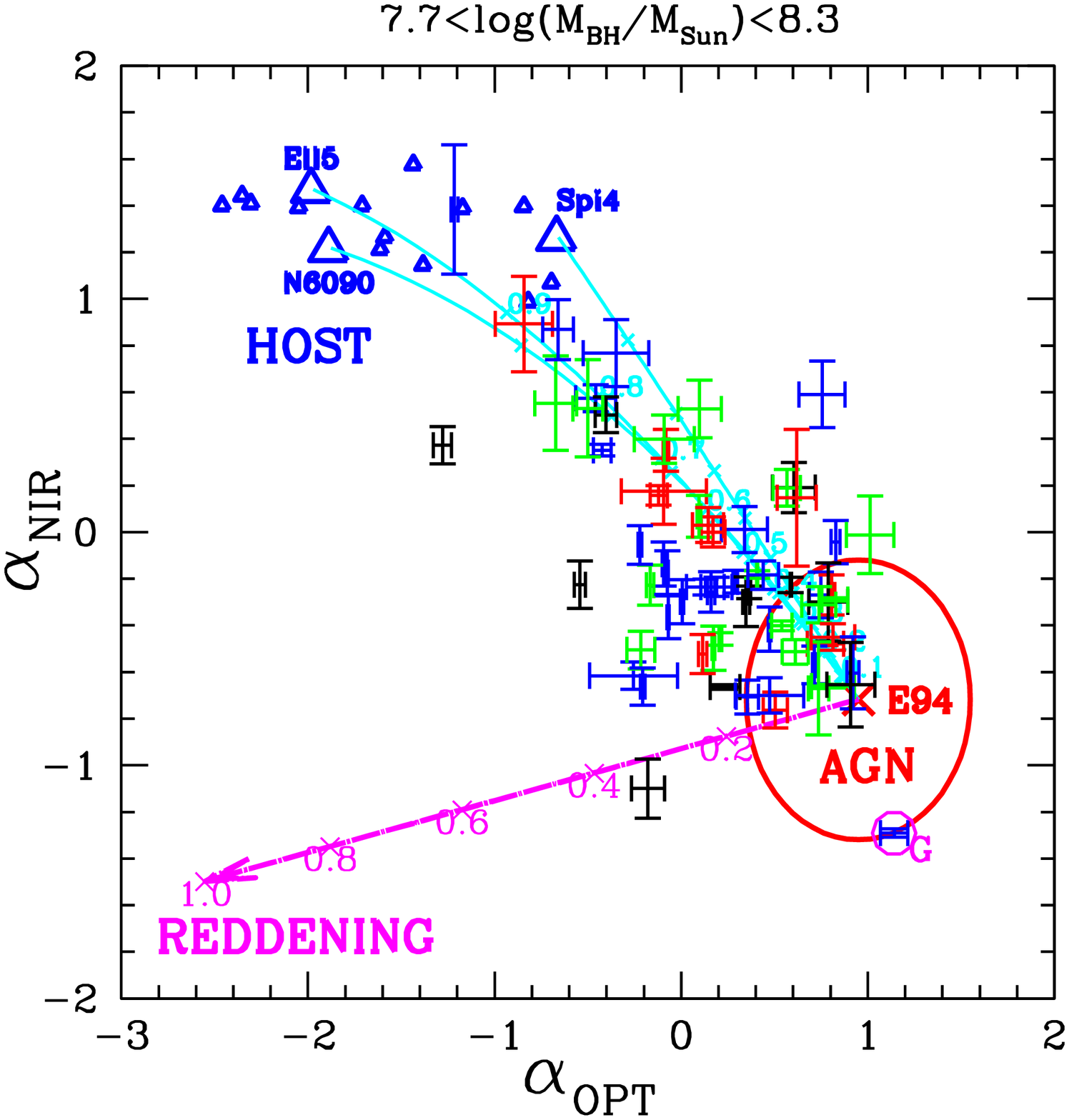}
\includegraphics[angle=0,width=0.245\textwidth]{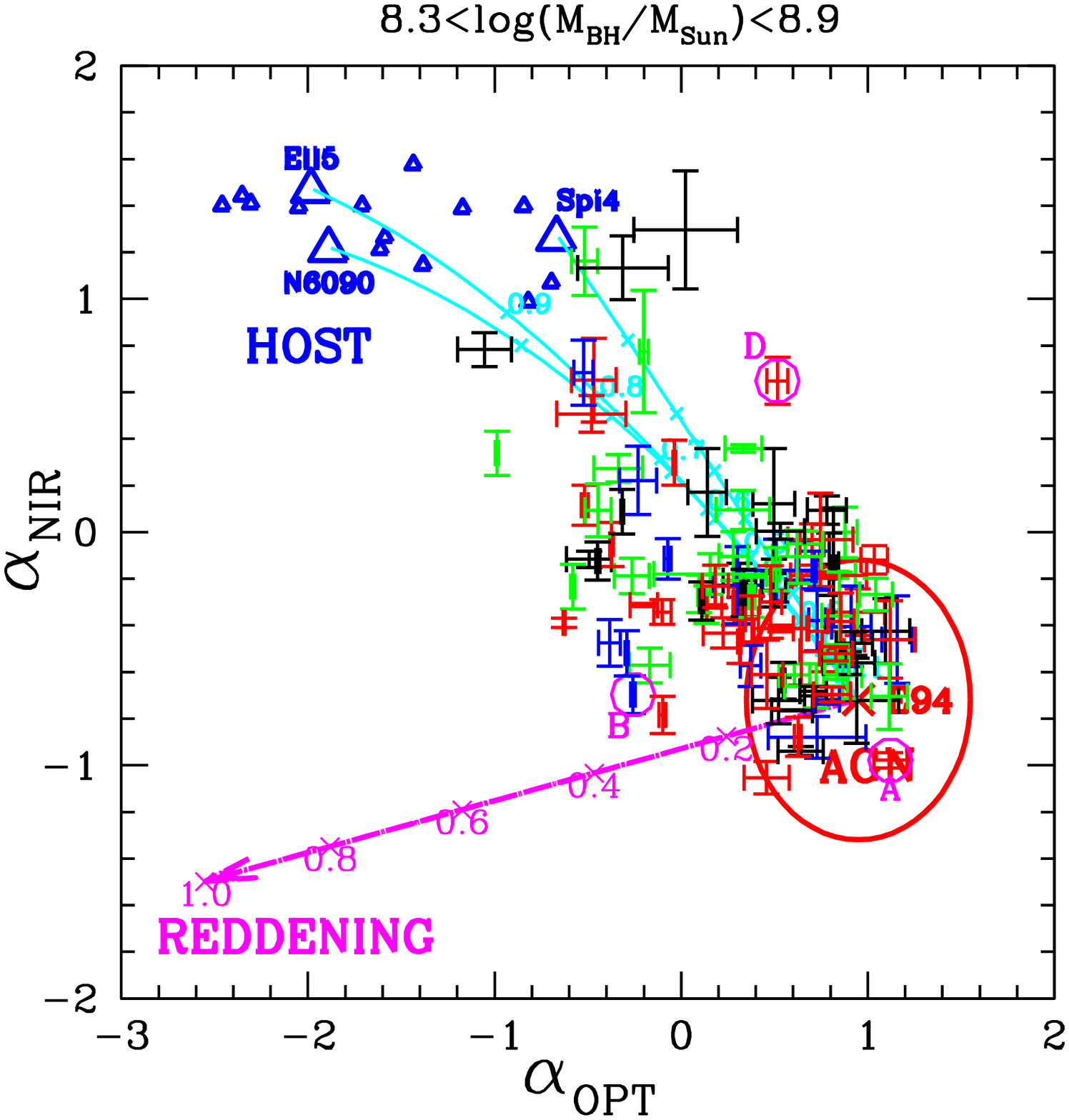}
\includegraphics[angle=0,width=0.245\textwidth]{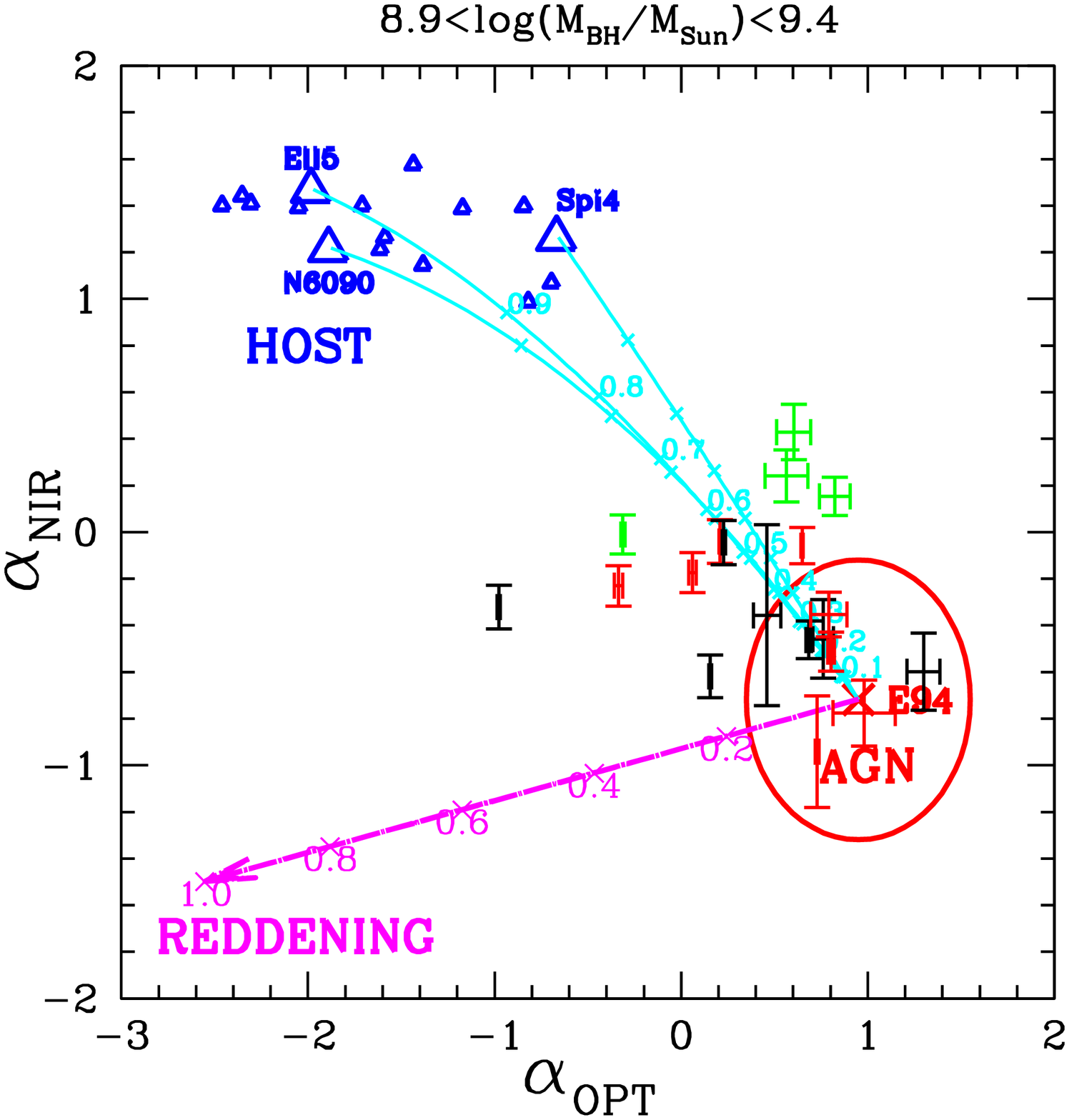}
\includegraphics[angle=0,width=0.245\textwidth]{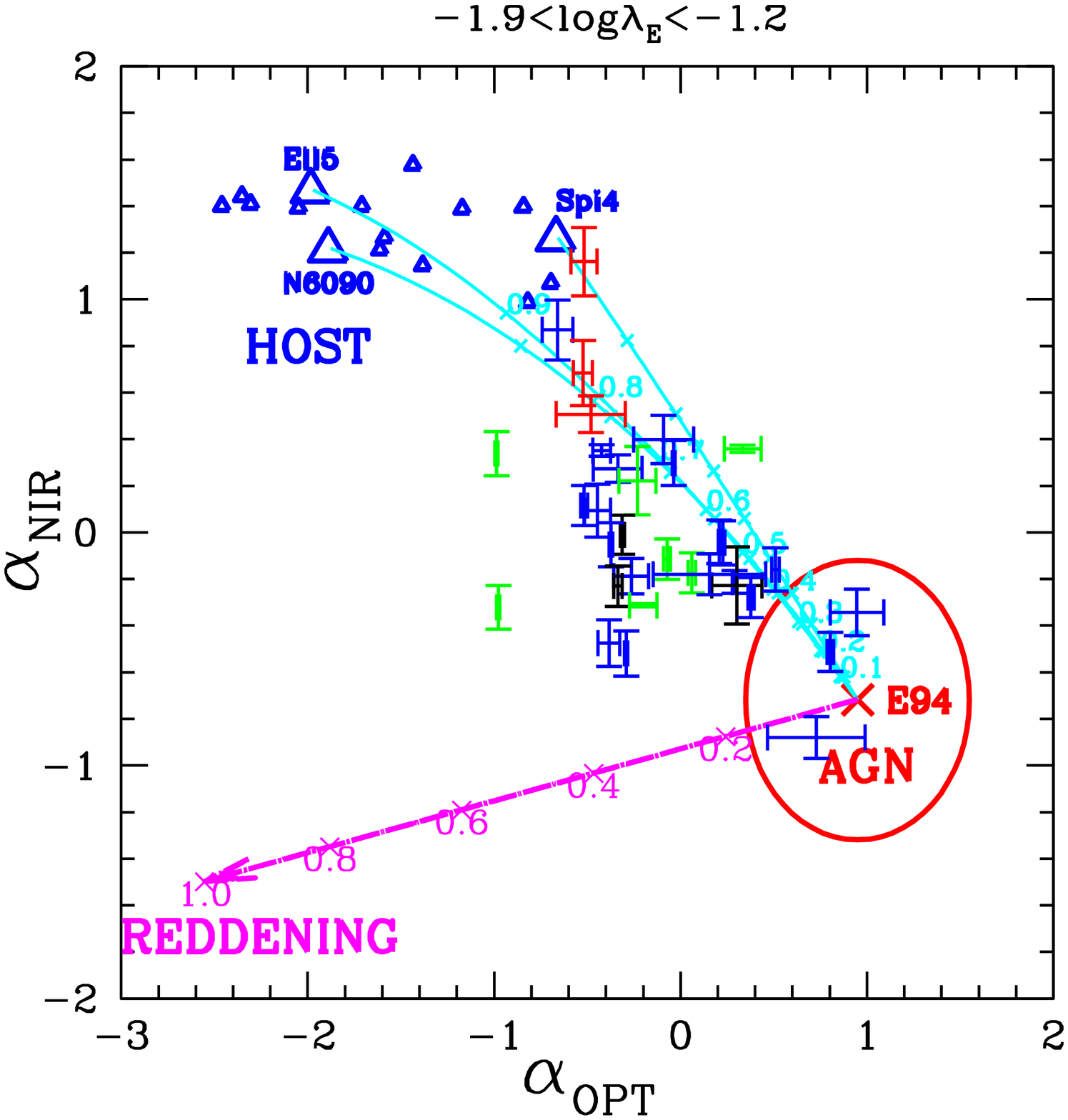}
\includegraphics[angle=0,width=0.245\textwidth]{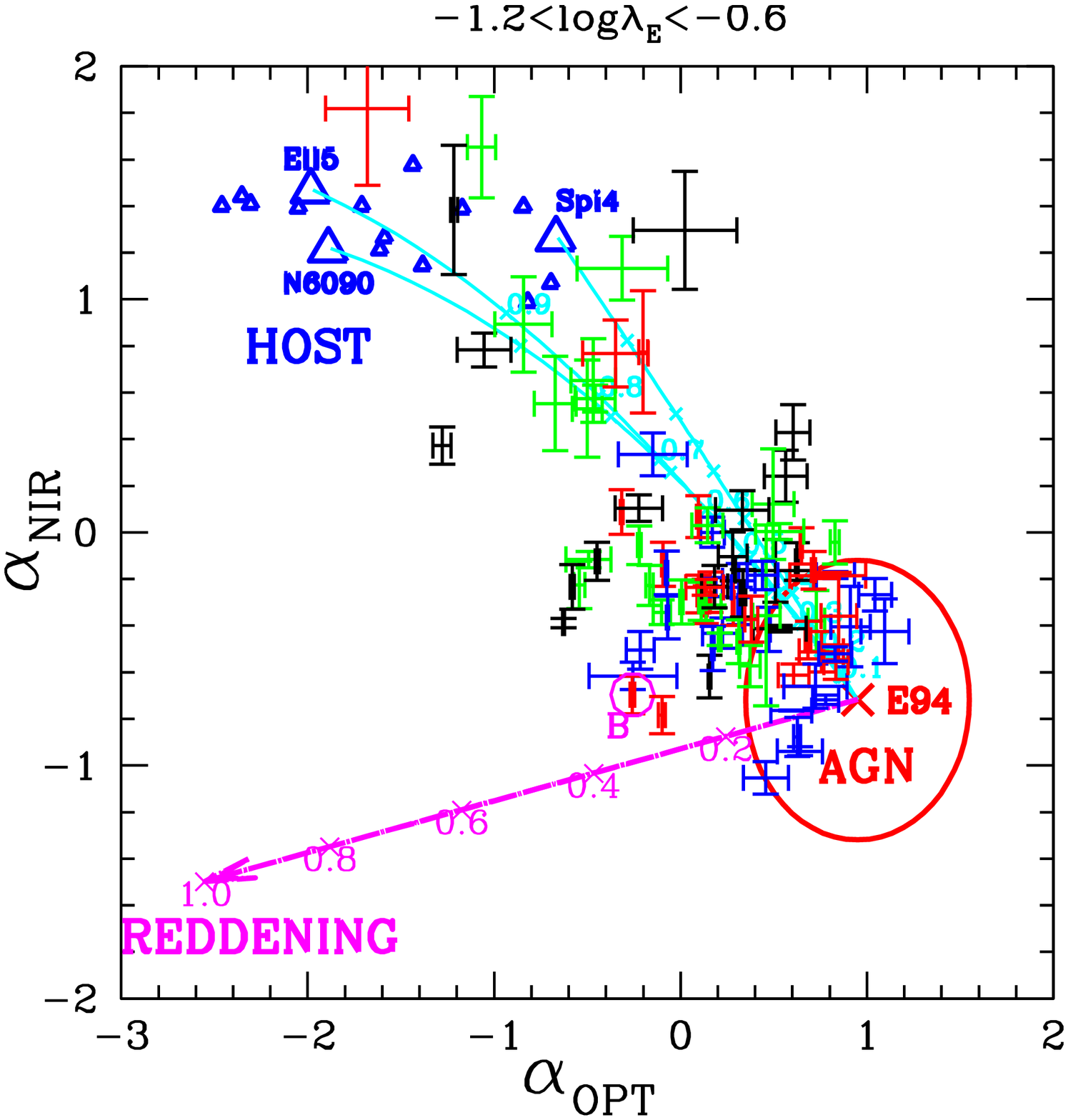}
\includegraphics[angle=0,width=0.245\textwidth]{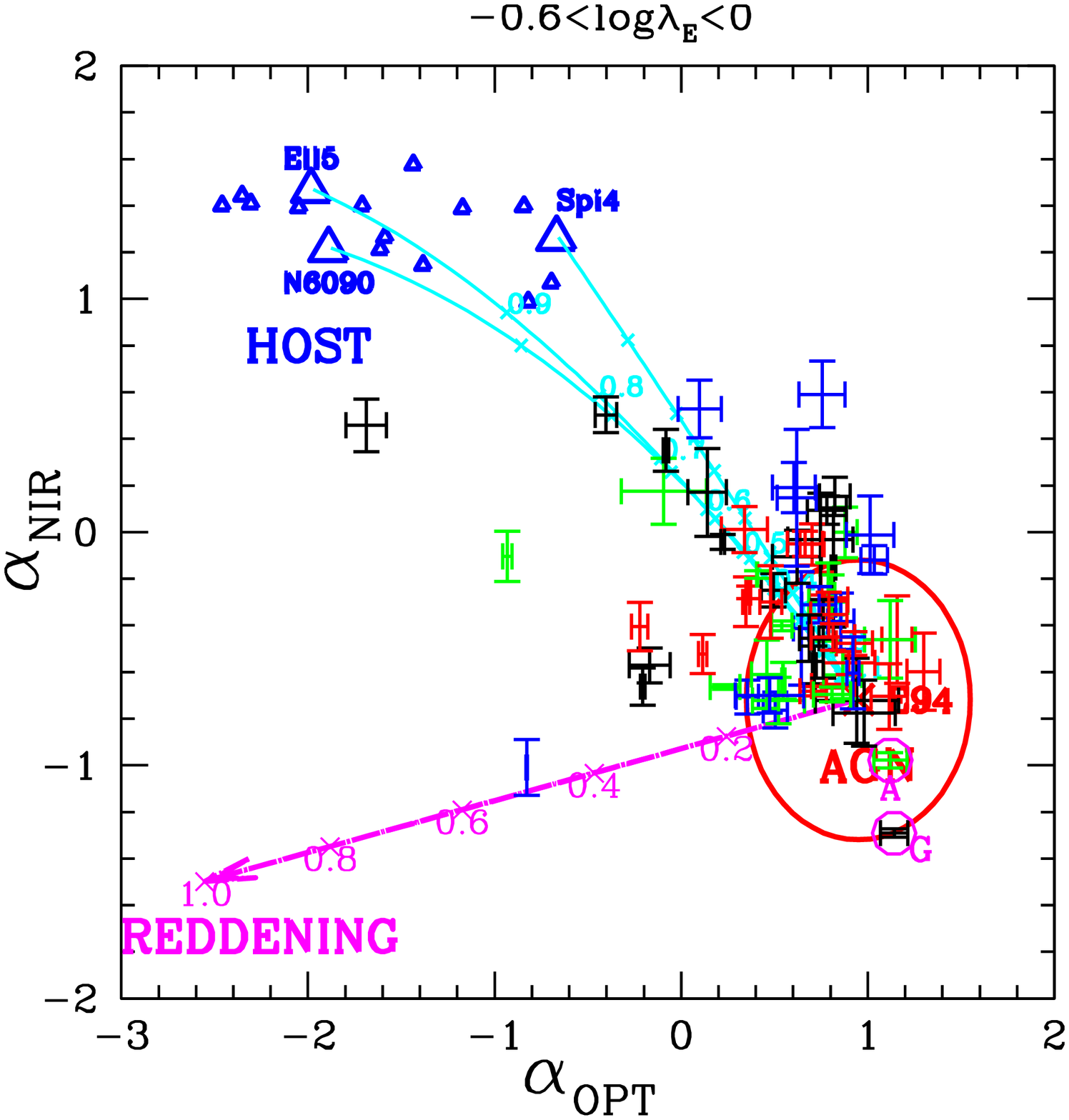}
\includegraphics[angle=0,width=0.245\textwidth]{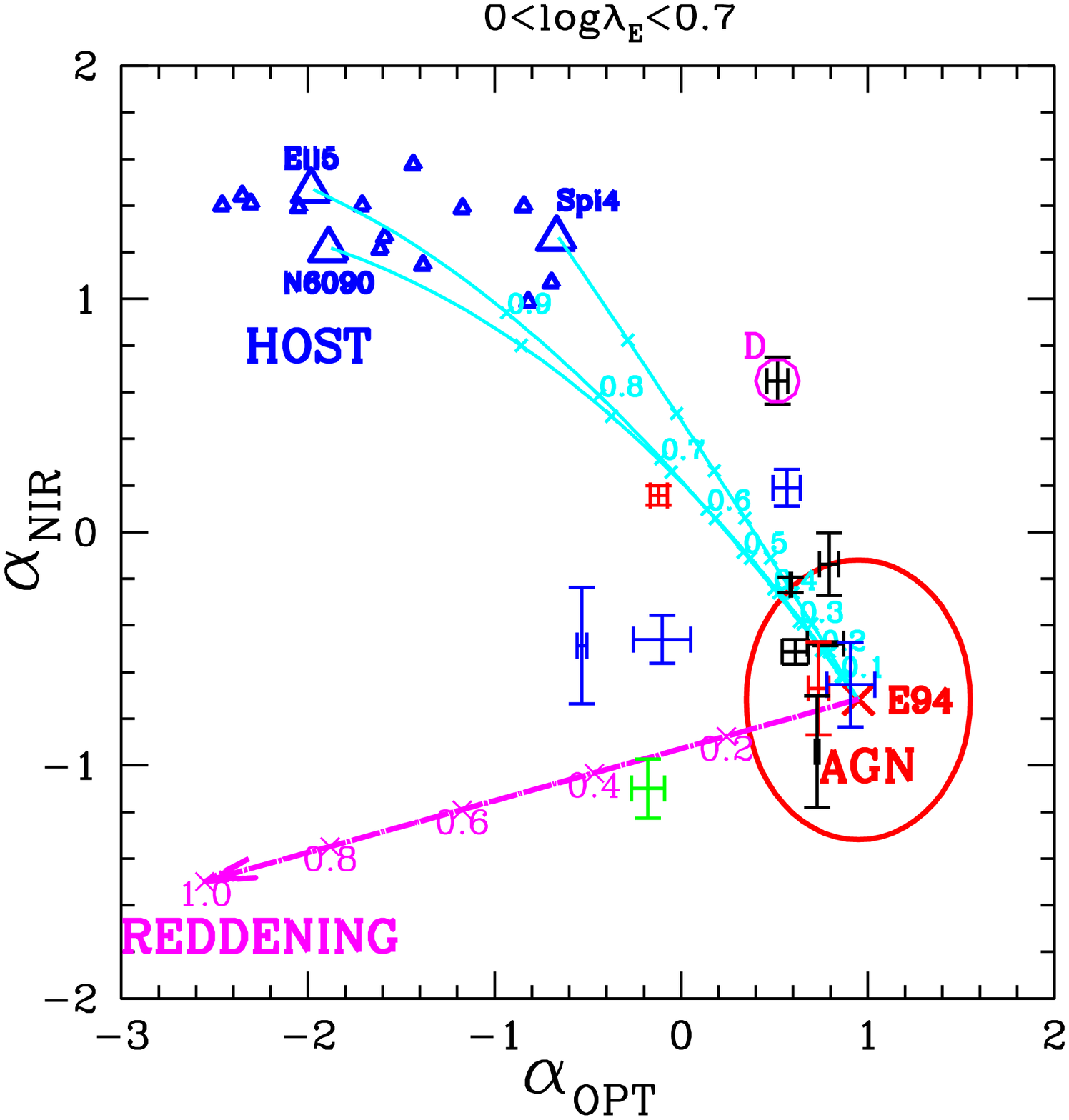}
\caption{$\alpha_{\rm{NIR}}$ v.s. $\alpha_{\rm{OPT}}$ plot for SS206
sample in $z$ bins [0 -- 1 -- 2 -- 3 -- 4.3] (top row), $\log
L_{bol}$ bins [44.2 -- 45.2 -- 45.8 -- 46.4 -- 47.3] (second row),
log($M_{BH}/M_{Sun}$) bins [7.1 -- 7.7 -- 8.3 -- 8.9 -- 9.4](third
row) and log$\lambda_E$ bins [-1.9 -- -1.2 -- -0.6 -- 0 -- 0.7].
Different color of points in each plot represent quasars in
different sub-bins, with sub-bin width 0.25 for $z$ and $\log
L_{bol}$ bins and sub-bin width 0.15 for $\log M_{BH}$ and $\log
\lambda_E$ bins, from low to high: black, red, green and blue. The
plots are color-coded as in Figure~\ref{slopezLpage}.
\label{slopezLmepage}}
\end{figure*}
%%%%%%%%%%%%%%%%%%

Paper II studied the evolution of the mean and dispersion of the SED
with physical parameters (redshift $z$, bolometric luminosity
$L_{bol}$, black hole mass $M_{BH}$, and Eddington ratio
$\lambda_E$\footnotemark) for the 407 radio quiet quasar in the
XC413 sample. Paper II showed that there is no obvious evidence for
evolution of the quasar SED shape with respect to these parameters.
The study was limited by the difficulties of host galaxy
subtraction. The conclusions are fully based on the assumption that
host galaxy correction according to the black hole mass and bulge
mass scaling relationship adding an evolutionary term is reliable.
The mixing diagram is a new tool to address this issue, with no need
to rely on the assumption that the host correction is properly done.
That is because the diagram itself can clearly show the contribution
from the host galaxy.

\footnotetext{$\lambda_E=\frac{L_{bol}}{L_{Edd}}=\frac{L_{bol}}{\frac{4\pi
Gcm_{p}}{\sigma_e}M_{BH}}
=\frac{L_{bol}}{1.26\times10^{38}(M_{BH}/M_{\odot})}$}

In order to search for quasar SED evolution with respect to physical
parameters, we plotted the mixing diagram for the XC413 sample in
bins of $z$, $\log L_{bol}$, and for the SS206 sample with two
additional parameters $\log M_{BH}$ and $\log\lambda_E$, because the
black hole mass estimates are only available for these 206 quasars
in XC413 (see \S~\ref{s:sample}). We divided the sample in four
bins: in $z$ [0 -- 1 -- 2 -- 3 -- 4.3], in $\log L_{bol}$ [44.2 --
45.2 -- 45.8 -- 46.4 -- 47.3], in $\log (M_{BH}/M_{\bigodot})$
[7.1--7.7--8.3--8.9--9.4], and in $\log\lambda_E$ [-1.9 -- -1.2 --
-0.6 -- 0 -- 0.7], respectively. For each physical parameter the
four bins have approximately equal bin size, so it is easy to
compare bins. The resulting mixing diagrams are shown in
Figures~\ref{slopezLpage} and Figure~\ref{slopezLmepage}.

To look for any SED evolution in smaller steps, we color coded the
quasars in each bin for four equal sub-bins
(Figure~\ref{slopezLpage} and Figure~\ref{slopezLmepage}). In each
$z$ and $\log L_{bol}$ mixing diagram, the black, red, green, and
blue points represent quasars with small to large $z$ and $\log
L_{bol}$, with the sub-bin size of 0.25. Similarly, in each $\log
M_{BH}$ and $\log\lambda_E$ mixing diagram, the black, red, green
and blue colors represent small to large values, with sub-bin size
of 0.15.

For the lowest bin of each parameter ($0<z<1$, $44.2<\log
L_{bol}<45.2$, $7.1<\log (M_{BH}/M_{\bigodot})<7.7$, $-1.9<\log
\lambda_E<-1.2$), almost all of the sources lie within the mixing
wedge defined by the AGN-host mixing curve, allowing for the
$1\sigma$ range of the E94 mean SED slope, mixing curve and the
reddening curve.

For high values of each parameter, the quasars
(Figure~\ref{slopezLpage} and Figure~\ref{slopezLmepage}) cluster
close to the quasar dominated region (within the red circle), while
in the lower value bins the quasars spread out along the mixing
curves toward the galaxy template locations. This effect is the
strongest in $\log L_{bol}$ bins. For different $\log L_{bol}$ bins,
the cluster of quasar locations clearly drifts along the mixing
curves, from completely outside the AGN-dominated circle at low
$\log L_{bol}$, with many sources lying near the pure
galaxy-dominated region, to almost completely inside the dispersion
circle at high $\log L_{bol}$. This is expected, as the galaxy
luminosity is generally no more than $10^{45}$~erg/s (Cirasuolo et
al. 2007). Thus for extremely high luminosity sources, the AGN
outshines the galaxy, especially in the optical. However, the
M-$\sigma$ relation puts a limit on how much a quasar can outshine
its host galaxy (Paper I, II).

Although almost all of the highest luminosity quasars
(Figure~\ref{slopezLpage}) lie within the AGN-dominated circle, they
are not centered at the E94 RQ mean. Instead they lie overwhelmingly
in the upper left quadrant of the dispersion circle, similar to the
R06 sample as shown in the left panel of Figure~\ref{slpR06E94}.
This suggests that some shift with respect to the E94 SED is present
in both spectral slopes. For these highest luminosity quasars, the
mean $\alpha_{\rm{OPT}}$ is 0.78, versus 0.95 for E94 mean SED, with
$\sigma=0.21$; and the mean $\alpha_{\rm{NIR}}$ is -0.44, versus
-0.72 for E94 mean SED, with $\sigma=0.30$. In XC413 the slopes of
the highest luminosity quasars are shifted by $\sim$1$\sigma$
relative to the E94 RQ mean SED. This may be an intrinsic shift, or
may indicate a non-negligible host galaxy component even in these
luminous quasars.

To compare in detail the highest luminosity quasars in XC413 and
E94, we checked the 8 E94 quasars which lie in the same highest
luminosity range (above $2\times10^{46}$~erg/s). The mean
$\alpha_{\rm{OPT, E94}}$ of these 8 high luminosity E94 quasars is
0.93 with $\sigma=0.26$, almost exactly the same as the optical
slope of E94 mean SED, and bluer ($\sim 1\sigma$) than the XC413
high luminosity quasars. Instead, the mean $\alpha_{\rm{NIR, E94}}$
of these 8 E94 quasars is -0.19, with $\sigma=0.38$, which is much
flatter than the NIR slope of E94 mean SED, and even flatter than
the XC413 high luminosity quasars. This difference is mainly due to
the two hot-dust-poor quasars in these 8 E94 quasars (Hao et al.
2011). From this comparison, we can only conclude that the highest
luminosity E94 quasars are bluer than the highest luminosity XC413
quasars. We are not sure if this result can be explained by
selection effects only.

In the higher $z$ bins, a population of outliers is present toward
the top right corner. These outliers are the hot-dust-poor (HDP)
quasars discussed in detail in Hao et al. (2010, 2011).  The
fraction of sources outside the mixing wedge is quite similar in the
top three $\log L_{bol}$, $\log M_{BH}$ and $\log \lambda_E$ bins.
This result agrees with the lack of evolution in HDP fraction with
$M_{BH}$ and $\lambda_E$ (Hao et al. 2010, 2011).

In first approximation, the contrast between nuclear AGN continuum
and host galaxy in B band (rest-frame) can be expressed in a single
formula (Merloni \& Heinz 2012): $$\frac{L_{AGN, B}}{L_{host,
B}}=\frac{\lambda_E}{0.1}\frac{(M_*/L_B)_{host}}{3(M_{\bigodot}/L_{\bigodot})}(B/T),$$
where $(M_*/L_B)_{host}$ is the mass-to-light ratio of the host
galaxy and (B/T) is the bulge-to-total galactic stellar mass ratio.
So for typical mass-to-light ratios and bulge-to-total galactic
stellar mass ratios, the contrast will be smaller if $\lambda_E$ is
smaller, hinted as shown in the bottom row of
Figure~\ref{slopezLmepage}, from left to right. When $\lambda_E$ is
getting larger, the quasars generally drift towards the quasar
dominated direction (smaller $f_g$).

A minority of XC413 quasars in each of the lower bins of $z$, $\log
L_{bol}$ lie in the highly reddened region ($E(B-V)>0.4$,
Figure~\ref{slopezLpage}). For example, for the lowest $z$ bin, the
fraction is 6\% (4 out of the 71 sources). In SS206
(Figure~\ref{slopezLmepage}), quasars in the highly reddened region
only exist in the lower bins of $z$ and $\log L_{bol}$. This effect
is not so evident in $\log M_{BH}$ and $\log \lambda_{E}$, where a
small fraction of highly reddened quasars appear in high $\log
M_{BH}$ or $\log \lambda_{E}$ bins. Extremely low Eddington ratio
AGN ($\lambda_E<10^{-4}$) tend to have red optical SED, unlike the
typical quasars ($\lambda_E>0.01$) with ``big-blue-bump'' (Ho et al.
2008, Trump et al. 2011). In SS206, we do not see any obvious trend
that small Eddington ratio quasars are more reddened for the typical
quasars. This is probably due to the small Eddington ratio range in
SS206 compared to the large difference between the low luminosity
AGN and the normal AGN.

%%%%%%%%%%%%%%%%%%%%%%%%%%%%%%%%%%%%%%%%%%%%%%%%%%%%%%
\subsection{Inferred Host Galaxy Fraction} \label{s:fg}
%%%%%%%%%%%%%%%
\begin{figure}
\epsfig{file=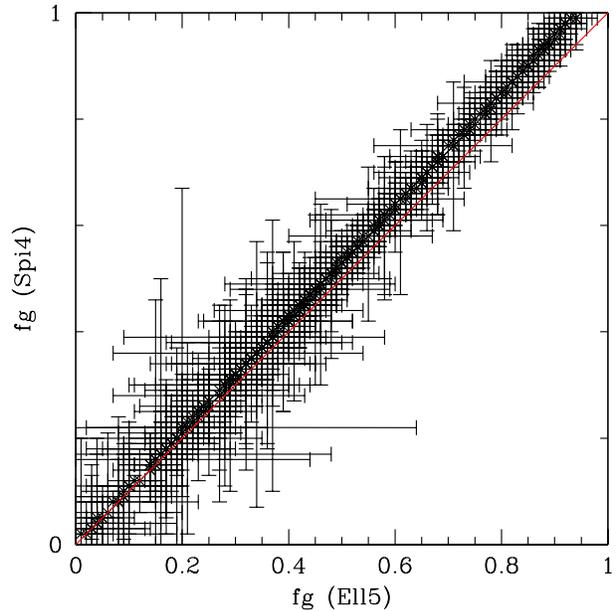, angle=0,width=\linewidth} \caption{The
$f_g$ value of the XMM-COSMOS sample using the 5 Gyr elliptical
galaxy template (Ell5) and the spiral galaxy (Spi4) from SWIRE
template library (Polletta et al. 2007). The red solid line shows
the one-to-one relation. \label{fgtmpcmp}}
\end{figure}
%%%%%%%%%%%%%%%

%%%%%%%%%%%%%%%
\begin{figure*}
\includegraphics[angle=0,width=0.31\textwidth]{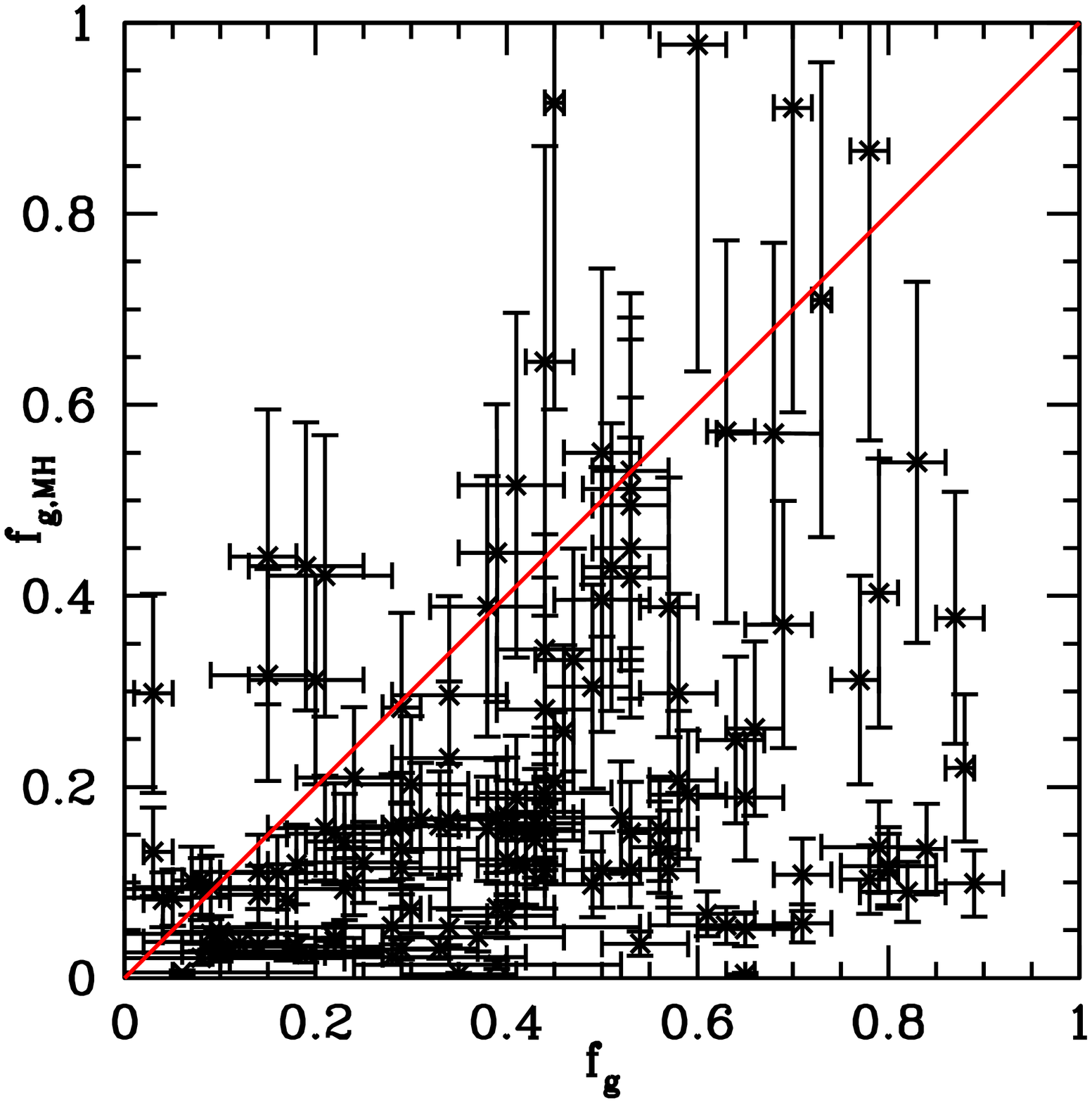}
\includegraphics[angle=0,width=0.31\textwidth]{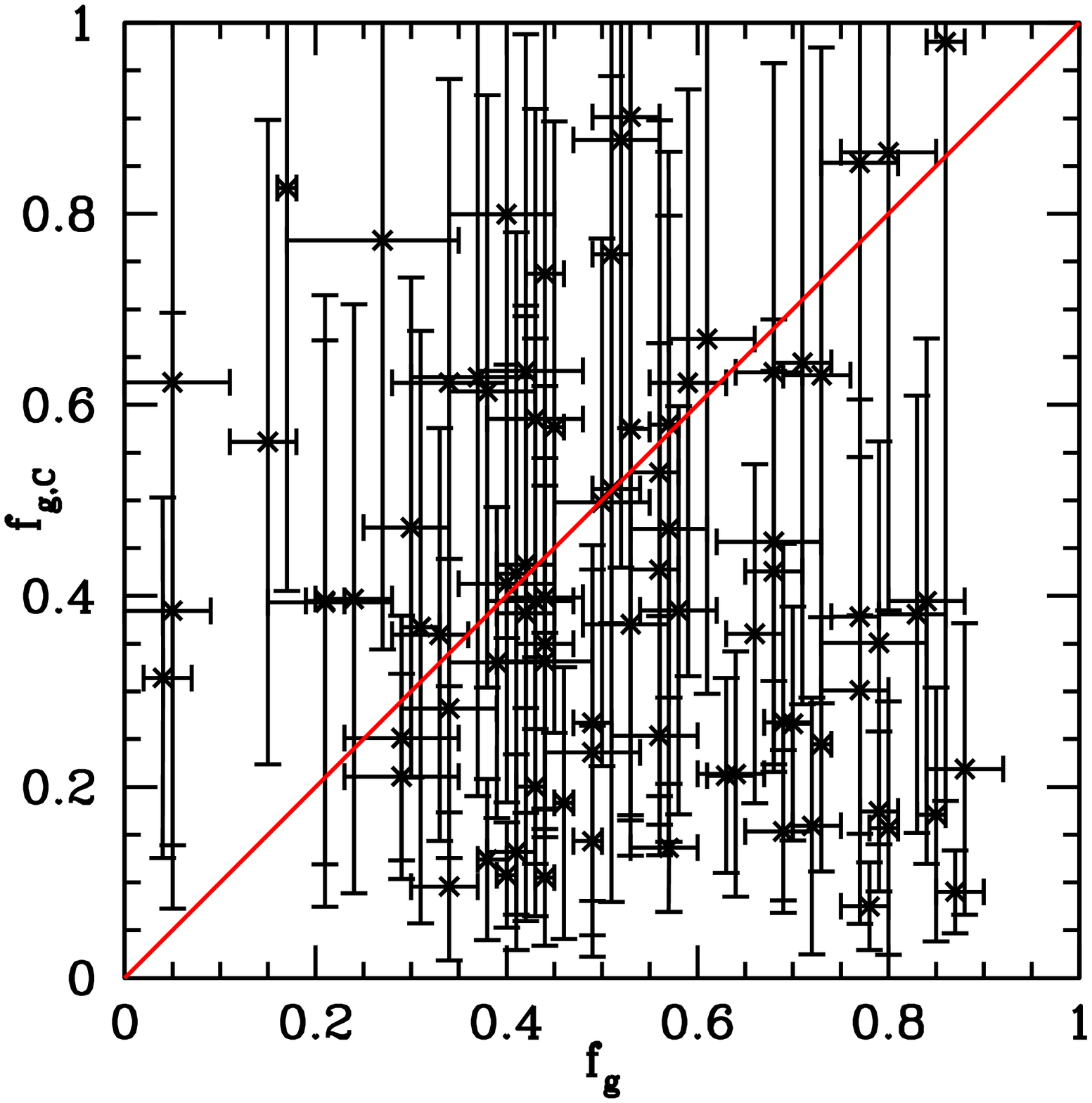}
\includegraphics[angle=0,width=0.31\textwidth]{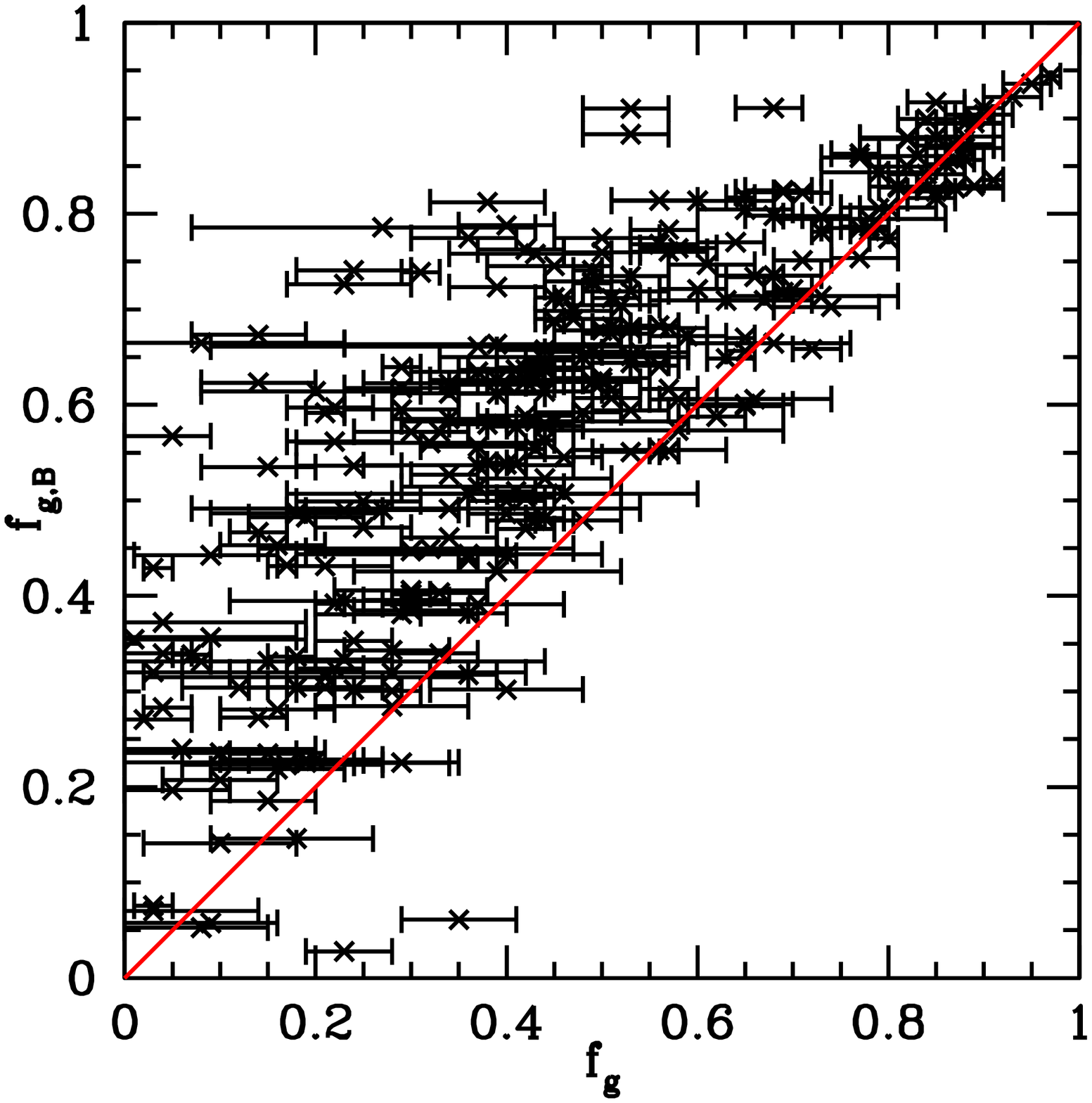}
\includegraphics[angle=0,width=0.31\textwidth]{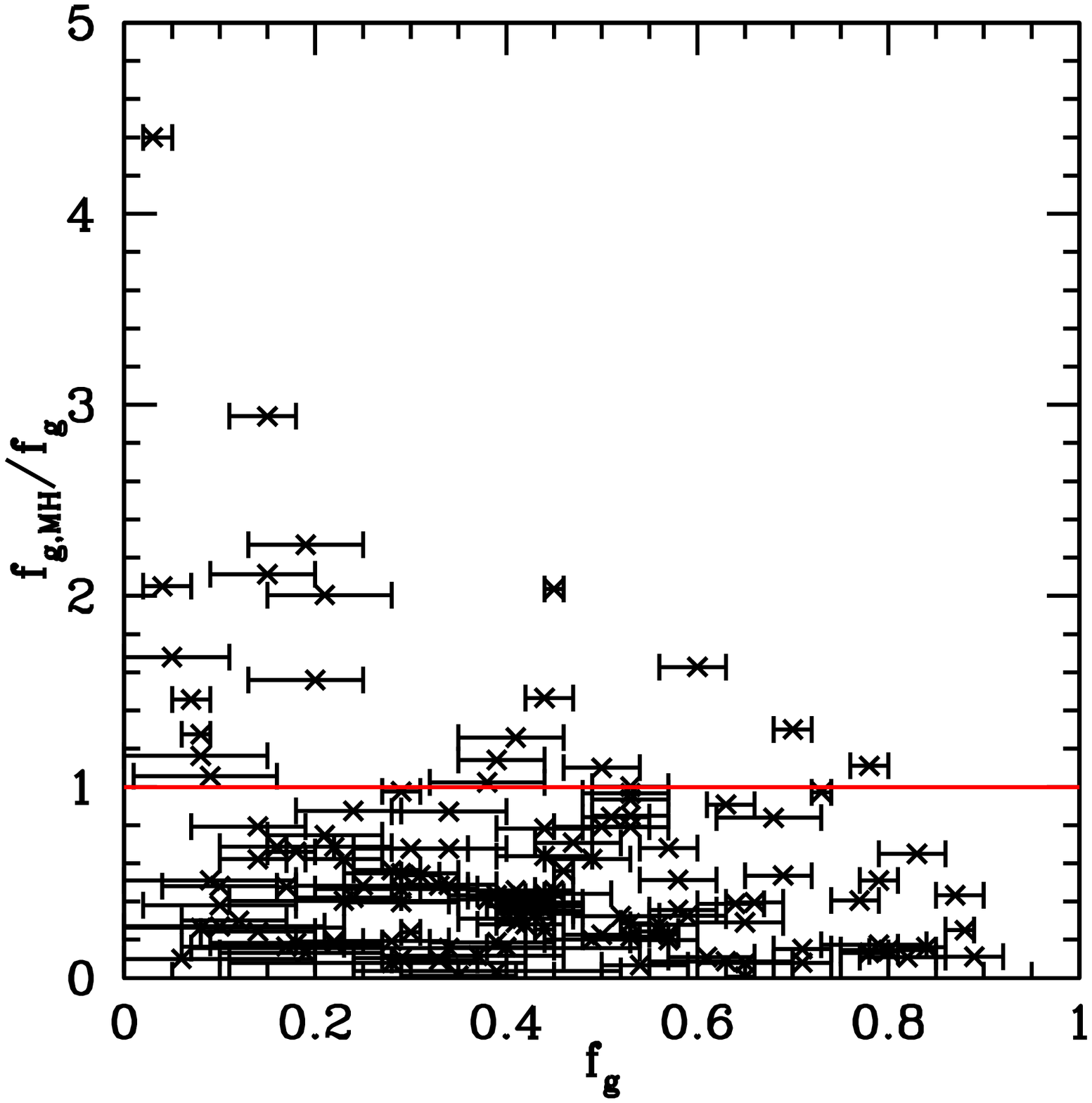}
\includegraphics[angle=0,width=0.31\textwidth]{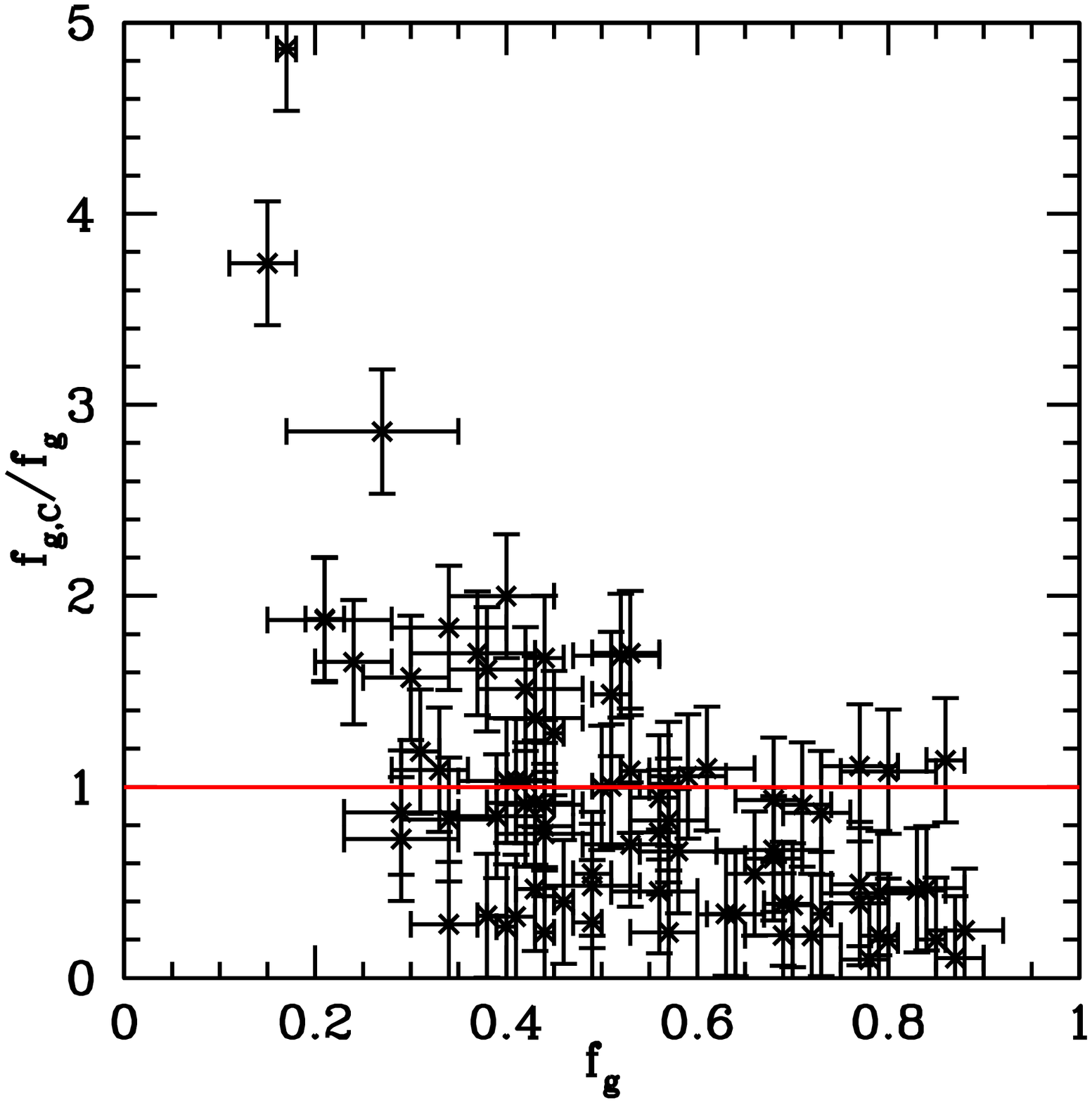}
\includegraphics[angle=0,width=0.31\textwidth]{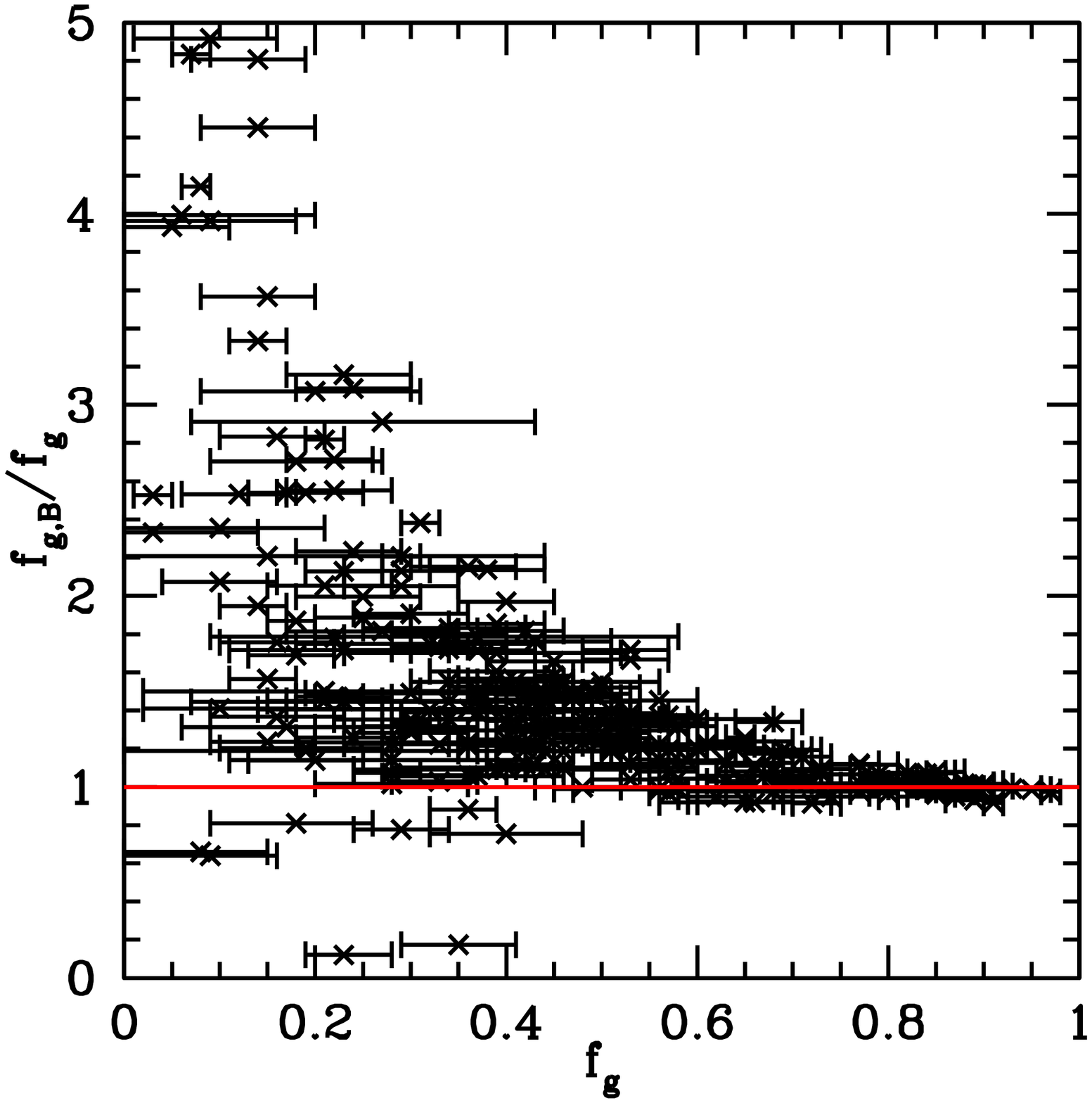}
\includegraphics[angle=0,width=0.31\textwidth]{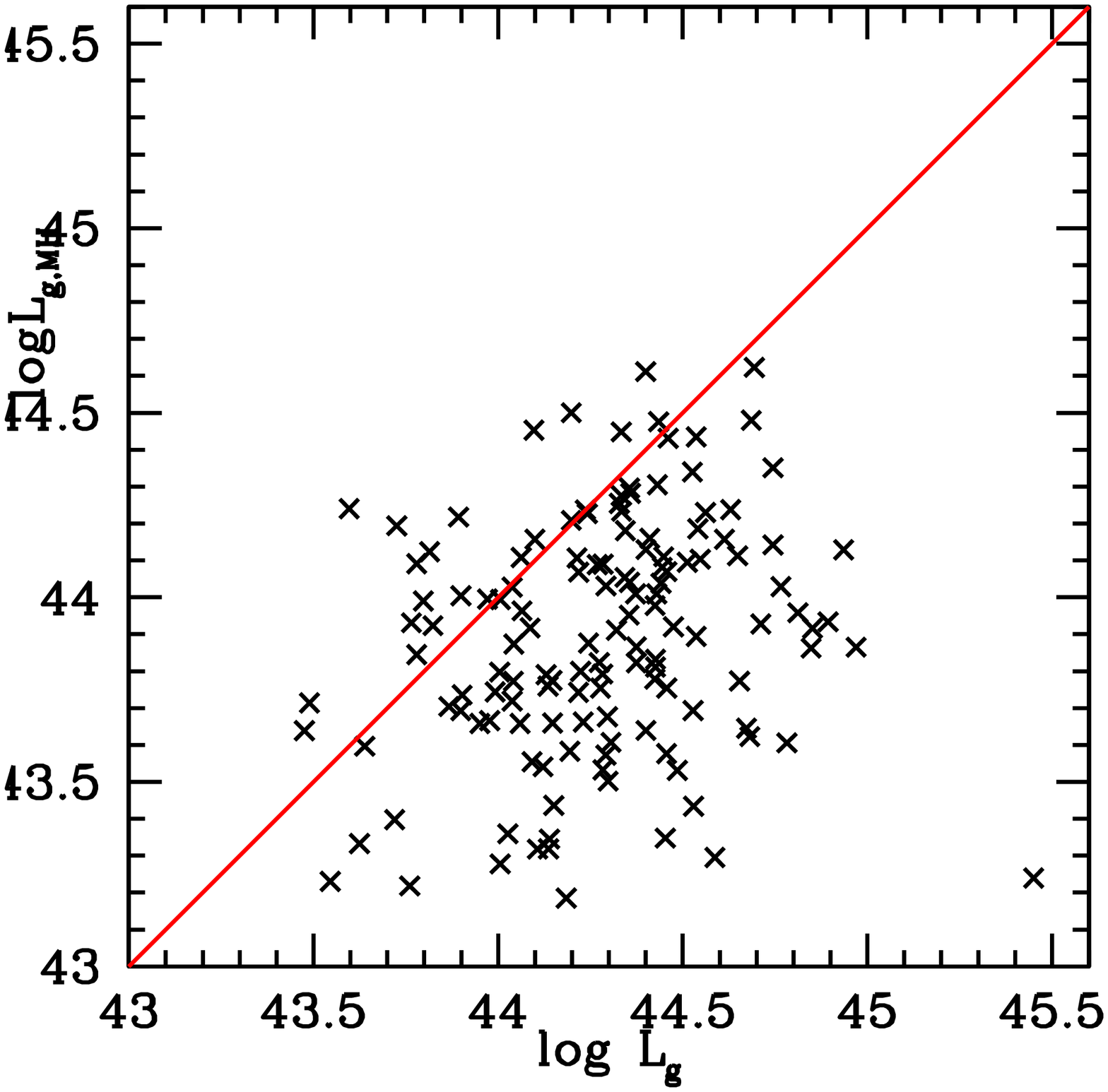}
\includegraphics[angle=0,width=0.31\textwidth]{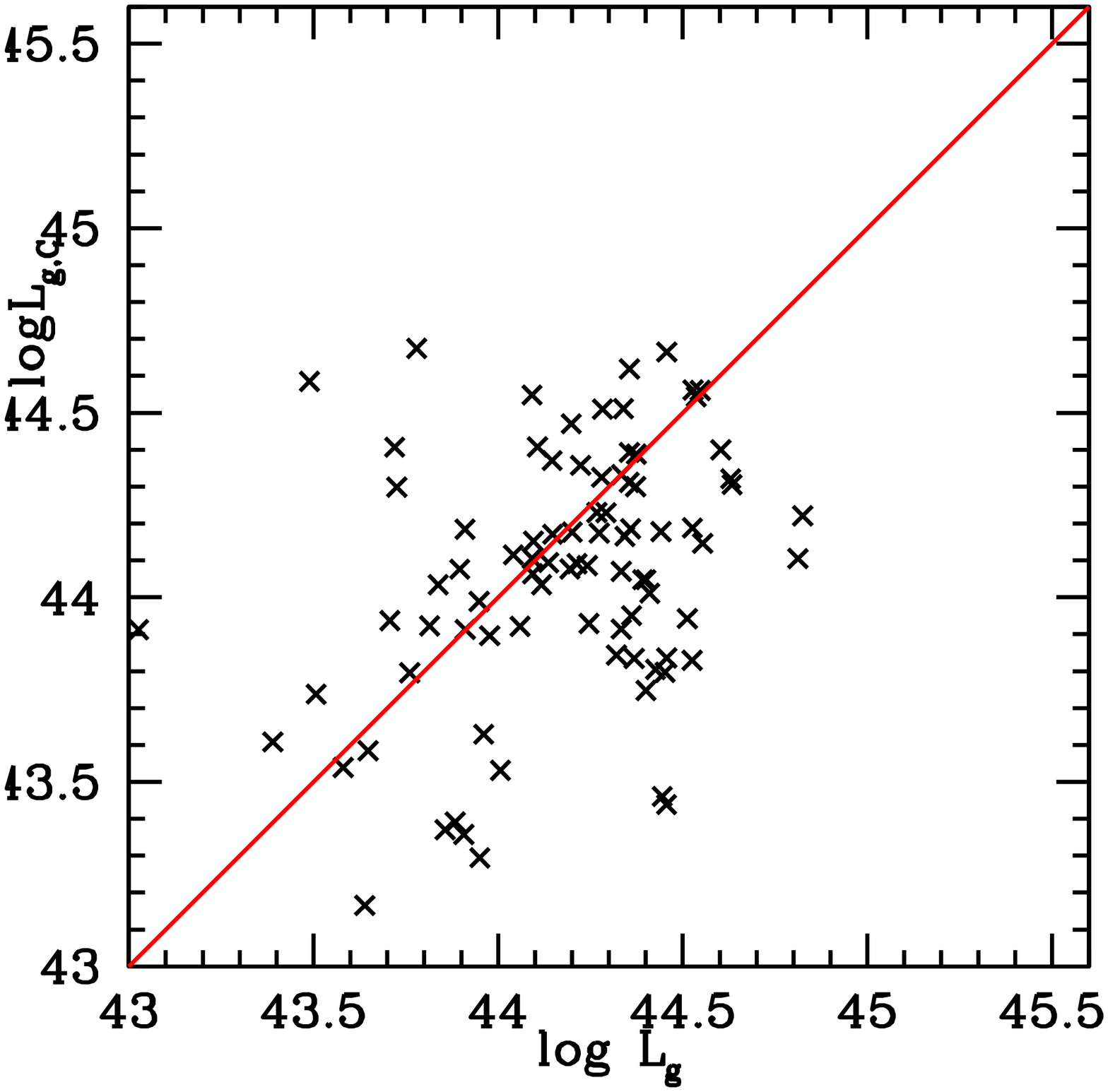}
\includegraphics[angle=0,width=0.31\textwidth]{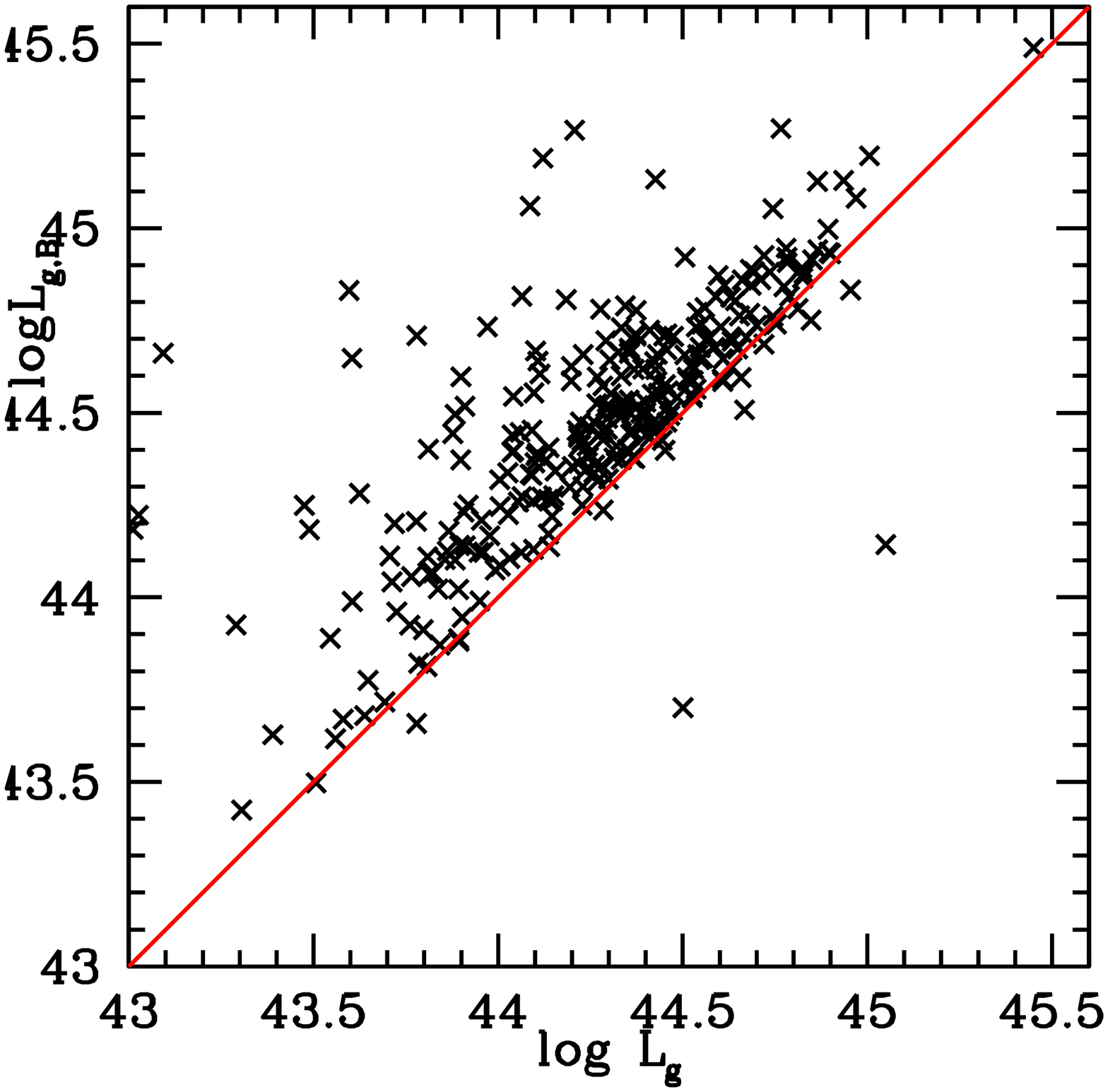}
\includegraphics[angle=0,width=0.31\textwidth]{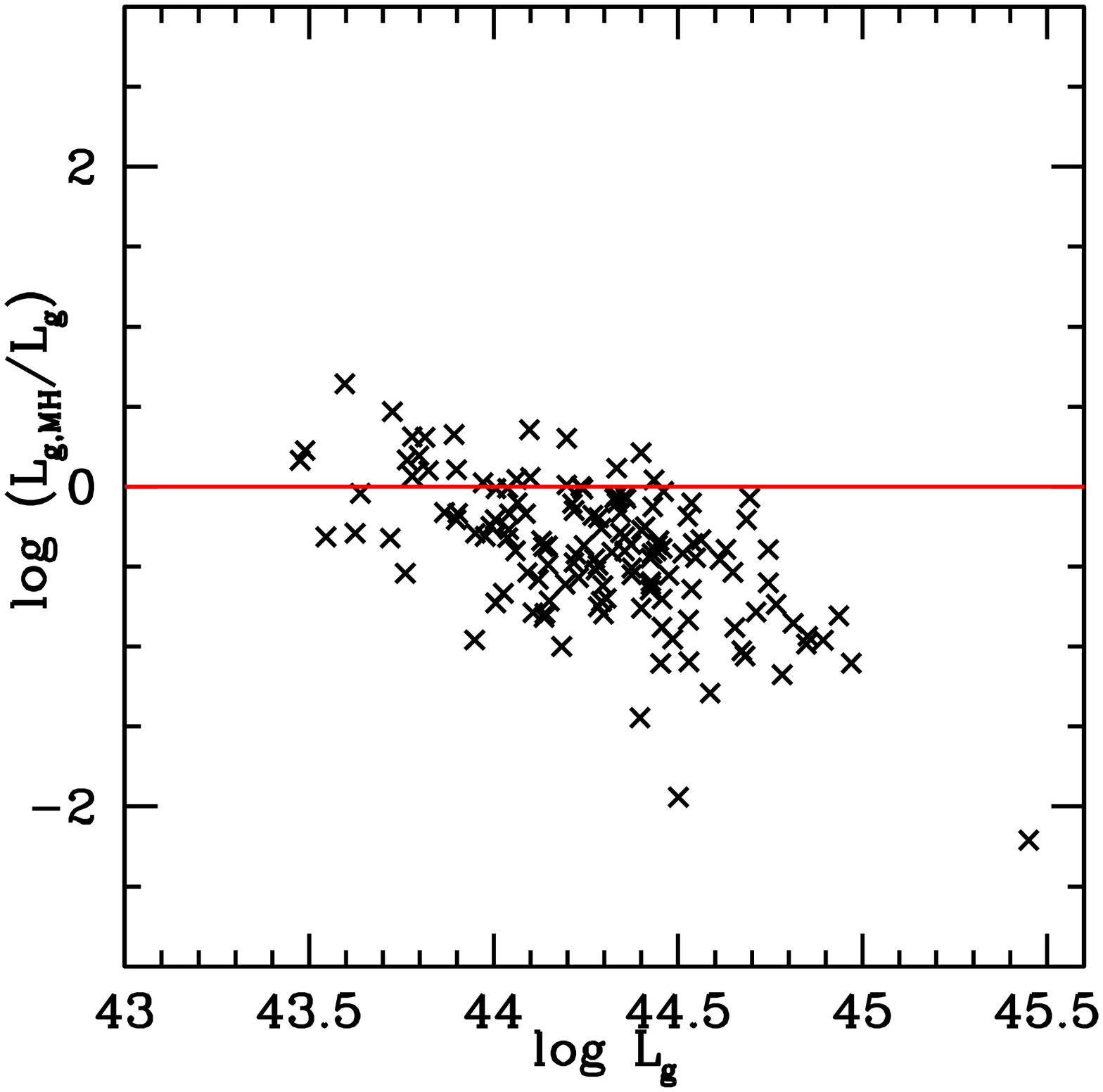}
\includegraphics[angle=0,width=0.31\textwidth]{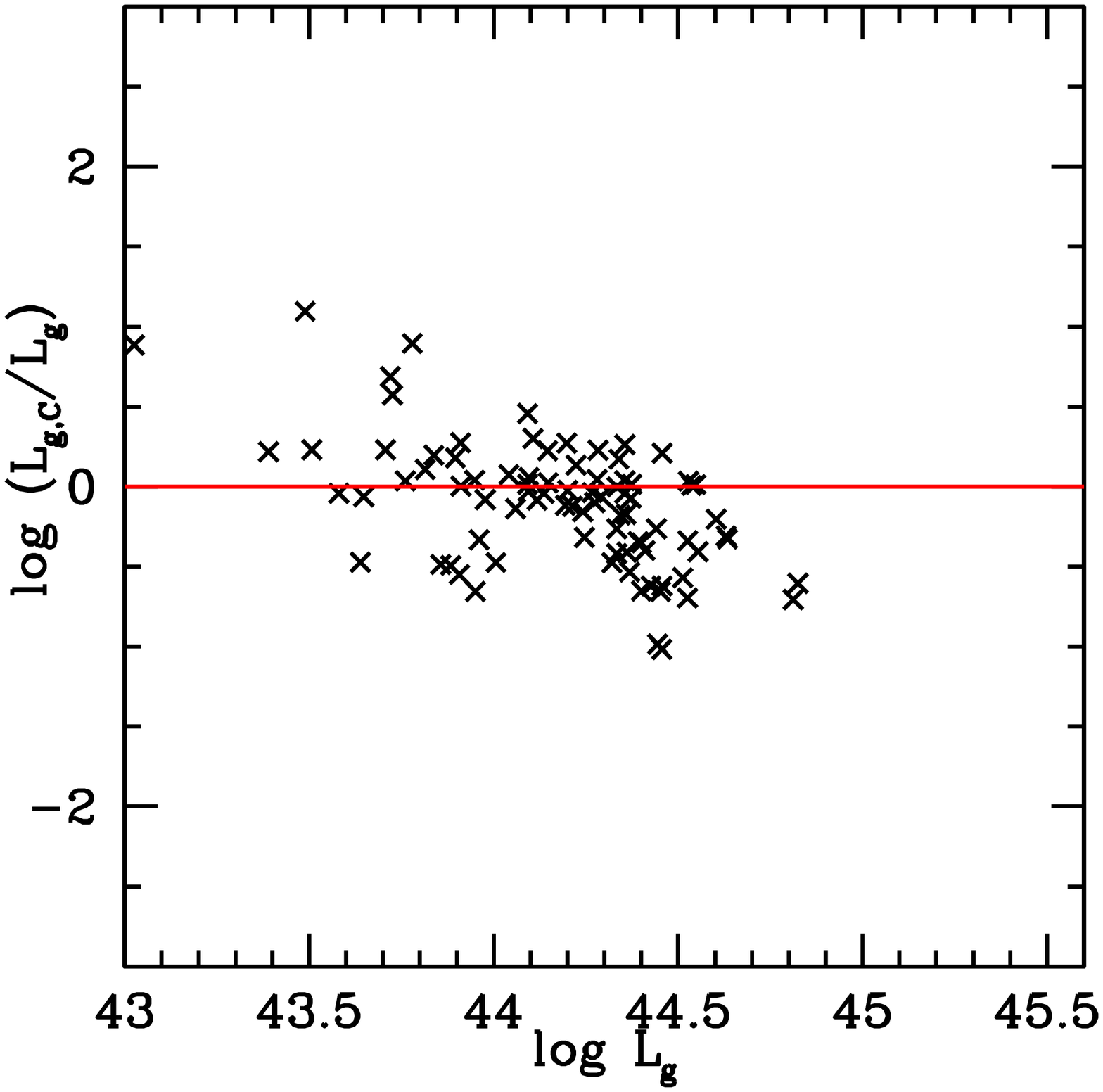}
\includegraphics[angle=0,width=0.31\textwidth]{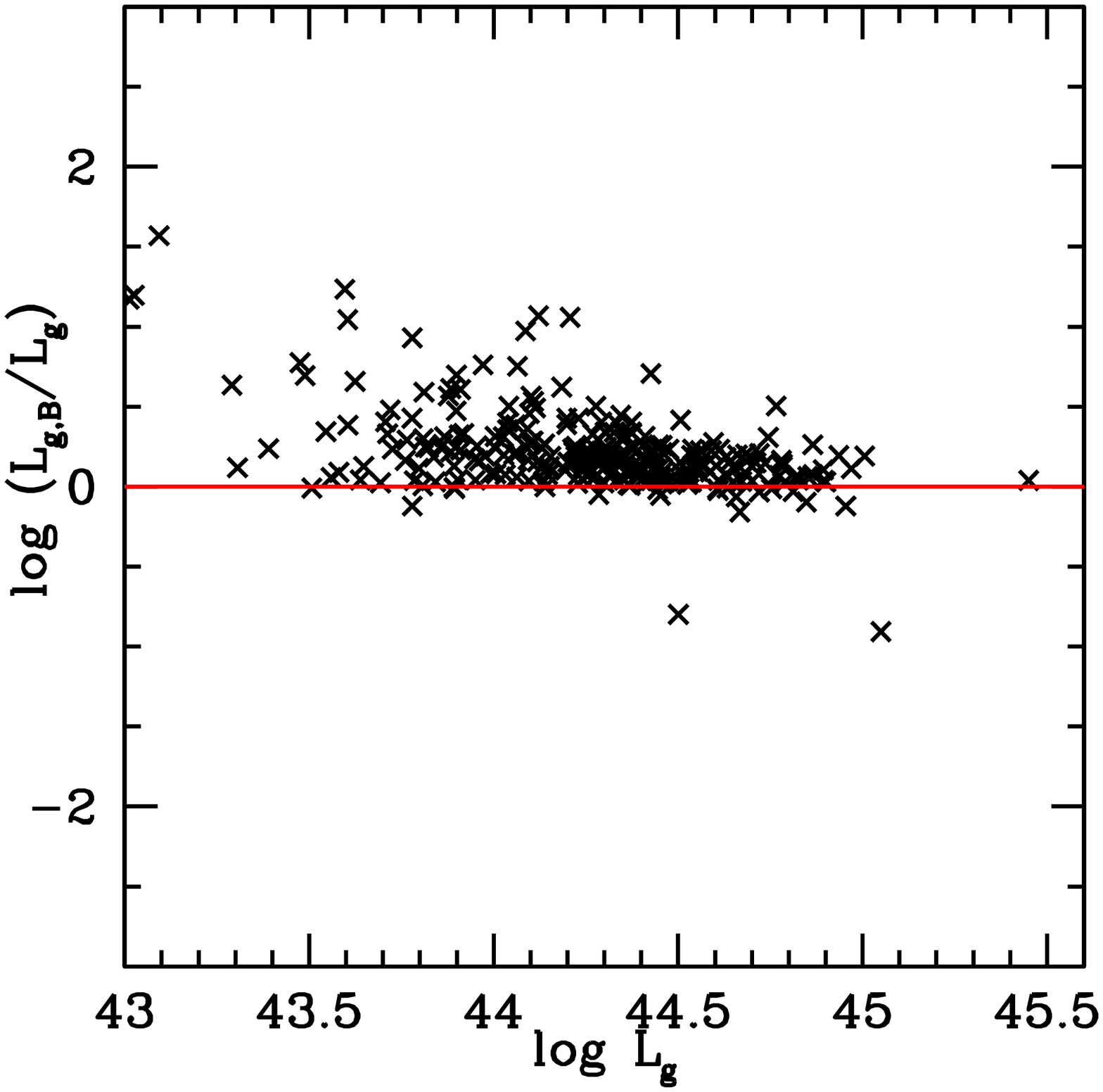}
\caption{The comparison of the galaxy fraction (top) and host galaxy
luminosity (bottom) of the XMM-COSMOS sample at rest frame $1~\mu$m:
(1) from the mixing diagram: $f_g, \log L_g$, using Ell5 mixing
curve, (2) from the Marconi \& Hunt (2003) scaling relationship
adding an evolutionary term (Bennert et al. 2010, 2011): $f_{g,MH},
\log L_{g,MH}$, (3) from the Hubble image decomposition (Cisternas
et al. 2011): $f_{g,C}, \log L_{g,C}$, and (4) from the SED fitting
(Bongiorno et al. 2012): $f_{g,B}, \log L_{g,B}$ . The red solid
lines show the one-to-one relation. \label{fgcomp}}
\end{figure*}
%%%%%%%%%%%%%%%

The mixing diagram provides a new estimate of the galaxy fraction
$f_g$ (\S~\ref{s:redvector}). The errors on the $f_g$ estimates are
caused by the error on the slopes, due to linear fitting of the
SEDs. Different galaxy templates also give slightly different $f_g$
values. Figure~\ref{fgtmpcmp} compares the values for two templates.
The differences are negligible and almost unbiased, compared to the
errors on $f_g$. The correlation coefficient is 1 (precise to the
4th place after decimal). If we fit a straight line, the best fit
slope is $1.06\pm0.001$, very close to 1.

We can compare $f_g$ with host galaxy fractions derived with three
other methods: using bulge - black hole scaling relations, direct
imaging and SED fitting. The three methods are briefly described
below:

(1) {\em Black hole mass - Galaxy bulge scaling relations:} For the
203 quasars in SS206, following Paper I, we used the relationship
between the black hole mass and near-infrared bulge luminosity
(Table 2 of Marconi \& Hunt, 2003) adding an evolutionary term
(Bennert et al. 2010, 2011) to estimate the host galaxy
contribution:
\begin{eqnarray}
\log(L_{J,Gal})&=&0.877\log(L_{bol})+3.545-0.877\log\lambda_E\nonumber\\
& &-1.23\log(1+z) \label{e:LJhost}
\end{eqnarray}

We used the Ell5 galaxy template to calculate the rest frame
$1~\mu$m host luminosity. In this band the differences among
different galaxy templates are small. With the host luminosity we
can calculate the galaxy fraction at rest frame $1~\mu$m
($f_{g,MH}$). The rest frame J band ($1.2~\mu$m) luminosity
$L_{J,gal}$ is used because this is the band closest to $1~\mu$m,
and is where the galaxy contribution peaks.

The small photometric errors in J imply that black hole mass
measurement errors dominate the error on $f_{g, MH}$.  Black hole
mass estimates from mass scaling relationships have an error $\Delta
M_{BH}/M_{BH}\sim40\%$ (Vestergaard \& Peterson 2006; Peterson
2010), so $\Delta f_{g,MH}/f_{g,MH}\sim35\%$, as $f_{g,MH}\propto
M_{BH}^{0.877}$ (according to Equation \ref{e:LJhost}).

(2) {\em Direct imaging:} For 94 low redshift ($z\lesssim1.2$)
quasars in the XMM-COSMOS sample, Cisternas et al. (2011) used the
Hubble images to decompose the AGN and galaxy emission and to
estimate the host galaxy fraction at 8140\AA\ (observed frame). We
transformed this galaxy fraction to the rest frame $1~\mu$m galaxy
fraction ($f_{g,C}$) using the Ell5 galaxy template.  As only the
best fit model of the host galaxy luminosity is given, we cannot
estimate the error on $f_{g,C}$ due to the fitting process.

However, the assumed template introduces an uncertainty.  The
observed F814W photometry point lies on the steep side of the galaxy
template for $z>0.1$. Hence, a small error in template slope (or,
effectively, in the age of the youngest stellar population in the
host) would lead to a large error in the host estimate at $1~\mu$m.
We can use this extrapolation uncertainty to estimate a minimum
error. To do so, we normalized the 16 SWIRE galaxy templates
(Polletta et al. 2007) at $1~\mu$m and measured the dispersion of
these different templates at the rest frame wavelength corresponding
to the observed 8140\AA. We use these dispersions as errors on the
host galaxy luminosities $L_{g, C}$ at $1~\mu$m for sources at
different redshifts. Therefore, the error on the galaxy fraction can
be estimated as $\Delta f_{g,C}/f_{g, C}=\Delta L_{g,C}/L_{g, C}$.
The error bar ranges from 0.02 to 0.77 with the median value of
0.21.

(3) With the multi-wavelength photometry data available, SED fitting
can be used to decompose the observed SEDs with some assumptions on
the intrinsic component SEDs. Bongiorno et al. (2012) used R06 with
SMC like dust-reddening (Prevot et al. 1984), and BC03 models with
Calzetti reddening (Calzetti et al. 2000) to fit the XMM-COSMOS
sources. We calculate the galaxy fraction and host galaxy luminosity
at rest-frame $1~\mu$m from their SED fitting and compare them with
the results derived directly from the mixing diagram.

The comparison of $f_g$ estimated from the mixing diagram with the
galaxy fraction from the other three methods $f_{g,MH}$ (from Paper
I), $f_{g,C}$ (from Cisternas et al. 2011) and $f_{g, B}$ (from
Bongiorno et al. 2012) is shown in the top row of Figure
\ref{fgcomp}. For ease of comparison, we also plot the ratio of the
$f_g$ values from the other three methods over $f_g$ from the mixing
diagram versus the $f_g$ from mixing diagrm in the second row.

The first two methods (scaling relationships and direct imaging
decomposition) give values which are poorly correlated with the
$f_g$ values from the mixing diagram. The correlation coefficient
for $f_g$ and $f_{g,MH}$ is 0.35, for $f_g$ and $f_{g,C}$ is -0.09.
$f_{g,MH}$ gives systematically smaller values than the other
methods. From equation~\ref{e:LJhost}, this effect is either due to
a systematic underestimate of black hole mass, or to an
over-estimate of the evolution of the scaling relationship, which
may be more likely (see e.g., Schramm \& Silverman 2013).

However, the $f_{g, B}$ values from the SED fitting are strongly
correlated with the $f_g$ values from the mixing diagram, although
with a shift in normalization. The correlation coefficient between
the two sets of values is 0.83. The host galaxy fraction from the
SED fitting is systematically slightly higher than the results of
the mixing diagram. This is probably due to the galaxy template
model employed. The SED fitting in Bongiorno et al. (2012) used BC03
models, which generally have smaller $\alpha_{\rm{NIR}}$ so the
mixing curves are shorter, leading to larger galaxy fractions (see
right panel of Figure~\ref{slp}).

The inferred $1~\mu$m host galaxy luminosities ($L_g$, $L_{g,MH}$,
$L_{g,C}$ and $L_{g,B}$) are also compared in Figure~\ref{fgcomp}
(bottom two rows). The correlation coefficient between $L_g$ and
$L_{g,MH}$ is 0.17, between $L_g$ and $L_{g,C}$ is 0.33 while that
between $L_g$ and $L_{g,B}$ is 0.75. For most cases the inferred
host galaxy luminosity $\nu L_{\nu}$ is less than $10^{44.6}$erg/s
(that is $M_{1\mu m}>-23$), a reasonable value, as $M_K^*\sim-23$ at
$0.25\leq z\leq 1.5$ (Cirasuolo et al. 2007).

Using the mixing diagram to estimate the host galaxy fraction
requires the following assumptions: 1) an intrinsic quasar SED
exists and is similar to E94 mean SED; 2)the chosen galaxy templates
are representative; 3) all the quasars have a similar reddening
curve which is SMC like. The first assumption is somewhat reasonable
based on the dependency studies of mean SEDs with physical
parameters (Paper II) and \S~\ref{s:intsdisp} in this paper. As
shown in Figure~\ref{fgtmpcmp}, choosing different galaxy templates
would give very similar results ($\lesssim 1\sigma$) even for the
host dominated sources. Therefore, the validity of the second
assumption will not affect the result much. In practice, one can
choose the proper galaxy templates that are closest to the
population in discussion or be more careful when citing the $f_g$
for galaxy dominated sources. The reliability of the third
assumption is hard to asses. For the currently commonly used
extinction curves (SMC, LMC and MW), the difference is small (see
\S~\ref{s:febv}).

Estimation of the host galaxy fraction using scaling relationships
has a large uncertainty due to the dispersion of the relationship
itself (e.g. Marconi \& Hunt 2003, Merloni et al. 2010) and
possibility of evolution in the relationship (e.g. Merloni et al.
2010, Schramm \& Silverman 2013). There are also significant
uncertainties of the $M_{BH}$ estimates (Vestergaard \& Peterson
2006; Peterson 2010).

Estimates of the host galaxy fraction using image decomposition
(e.g. Cisternas et al. 2011) is observationally limited to moderate
redshifts. Most importantly, this method leads to large
uncertainties, because the ratio of the host galaxy to AGN
luminosity is a strong function of the wavelength. The uncertainties
of the intrinsic SED shapes in both the host and the quasar will
lead to large uncertainty in the fraction if we transfer from the
observed wavelength to another wavelength we are interested in
(Paper I).

From the mixing diagram we can easily derive the host galaxy
fractions at $1~\mu$m and obtain reasonably consistent values with
the results from SED fitting (see rightmost panel of
Figure~\ref{fgcomp}). The obvious advantage of the use of the mixing
diagram is that it is simple to construct and is directly derived
from the photometry. The SED fitting uses multi-wavelength data over
a larger frequency range, but is hard to estimate exactly how the
results depend on the number of different components and the assumed
component templates.

The $f_g$ calculated from the mixing diagram is thus useful and
reliable compared to other methods.

\subsection{Inferred Reddening}
\label{s:febv}
%%%%%%%%%%%%%%%
\begin{figure*}
\includegraphics[angle=0,width=0.32\textwidth]{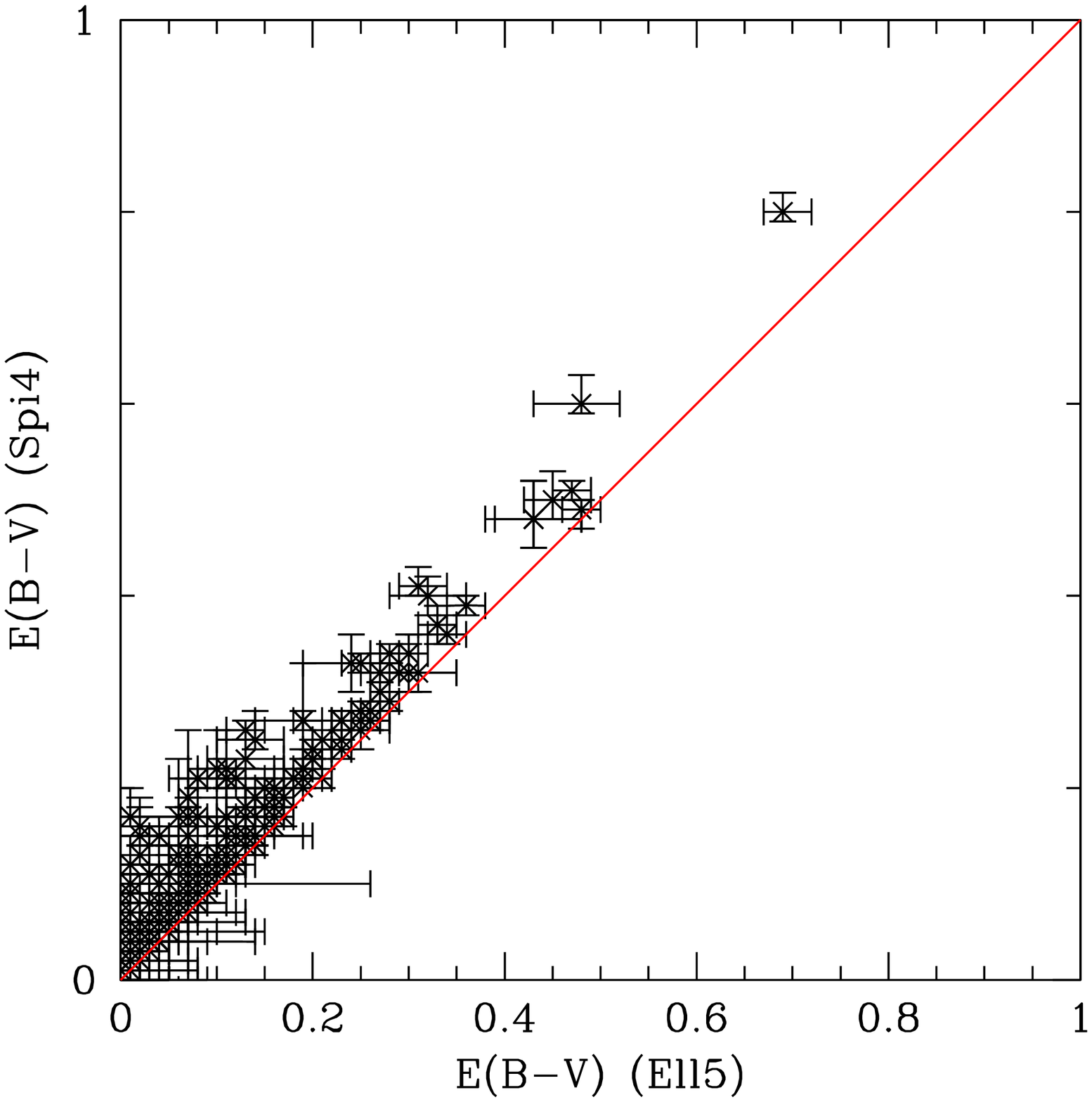}
\includegraphics[angle=0,width=0.32\textwidth]{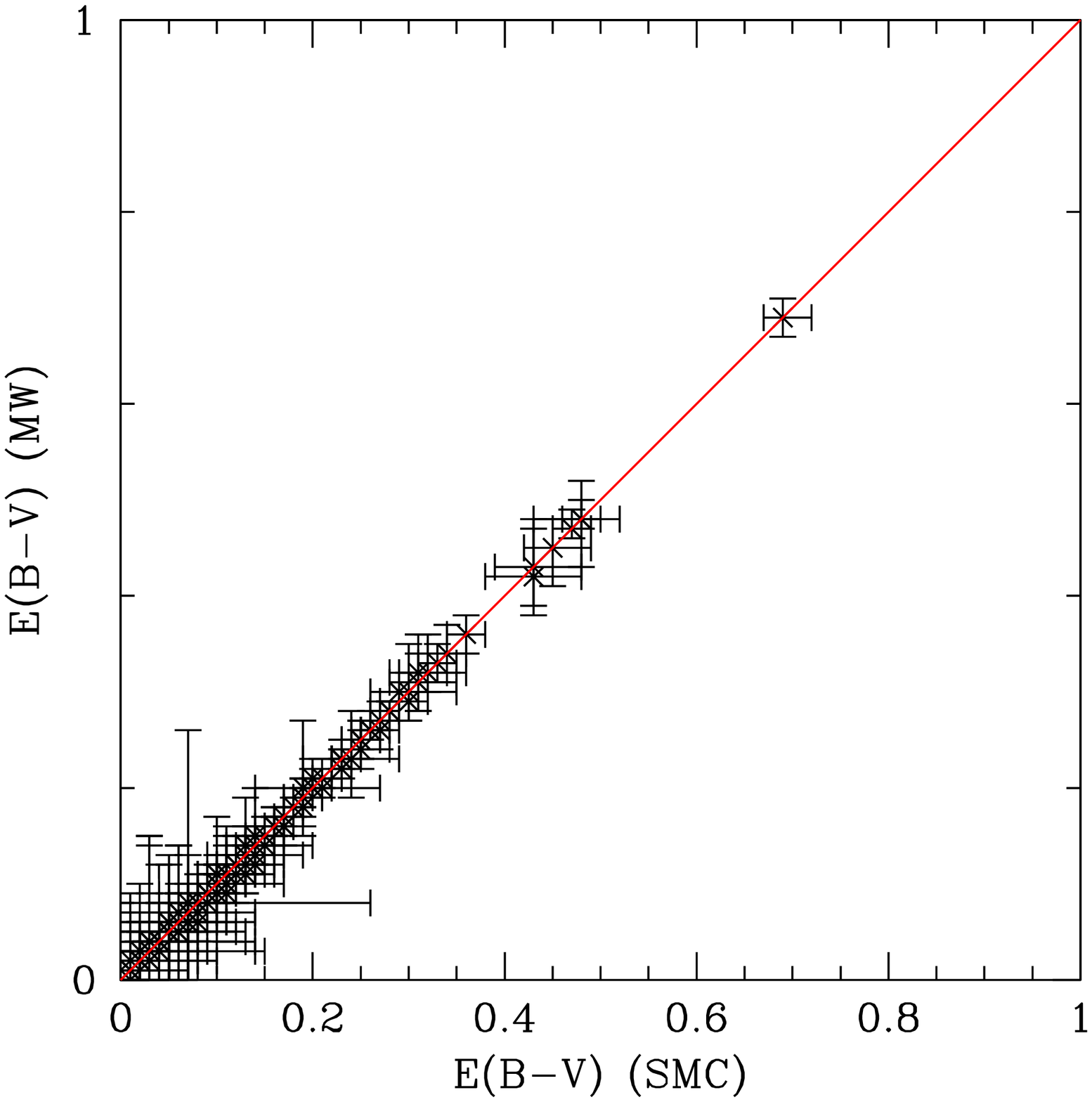}
\includegraphics[angle=0,width=0.32\textwidth]{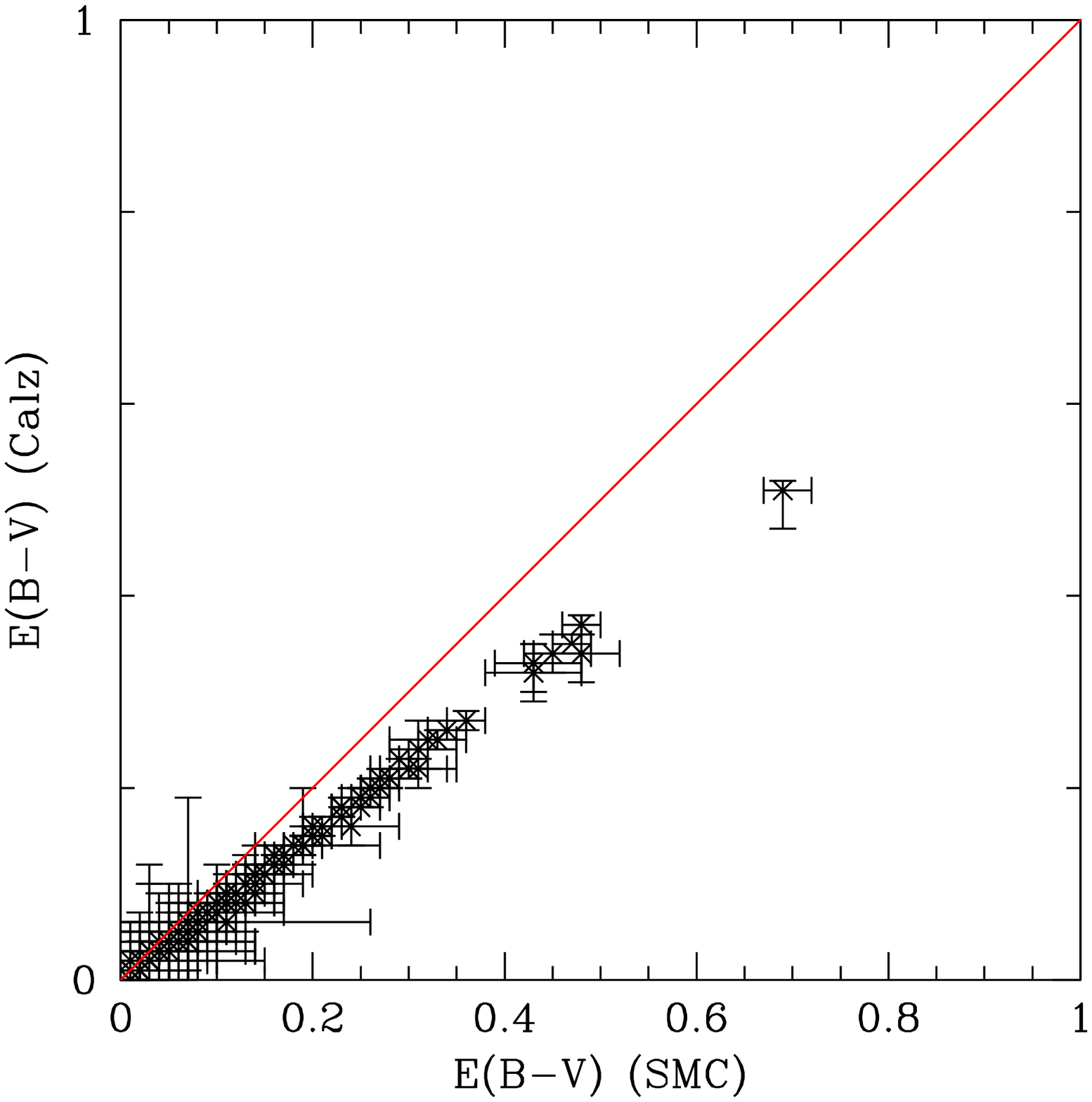}
\caption{The $E(B-V)$ estimates of the XMM-COSMOS sample
(1)\emph{left:} using the 5 Gyr elliptical galaxy template (Ell5)
and the spiral galaxy (Spi4) from SWIRE template library (Polletta
et al. 2007) and using the SMC reddening law for both axis; (2)
\emph{center:} using the SMC and MW reddening law respectively and
using the Ell5 galaxy template for both axis; (3) \emph{right:}
using the SMC and Calzetti et al. (2000) reddening law respectively
and using the Ell5 galaxy template for both axis;. The red solid
line shows the one-to-one relation. \label{febvtmpcmp}}
\end{figure*}
%%%%%%%%%%%%%%%

%%%%%%%%%%%%%%%
\begin{figure*}
\includegraphics[angle=0,width=0.49\textwidth]{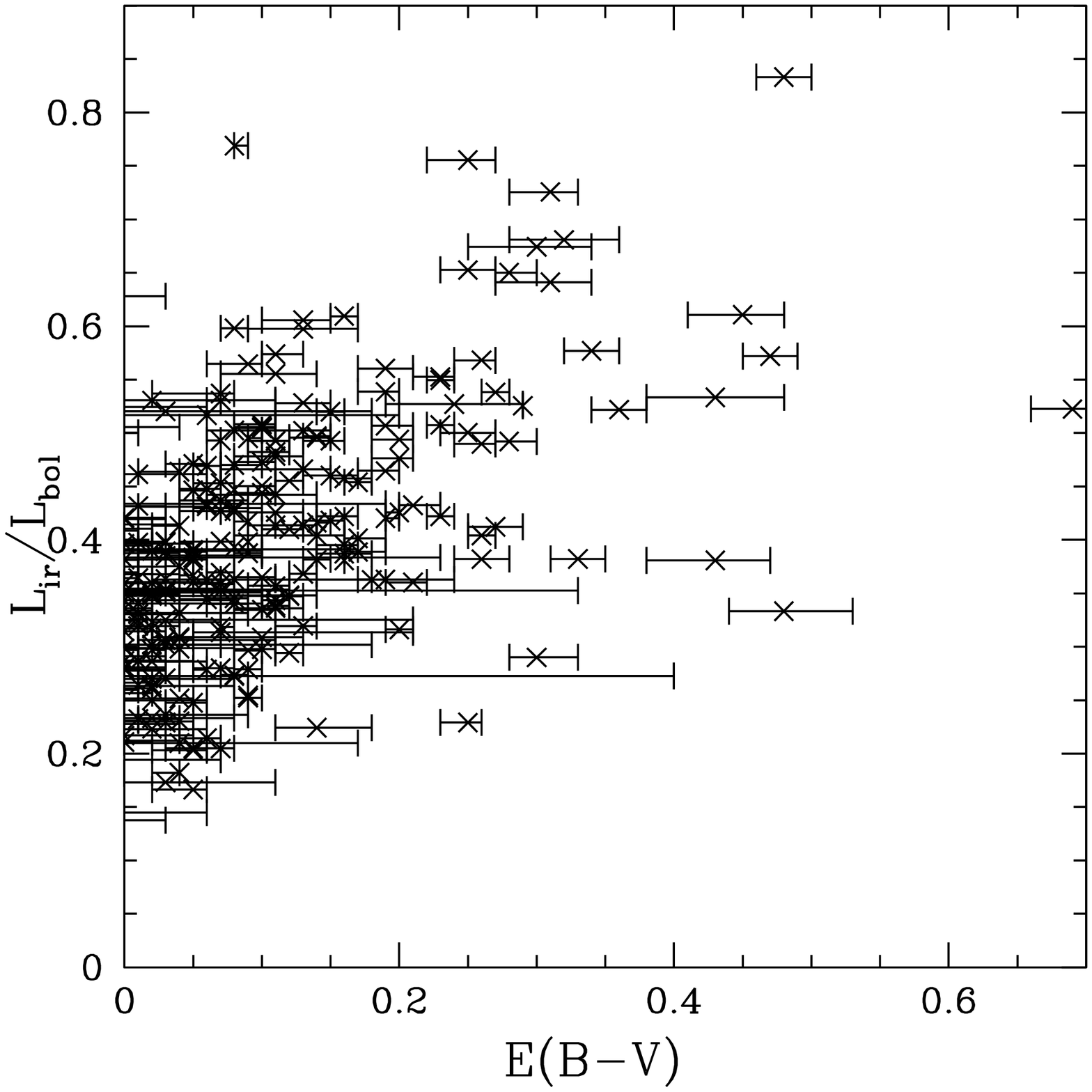}
\includegraphics[angle=0,width=0.49\textwidth]{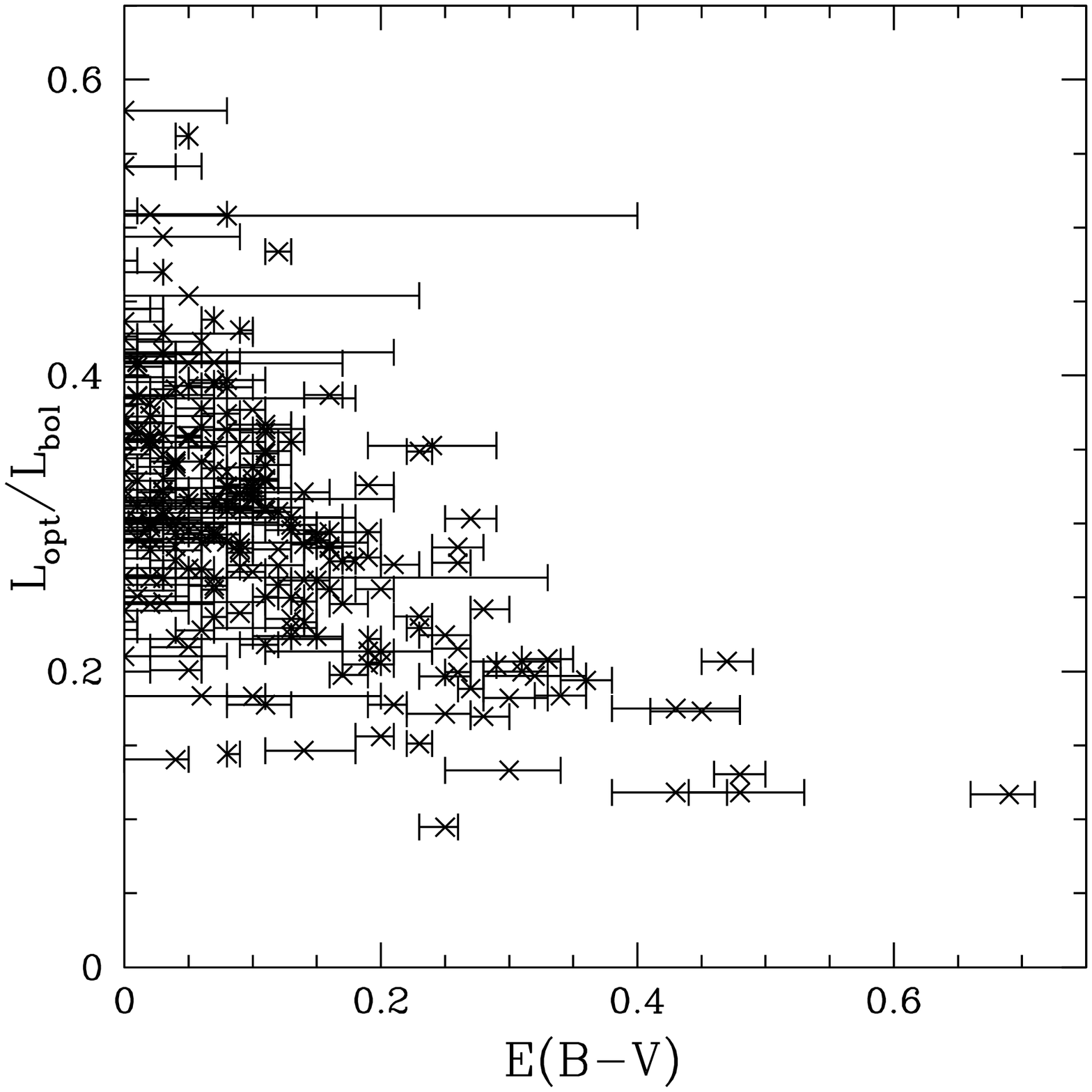}
\caption{The NIR luminosity fraction versus $E(B-V)$ (left) and the
optical luminosity fraction versus $E(B-V)$ (right). Here we use the
5 Gyr elliptical galaxy template (Ell5) from SWIRE template library
(Polletta et al. 2007) and the SMC reddening law to derive the
$E(B-V)$ estimates. \label{febvfr}}
\end{figure*}
%%%%%%%%%%%%%%%

%%%%%%%%%%%%%%%
\begin{figure*}
\includegraphics[angle=0,width=0.49\textwidth]{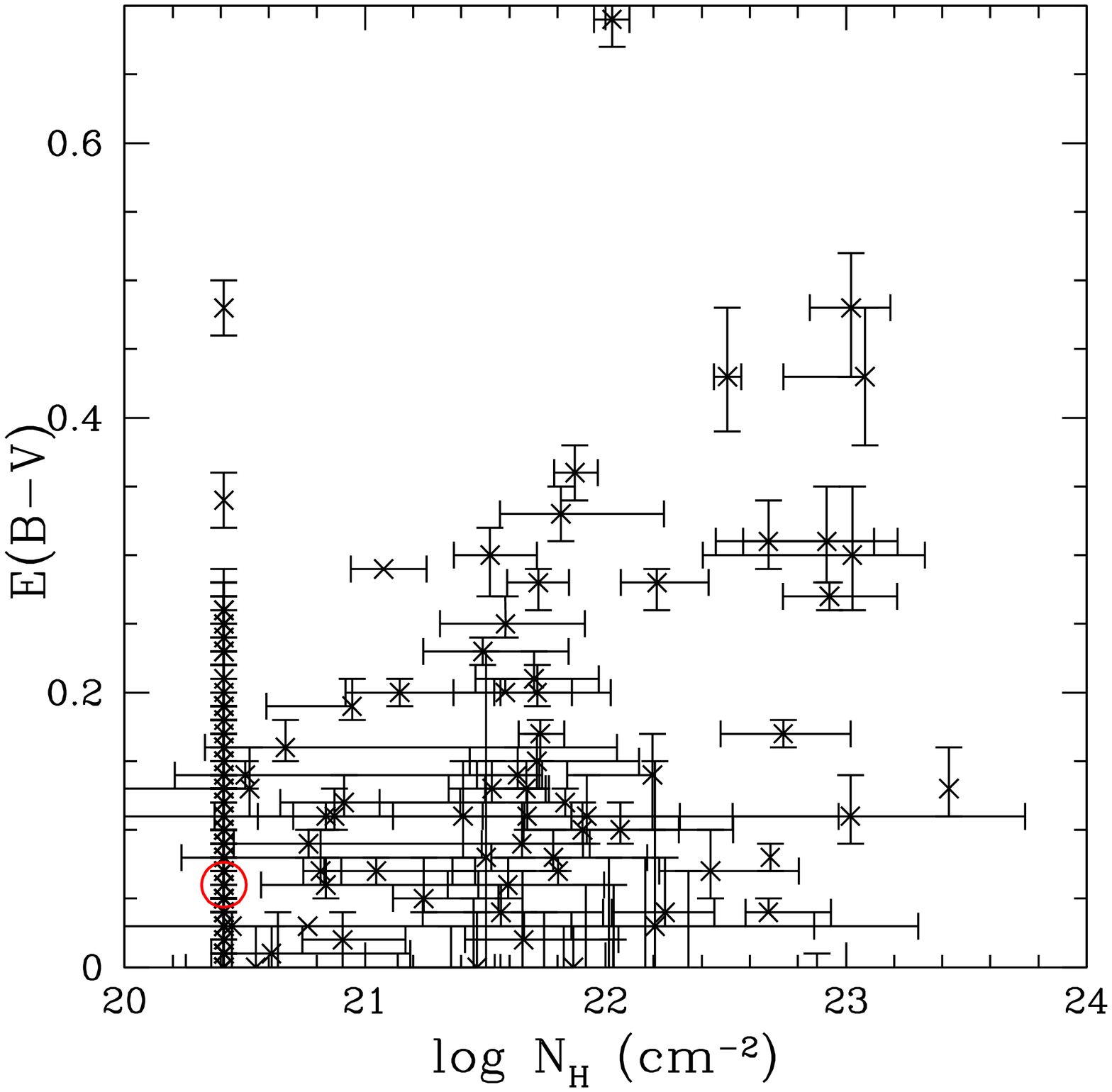}
\includegraphics[angle=0,width=0.49\textwidth]{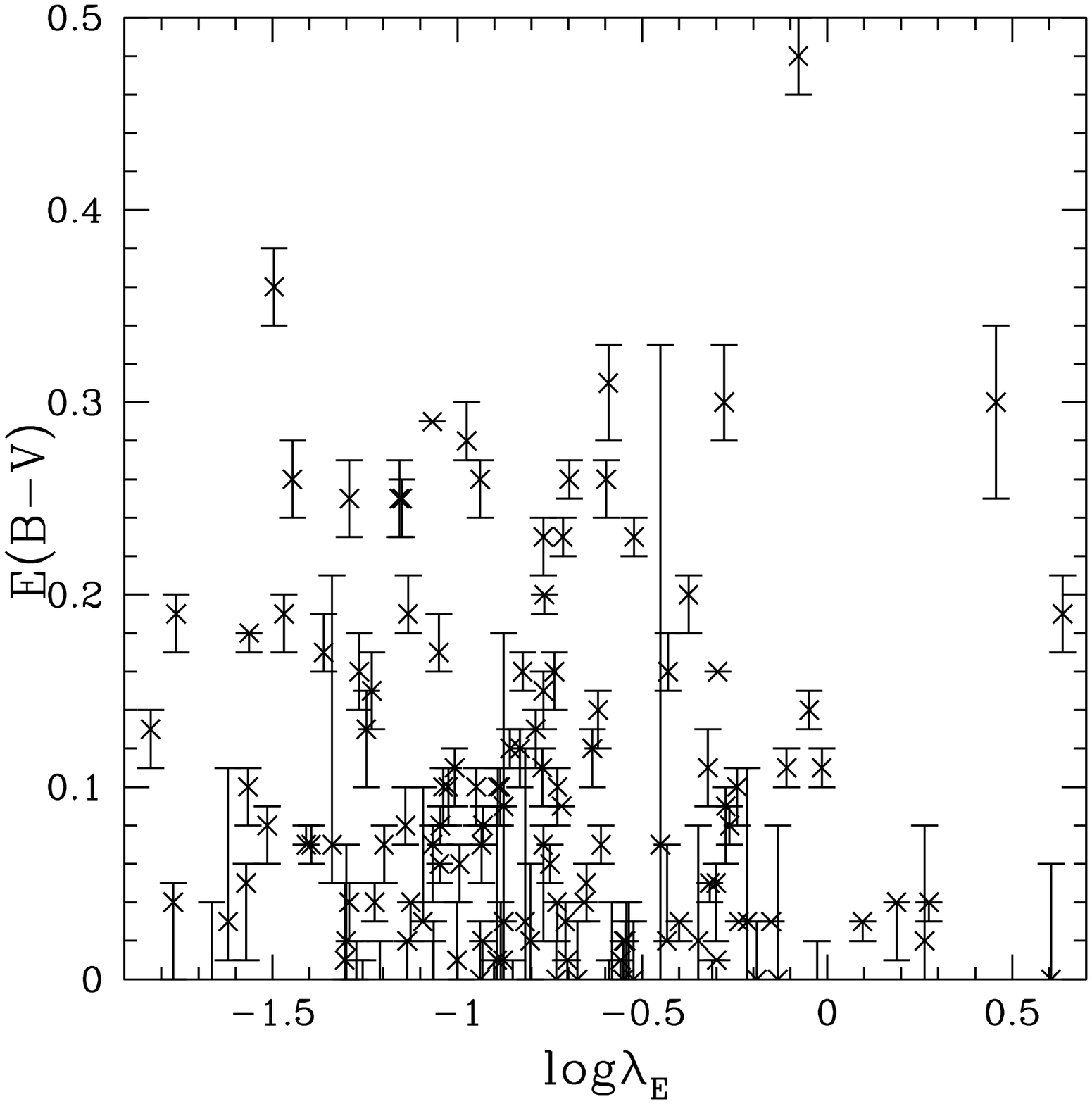}
\caption{$E(B-V)$ versus the neutral Hydrogen column density
$\rm{N_H}$ (left) and $E(B-V)$ versus Eddington Ratio
$log\lambda_E=log (L_{bol}/L_{Edd})$ (right). Here we use the 5 Gyr
elliptical galaxy template (Ell5) from SWIRE template library
(Polletta et al. 2007) and the SMC reddening law to get the $E(B-V)$
estimates. The red circle in the left panel shows the median
$E(B-V)$(=0.06) of the AGN with no intrinsic $\rm{N_H}$.
\label{febvnh}}
\end{figure*}
%%%%%%%%%%%%%%%

In addition to the galaxy fraction estimation, from the mixing
diagram we could get an estimation of the $E(B-V)$ value from the
position of the source on the mixing diagram (\S~\ref{s:redvector}).
The errors on the $E(B-V)$ estimates are also caused by the error on
the slopes due to linear fitting of the SEDs similar to the $f_g$
estimates. Different galaxy templates give different $f_g$ values.
the left panel of Figure~\ref{febvtmpcmp} compares the values
$E(B-V)$ for two templates. The correlation coefficient is 0.96 and
if a straight line is fitted, the slope is $0.95\pm0.01$, very close
to 1. Compared to $f_g$ estimates, the $E(B-V)$ estimates are more
affected by which galaxy template is chosen. This is expected as the
galaxy templates are distributed in a sparse region on the upper
left part of the mixing diagram. Thus, the mixing curves for
different templates would spread out in the large $f_g$ direction,
leading to large difference of the $E(B-V)$ estimates for the same
quasar.

Different reddening laws used in the mixing diagram will lead to
different $E(B-V)$ estimates (middle and right plots of the
Figure~\ref{febvtmpcmp}). As shown in Table~\ref{t:redv} and
Figure~\ref{slp}, the SMC and LMC reddening laws lead to the same
reddening vector. So there are no differences between the $E(B-V)$
estimates given by these two reddening laws. The MW reddening vector
is quite close to the SMC reddening vector leading to similar
results in the $E(B-V)$ estimates (center panel in
Figure~\ref{febvtmpcmp}). The correlation coefficient between the
SMC and MW $E(B-V)$ values is 0.999. If a line is fitted, the slope
is $0.99\pm0.02$ and the intersection is 0.0002. So the SMC, LMC and
MW reddening laws give the same $E(B-V)$ estimates. The $E(B-V)$
estimates derived from the Calzetti et al. (2000) reddening law are
different, especially for large $E(B-V)$. However, when compared to
results from the SMC law (right panel of Figure~\ref{febvtmpcmp}),
the correlation coefficient is 0.998 and the slope is $0.74\pm0.02$.
The estimates of $E(B-V)$ derived from different reddening laws are
all tightly correlated.

$E(B-V)$ is estimated by applying a standard extinction law to an
assumed intrinsic optical-to-NIR quasar SED template (e.g. Vasudevan
et al. 2009, Glikman et al. 2012). Here the SMC extinction curve is
chosen because the extinction curve of quasars is generally believed
to be better described by the SMC type (Hopkins et al. 2004;
Gallerani et al. 2010). The $E(B-V)$ estimate derived from the
mixing diagram is equivalent to assuming the E94 template as an
intrinsic quasar template and applying the SMC reddening law. As the
E94 template is the mean SED of the bright quasar sample, and for
each quasar in the E94 sample the possible reddening is not
corrected, we expect the E94 template to be slightly redder than the
intrinsic quasar SED. In this case, the $E(B-V)$ estimation derived
from the mixing diagram should be a lower limit. As a fraction of
the quasars lies in the upper right corner beyond the mixing curve
leading to negative $E(B-V)$ values, we ignore these quasars from
further discussion in this section. If different galaxy templates
with younger stellar populations are chosen, these sources could lie
within the mixing region with positive $E(B-V)$ estimates. The size
of the galaxy fraction clearly depends on the mixing curve chosen to
derive the $E(B-V)$ values.

Other than estimate the $E(B-V)$ from the optical-to-NIR SED, Balmer
decrements have been used historically to estimate the reddening
along the line of sight of quasars (e.g., Maiolino et al. 2001, Xiao
et al. 2012). However, this method requires spectra that include
both the H$\rm{\alpha}$ and H$\rm{\beta}$ lines, which is not
suitable for the XMM-COSMOS sample, because most of the quasars are
at redshifts around 1--2. Besides, Glikman et al. (2012) argued that
using the optical-to-NIR SED to estimate the reddening is much more
reliable than the Balmer decrements estimation.

Other independent estimates of $E(B-V)$ are very difficult. The
galaxy inclination derived from HST images or the total dust masses
estimated from the infrared luminosity might give a hint to how much
reddening we would expect, but to get $E(B-V)$ estimates by these
methods would require lots of assumption on the gas and dust content
of the host galaxy. Thus, it is very difficult to compare the
$E(B-V)$ values derived from the mixing diagram with those from
other measurements to test the reliability of the mixing diagram. In
general, we would expect that for quasars with high $E(B-V)$ values,
the infrared bump would be more prominent and the `big-blue-bump'
would be less prominent. We check the correlation of the NIR
luminosity fraction ($L_{ir}/L_{bol}$, where $L_{ir}$ is the
luminosity integrated from rest-frame 24$~\mu$m to $1~\mu$m, Paper
II) and the optical luminosity fraction ($L_{opt}/L_{bol}$, where
$L_{opt}$ is the luminosity integrated from rest-frame $1~\mu$m to
912\AA\, Paper II) with $E(B-V)$ respectively (Figure~\ref{febvfr}).
In Figure~\ref{febvfr}, we compare the optical and NIR luminosity
fraction with the $E(B-V)$ values derived from the Ell5 mixing curve
as an example. For the 226 quasars with positive $E(B-V)$ values
from the Ell5 mixing curve, the correlation coefficient for the NIR
luminosity fraction with $E(B-V)$ is 0.54 and for the optical
luminosity fraction with $E(B-V)$ is -0.62. So the optical and NIR
luminosity fractions with $E(B-V)$ are correlated as expected.

The neutral Hydrogen column density ($\rm{N_H}$) estimated from the
X-ray spectrum is usually used as an indicator of the absorber.
However, the optical and X-ray obscuration are caused by different
physical processes and thus can be very different in an object (e.g.
Crenshaw \& Kraemer 2001). We compare the estimated $E(B-V)$ values
from mixing diagram with the X-ray $\rm{N_H}$ values (Mainieri et
al. 2007) for the XMM-COSMOS sample (Figure~\ref{febvnh} left). For
the 413 quasars in XMM-COSMOS sample, 378 quasars have good enough
XMM spectra to make a fit. In 273 out of the 378 cases, no intrinsic
$\rm{N_H}$ is necessary from the spectrum, so the $\rm{N_H}$ value
is set to the Galactic $\rm{N_H}$ in the COSMOS region ($\rm{log
N_H=20.413~cm^{-2}}$). Using the 205 quasars with a $\rm{N_H}$
estimate and positive $E(B-V)$ give a correlation coefficient of
0.40, which corresponds to a significant correlation at $>5\sigma$
level. Figure~\ref{febvnh} (left) shows a clear correlation with
some potentially interesting outliers, e.g. objects with no
intrinsic $\rm{N_H}$ and high $E(B-V)$.

Low accretion rate (Eddington ratio $\lambda_E\lesssim10^{-4}$)
quasars are thought to have more reddened `big-blue-bump' (e.g. Ho
2008, Trump et al. 2011). We compare the estimated $E(B-V)$ versus
the Eddington ratio ($\lambda_E$) in Figure~\ref{febvnh} (right) to
see if there is a similar trend in XMM-COSMOS sample.  The
correlation coefficient between $E(B-V)$ and log$\lambda_E$ is
-0.035 for the 119 quasars with log$\lambda_E$ estimates and
positive $E(B-V)$ estimates, thus no correlation is observed. The
studies of Fabian et al. (2008, 2009) identify the effective
Eddington limit for dusty gas in the $\rm{N_H-\lambda_E}$ plane, and
therefore causing a `forbidden region' in the $\rm{N_H-\lambda_E}$
space within which absorbing dusty gas clouds are unstable to
radiation. Vasudevan et al. (2009) shows a similar `forbidden
region' in the upper right corner of the $E(B-V)-\lambda_{\rm{E}}$
plane. In the right panel of Figure~\ref{febvnh}, we can see a
similar lack of high accretion rate and high $E(B-V)$ objects.

%%%%%%%%%%%%%%%%%%%%%%%%%%%%%%%%%%%%%%%%%%%%%%%%%%%%%%
\subsection{Mixing Diagram Outliers} \label{s:outliers}
%%%%%%%%%%%%%%%%
\begin{figure*}
\includegraphics[angle=0,width=0.45\textwidth]{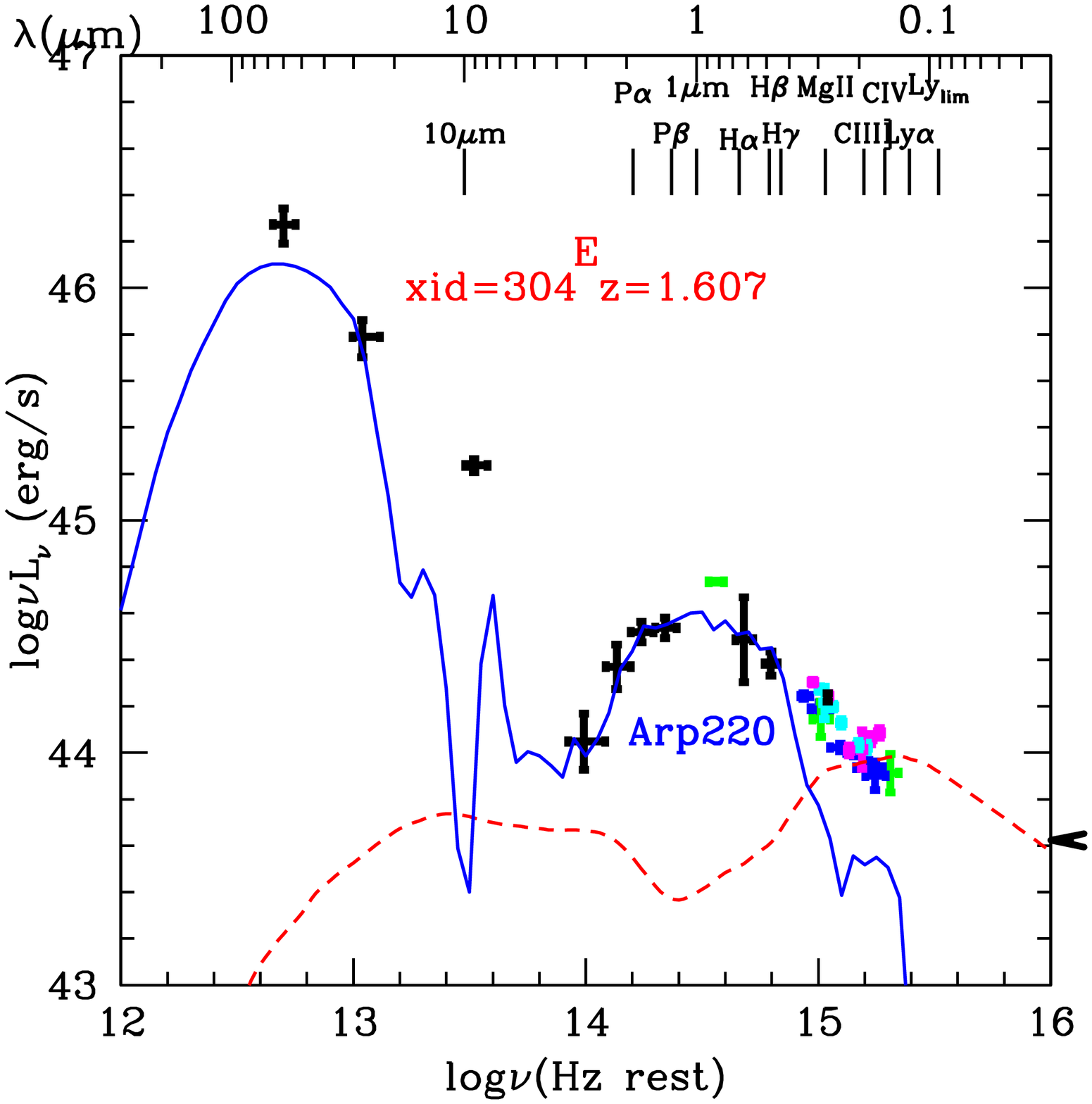}
\includegraphics[angle=0,width=0.45\textwidth]{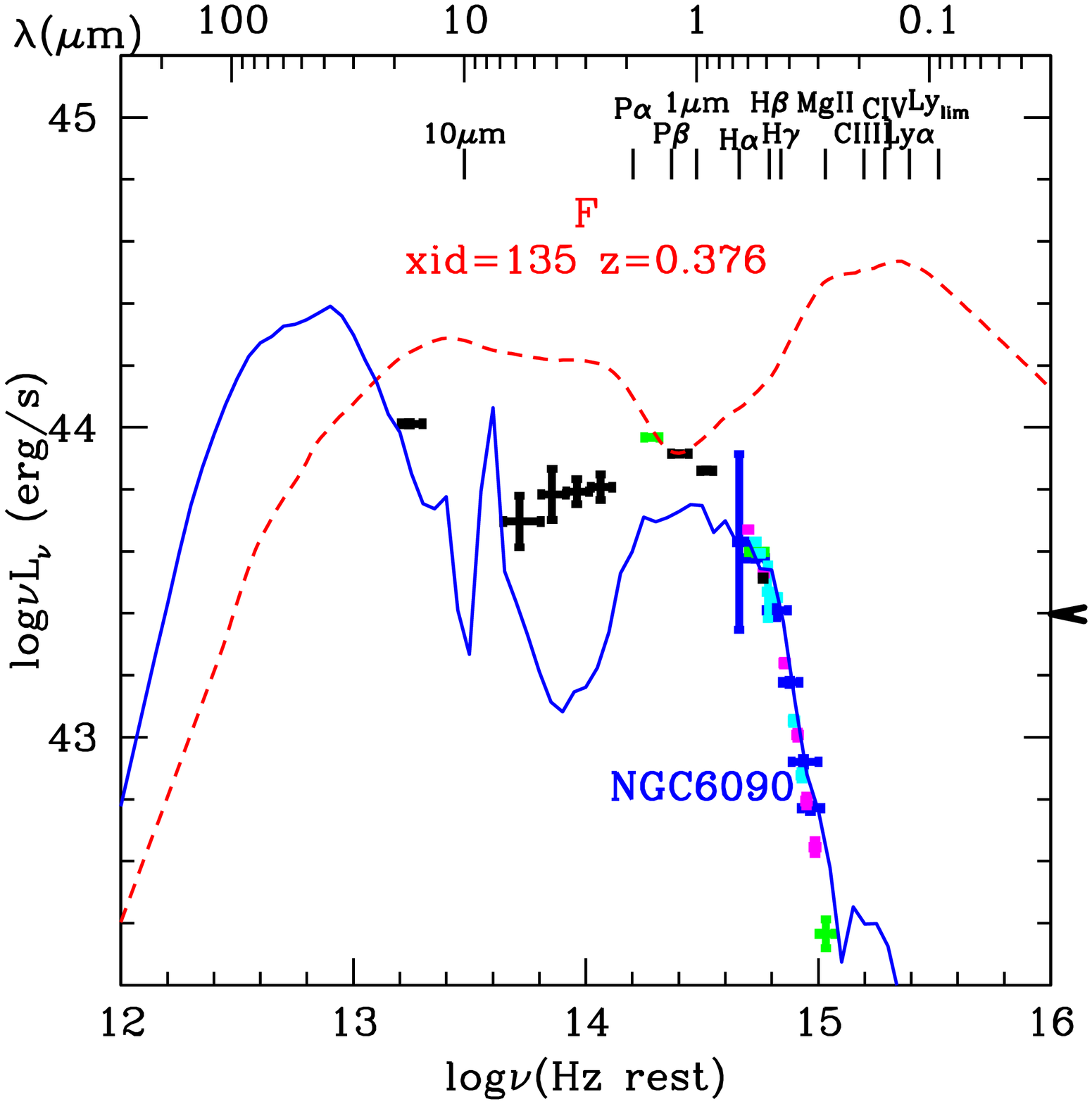}
\includegraphics[angle=0,width=0.45\textwidth]{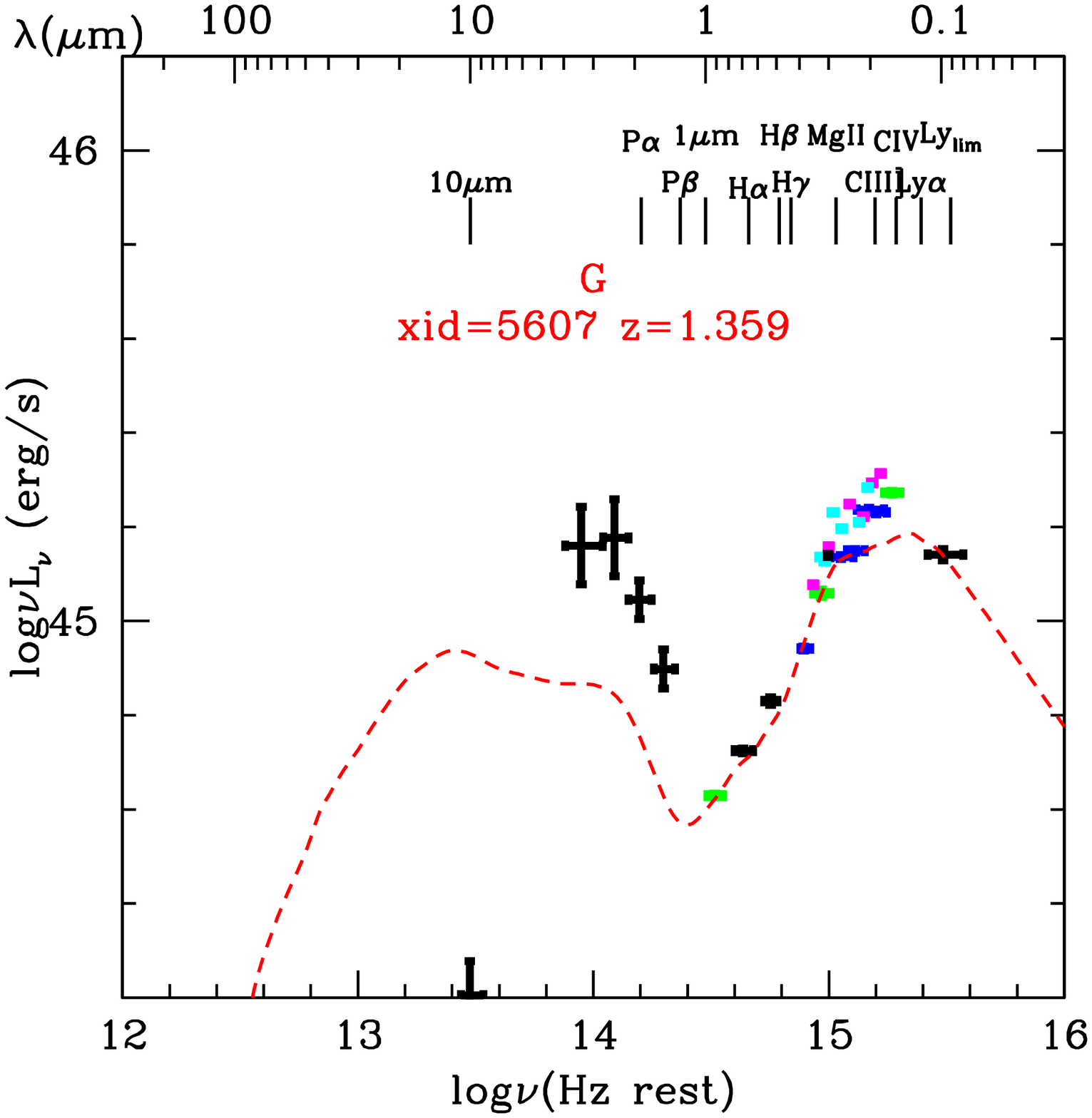}
\includegraphics[angle=0,width=0.45\textwidth]{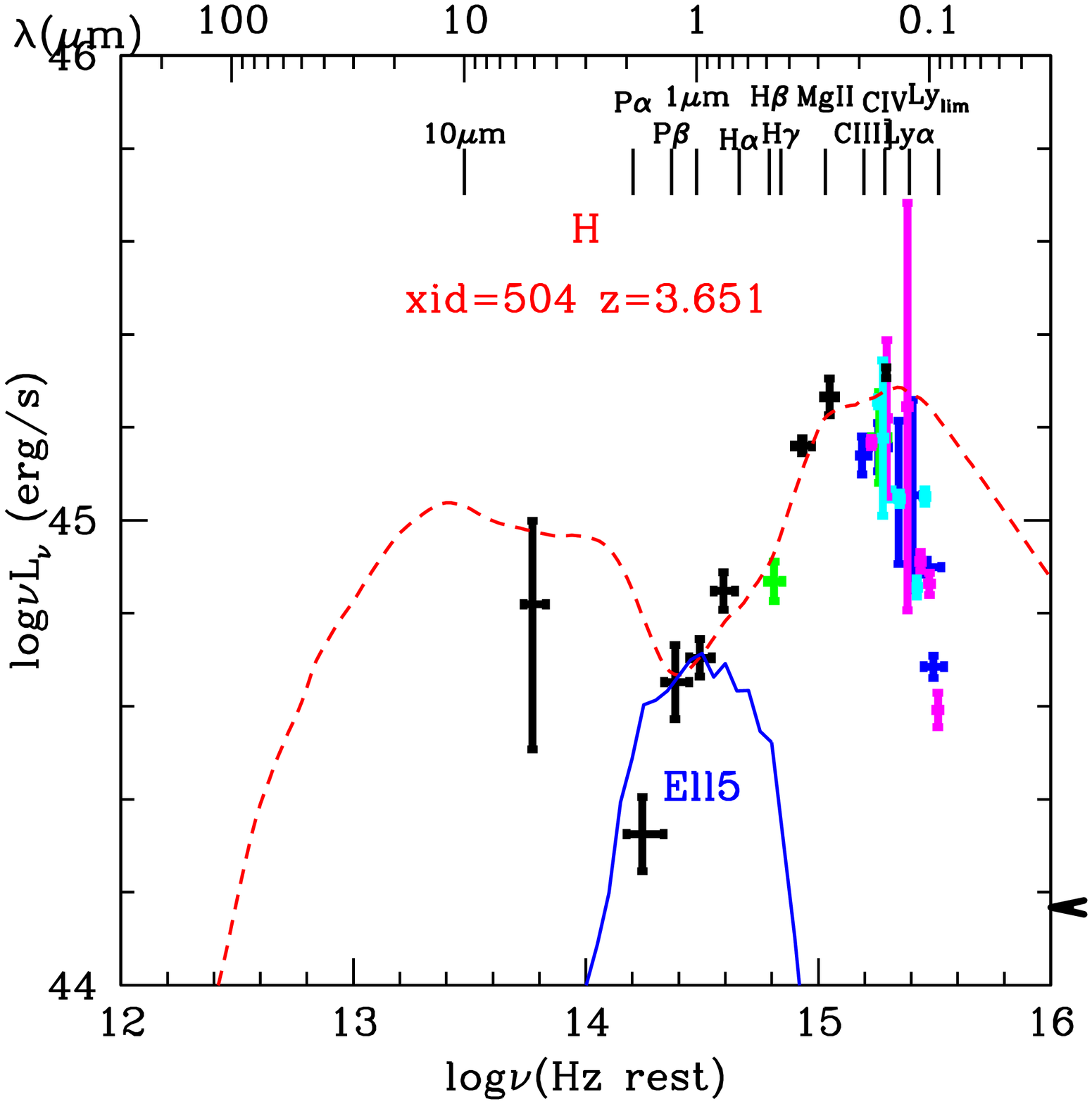}
\caption{Extreme examples of SEDs: {\em top left:} E. a close analog
of a ULIRG SED, with prominent infrared emission; {\em top right:}
F. no big blue bump, probably due to extreme reddening; {\em bottom
left:} G. big near-infrared bump, due to rich hot dust emission;
{\em bottom right:} H. hot-dust-poor quasar, strong big blue bump
but no $1~\mu$m inflection due to a weak near-IR bump. The red
dashed line is the E94 RQ mean SED. The blue lines are the galaxy
templates (Polletta et al. 2007). The data points in the SED are
color-coded as in Elvis et al. (2012). From low frequency to high
frequency, the black data points are: 24$~\mu$m, 8$~\mu$m,
5.7$~\mu$m, 4.5$~\mu$m, 3.6$~\mu$m, K-band, H-band, J-band, the NUV
and FUV. The blue data points are the Subaru broad bands
($\rm{B_J}$, g, r, i, z) from 2005. The green data points are the
(CFHT) K-band, and the (CFHT) u band and i band. The purple data
points are the 6 Subaru intermediate bands for season 1 (2006)
(IA427, IA464, IA505, IA574, IA709, IA827).  The cyan data points
are the 5 Subaru intermediate bands for season 2 (2007) (IA484,
IA527, IA624, IA679, IA738, IA767). The arrow on the right show the
X-ray luminosity at 2keV.\label{extremeseds}}
\end{figure*}
%%%%%%%%%%%%%%%%

There are sources lying outside the mixing wedge that are outliers
with respect to the bulk of the type 1 AGN population. As noted
above (\S~\ref{s:evolution}), the mixing diagram has already been
successfully used to identify a population of HDP quasars lacking
the characteristic maximally hot dust of AGN (Hao et al. 2010,
2011).

The four extreme examples of SEDs singled out in Paper I (A, B, C,
D) are also marked in Figure~\ref{slpxmm}. They lie at the four
corners of the mixing diagram. Figure~\ref{extremeseds} displays the
SEDs of four additional outliers (E, F, G, H) which are discussed
briefly below. These four quasars lies in the furthest corners of
the mixing diagram. A detailed discussion will be deferred to later
papers.

\begin{itemize}
  \item {\em A Newborn quasar?} Object E (XID=304, COSMOS\_J 095931.58+021905.52, z=1.607)
  has an SED well fit by the ULIRG Arp 220 SED (Polletta et al. 2007) at
  $\lambda>0.40~\mu$m. However, in the UV (at $\lambda<0.40~\mu$m), a weak
  quasar component emerges, as do the broad emission lines that identify
  it as a type 1 AGN. This object has a luminosity in the ULIRG regime
  (the bolometric luminosity integrated in 24$~\mu$m -- 40 keV range
  is 10$^{12.2}L_{\bigodot}$) and appears to be a
  composite quasar/starburst.  The rarity of objects like E in
  XMM-COSMOS argues for a short-lived phase.  Object E is thus
  a good candidate for a newly born quasar, or at the beginning
  of the ``buried quasar stage'', where the quasar
  emerges during a merger triggered starburst (Hopkins et al.
  2006). The obscured starburst activity still dominates the SED and the
  quasar is still too weak to quench the starburst activity.

  \item {\em A Weak Big Blue Bump Quasar?} Object F (XID=135, COSMOS\_J095848.21+022409.3,
  z= 0.376) shows a two dex drop in the u-band compared to the E94 RQ mean SED. An
  extinction of $E(B-V)=0.8$ could be applied. This source is
  classified as type 1 AGN because a strong broad H${\alpha}$
  line (FWHM$\sim 5000~km/s$) is present in the optical
  spectrum. There may be strong differential reddening between the continuum and the broad
  line emitting region.  Alternatively, an NGC~6090 template fits the optical/UV
  SED well. Is then the UV `big blue bump' intrinsically weak in this object? The
  high X-ray flux relative to the optical would make for a truly unusual SED in
  the extreme UV.

  \item {\em A ``Blow-out'' Phase Quasar?} Object G (XID = 5607,
  COSMOS\_J 095743.33+024823.8, z=1.359) is well fitted by the E94 RQ mean
  SED in the optical/UV, but shows an unusually strong near-infrared bump,
  two times brighter than the E94 RQ mean SED at 3$~\mu$m, indicating an
  unusually rich hot dust component. Such a quasar could be a good
  candidate for objects at the end of the ``buried
  quasar stage'' or the beginning of the ``blow out phase'', where the quasar
  emerges from its dusty cocoon and begins to dominate the SED (Hopkins et al. 2006).
  The properties of these quasars still need to be investigated.

  \item {\em Hot Dust Poor Quasar} Object H (XID=504, COSMOS\_J 095931.01+021333.0,
  z=3.651) is located in the
  upper right corner, furthest from the E94 mean SED template in the mixing
  diagram of the XC413 sample. The SED of object H has a typical strong big blue bump but weak
  infrared emission. It is another hot-dust-poor quasar, similar to source D described in
  Paper I, and discussed in detail in Hao et al. (2010). These could
  be sources that have used up or blown-out most their dust and gas.
\end{itemize}

%%%%%%%%%%%%%%%%%%%%%%%%%%%%%%%%%%%%%%%%%%%%%%%%%%%%%%%%%%%%%%%%%%%%%%%%
\section{Discussion and Conclusions}

Making use of the strong SED shape differences around 1$~\mu$m for
galaxies and quasars, we defined the quasar-galaxy mixing diagram: a
plot of the 1-3$~\mu$m SED slope versus the 0.3-1$~\mu$m SED slope.
This diagram allows us to easily distinguish among quasar-dominated,
galaxy-dominated and reddening-dominated SEDs without making strong
model assumptions.

This mixing diagram, when applied to the XMM-COSMOS sample shows
that $\sim$90\% of the quasar SEDs can be explained by the
combination of (1) an E94-like mean SED, (2) a host galaxy SED and
(3) reddening. The mixing diagram is a very useful tool and, as we
have outlined, has various applications.

Changes in the quasar SED shape with respect to the physical
parameters $z$, $L_{bol}$, $M_{BH}$ and $\lambda_E$ were sought. At
high $z$, $\log L_{bol}$, $\log M_{BH}$ and $\log \lambda_{E}$, the
XMM-COSMOS quasars cluster close to the E94 mean, with a slight
offset, which could be due to either an intrinsic SED change, or a
small but not negligible host galaxy component.  Lower $z$,
$L_{bol}$, $M_{BH}$ and $\lambda_{E}$ sources spread along the E94
mean SED - host mixing curves. The mixing diagram allows estimates
of the galaxy fraction and the reddening for each AGN. Reddening of
$E(B-V)>0.4$ is seen mainly among low $z$, $L_{bol}$ objects.

Most importantly, the mixing diagram can give a reliable estimate of
the 1$~\mu$m host galaxy fraction or luminosity and the $E(B-V)$.The
galaxy fractions estimated from the mixing diagram were compared
with those estimated from the black hole mass - bulge mass scaling
relationship adding an evolutionary term, from direct Hubble image
decomposition and from SED fitting. The host fraction estimated from
the scaling relationship and the image decomposition show weak
correlation with the galaxy fraction from the mixing diagram, though
all have large errors. The black hole mass - bulge method gives
systematically smaller galaxy fractions. But the galaxy fractions
from the mixing diagram are consistent with the results from the SED
fitting. The mixing diagram appears to be a useful and reliable tool
to estimate the host galaxy fraction and luminosity at 1$~\mu$m.

The reddening ($E(B-V)$) estimated from the mixing diagram were
correlated with the NIR luminosity ratio ($L_{ir}/L_{bol}$) and OPT
luminosity ratio ($L_{opt}/L_{bol}$). A significant correlation is
found for $E(B-V)$ versus $\rm{N_H}$, although with a large spread.
The derived $E(B-V)$ and $\lambda_{E}$ are not significantly
correlated. A `forbidden region' in the $E(B-V)$ versus
$\lambda_{E}$ space is seen as in Vasudevan et al. (2009).

The mixing diagram can be used also to identify outliers. As these
AGN are rare in a deep X-ray selected sample, they may represent
different short-lived stages of the quasar-galaxy co-evolution.

The mixing diagram can clearly distinguish among the
quasar-dominated, host-dominated and reddening-dominated SEDs. Thus
different phases of galaxy formation and evolution would locate in
different regions of the diagram. A complete evolutionary track of
the quasar-galaxy co-evolution cycle can, in principle, be drawn on
the mixing diagram, by analogy to tracks in the HR diagram in
stellar astrophysics. Numerical simulations have reproduced quasars
at various redshifts from hierarchical assembly in the $\Lambda$CDM
cosmology (Hopkins et al. 2006; Li et al. 2007), but have not
addressed how the resulting SEDs change.

There are various different galaxy formation and evolution models.
Two representatives would be 1) the ``cosmic cycle'' (Hopkins et al.
2006) for galaxy formation and evolution, which are regulated by
black hole growth in mergers; 2) the galaxy evolution triggered by
self-regulated baryonic process (Granato et al. 2004). The main
difference between these two models is in the beginning phase: 1) in
the merger-driven model (Hopkins et al. 2006), star-formation is
enhanced by the merging of two late-type galaxies; 2) in the
anti-hierarchical baryon collapse model (Granato et al. 2004) the
proto-spheroidal galaxies formed in the virialized dark matter halo
have high star-formation rate (Mao et al. 2007, Cai et al. 2013).
The following black hole growth (Lapi et al. 2006, Hopkins et al.
2006) and galaxy evolution in both models are similar to each other
with some difference in timescales of different phases. Thus in the
mixing diagram, the evolutionary tracks between different models
would be very similar in most regions.

\begin{figure}
\vspace{-2cm} \epsfig{file=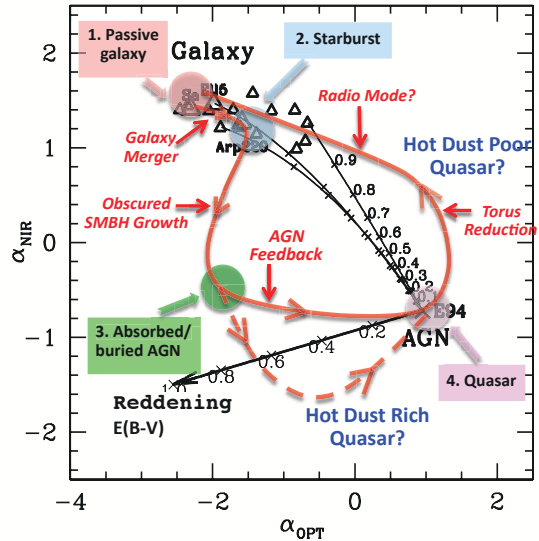, angle=0,width=\linewidth}
\vspace{-2.5cm} \caption{The evolution view of the mixing diagram.
Four different phases of the ``cosmic cycle'' are shown as colored
circles. The red line shows the evolution track of an AGN life
cycle. \label{evlcyc}}
\end{figure}

A sketch of a possible evolutionary track is shown in
Figure~\ref{evlcyc}. Mergers drive a galaxy (1, red) into the
starburst region (2, blue). Here, the SMBH grows by accretion. The
quasar emission gradually comes to dominate the luminosity, but is
`buried' by gas and dust, so the source moves downward in the mixing
diagram for the phase of obscured quasar activity (3, green).
Sources in this stage would be identified as type~2 AGN, not
included in the XC413 sample. At the end of this buried quasar
phase, hot dust rich (HDR) quasars - the outliers with much stronger
hot dust emission than typical quasar and broad emission lines -
would be found at the very bottom of the mixing diagram. At this
stage, feedback from the SMBH expels enough interstellar medium, and
the obscuring ``torus'' and the broad line region emission become
visible, and the object gradually moves either from a `buried' or
`HDR' quasar to the typical quasar region (4, purple) if the ratio
between AGN and host galaxy luminosity is high. Lower luminosity AGN
would move near the mixing curves. As the SMBH continues to accrete,
the gas and dust is either used up as a reservoir, or expelled. The
dust covering factor reduces, and the source moves up to the HDP
quasar region, before finally becoming quiescent once more. The
length of the timescale of each stage may be reflected by the number
of sources in each region on the mixing diagram in a complete
sample.

%%%%%%%%%%%%%%%
%%%%%%%%%%%%%%%%

A family of possible evolutionary cycles could be drawn, varying the
parameters of the initial merging (e.g. mass, gas fraction,
accretion rate). A quantitative picture of the cosmic cycle (e.g.
the duration of the duty cycle in each phase, the dependence on the
initial conditions, etc.) could thus be obtained from the density of
objects around the mixing diagram. These results, in turn, could put
constraints on the physics adopted to model AGN/galaxy coevolution
in numerical simulations.  We plan to address the quantitative
evolution of quasar-galaxy SEDs in the mixing diagram, over the
complete cosmic cycle in later papers, including in the analysis of
also type 2 AGN.

However, we have to note that Figure~\ref{evlcyc} is just an
idealized illustration. The tracks of the evolution of sources could
be very complicated and sources could evolve in various direction in
the mixing region. Bongiorno et al. (2012) plotted all the
XMM-COSMOS sources in the mixing diagram and there is no obvious
accretion rate distribution correlated with different regions on the
mixing diagram observed.

For a longer term study of the full evolutionary picture, the mixing
diagram definition could be extended to other wavelengths. For
example, we could investigate the optical to ultraviolet SED with
respect to the near infrared SED for the extinction law; we could
study the radio and far-infrared SED with respect to
optical/near-infrared for the radio-loudness; we could check the
ultraviolet SED with respect to X-ray for the $\alpha_{\rm{OX}}$.
The multiwavelength analysis of the AGN emission could not only
significantly improve our understanding of the SMBH accretion, the
AGN structure and the unification of AGN, but also would help us
understand the role of the SMBH in the co-evolution cosmic cycle.

%%%%%%%%%%%%%%%%%%%%%%%%%%%%%%%%%%%%%%%%%%%%%%%%%%%%%%%%%%%%%%%%%%%%%%%%
\section*{Acknowledgments}

HH thanks Belinda Wilkes, Martin J. Ward and Zhenyi Cai for valuable
discussions. This work was supported in part by NASA {\em Chandra}
grant number G07-8136A (HH, ME, CV). Support from the Italian Space
Agency (ASI) under the contracts ASI-INAF I/088/06/0 and I/009/10/0
is acknowledged (AC and CV). MS acknowledges support by the German
Deutsche Forschungsgemeinschaft, DFG Leibniz Prize (FKZ HA
1850/28-1). KS gratefully acknowledges support from Swiss National
Science Foundation Grant PP00P2\_138979/1.

%%%%%%%%%%%%%%%%%%%%%%%%%%%%%%%%%%%%%%%%%%%%%%%%%%%%%%%%%%%%%%%%%%%%%%%%

\label{lastpage}

%%%%%%%%%%%%%%%%%%
%%%%%%%%%%%%%%%%%%

\begin{thebibliography}{}

\bibitem[Alongi et al. (1993)]{alongi93} Alongi, M., Bertelli, G., Bressan, A.,
Chiosi, C., Fagotto, F., Greggio, L., Nasi, E., 1993, A\&AS, 97, 851

\bibitem[Barvainis (1987)]{barvainis87} Barvainis, R. 1987, ApJ,
320, 537

\bibitem[Bennert et al. (2010)]{bennert10} Bennert, V. N., Treu,
T., Woo, J.-H., Malkan, M. A., Le Bris, A., Auger, M. W., Gallagher,
S., Blandford, R. D., 2010, ApJ, 708, 1507

\bibitem[Bennert et al. (2011)]{bennert11} Bennert, V. N., Auger,
M. W., Treu, T., Woo, J.-H., Malkan, M. A., 2011, ApJ, 742, 107

\bibitem[Bongiorno et al. (2007)]{Bongiorno07} Bongiorno, A., et al., 2007, A\&A, 472, 443

\bibitem[Bongiorno et al. (2012)]{Bongiorno12} Bongiorno, A., et
al., 2012, MNRAS, 427, 3103

\bibitem[Bouwens et al. (2012)]{bouwens12} Bouwens, R. J., et al.,
2012, ApJ, 754, 83

\bibitem[Brandt\& Hasinger (2005)]{Brandt05} Brandt, W. N. \&
Hasinger, G. 2005, ARA\&A, 43, 827

\bibitem[Bressan et al. (1993)]{bressan93} Bressan, A., Fagotto, F.,
Bertelli, G., Chiosi, C., 1993, A\&AS, 100, 647

\bibitem[Brusa et al. (2007)]{brusa07} Brusa, M., et al., 2007, ApJS,
2007, 172, 353

\bibitem[Brusa et al. (2010)]{brusa10} Brusa, M., et al., 2010, ApJ,
716, 348

\bibitem[Bruzual \& Charlot (2003)]{bruzual03} Bruzual, G. \&
Charlot, S., 2003, MNRAS, 344, 1000

\bibitem[Cai et al. (2013)]{cai13} Cai, Z., et al., 2013, ApJ, 768,
21

\bibitem[Calzetti et al. (2000)]{calzetti00} Calzetti, D., Armus, L.
Bohlin, R. C., Kinney, A. L., Koornneef, J., Storchi-Bergmann, T.,
2000, ApJ, 533, 682

\bibitem[Capak et al. (2007)]{capak07} Capak, P., et al., 2007, ApJS, 172, 99

\bibitem[Cappelluti et al. (2007)]{cappelluti07} Cappelluti, N., et al., 2007,
ApJS, 172, 341

\bibitem[Cappelluti et al. (2009)]{cappelluti09} Cappelluti, N., et al.,
2009, A\&A, 497, 635

\bibitem[Cardelli et al. (1989)]{cardelli89} Cardelli, J. A.,
Clayton, G. C., Mathis, J. S., 1989, ApJ, 345, 245

\bibitem[Chabrier (2003)]{chabrier03} Chabrier, G., 2003, PASP, 115,
763

\bibitem[Cirasuolo et al. (2007)]{cirasuolo07} Cirasuolo, M. et al.
2007, MNRAS, 380, 585

\bibitem[Cisternas et al. (2011)]{cisternas11} Cisternas, M. et al.
2011, ApJ, 726, 57

\bibitem[Civano et al. (2011)]{civano11} Civano, F., et al., 2011,
ApJ, 741, 91

\bibitem[Coleman et al. (1980)]{coleman80} Coleman, G. D., Wu, C.
C., \& Weedman, D. W., 1980, ApJS, 43, 393

\bibitem[Crenshaw \& Kraemer (2001)]{crenshaw01} Crenshsw D. M. \&
Kraemer S. B. 2001, ApJ, 562, L29

\bibitem[Elvis et al. (1994)]{elvis94} Elvis, M. et al., 1994, ApJS, 95, 1

\bibitem[Elvis et al. (2012)]{elvis11} Elvis, M. et al., 2011, ApJ,
accepted, astro-ph/1209.1478

\bibitem[Fabian et al. (2008)]{fabian08} Fabian, A. C., Vasudevan, R. V., \& Gandhi,
P., 2008, MNRAS, 385, L43

\bibitem[Fabian et al. (2009)]{fabian09} Fabian, A. C., Vasudevan, R. V., Mushotzky, R. F.,
Winter, L. M., \& Reynolds, C. S., 2009, MNRAS, 394, 89

\bibitem[Ferrarese \& Merrit (2000)]{ferrarese00} Ferrarese, L. \& Merrit, D. 2000 ApJ, 539, L9

\bibitem[Fitzpatrick (1999)]{fitzpatrick99} Fitzpatrick, E. L. 1999,
PASP, 111, 63

\bibitem[Fagotto et al. (1994a)]{fagotto94a} Fagotto, F., Bressan,
A., Bertelli, G., Chiosi, C., 1994a, A\&AS, 104, 365

\bibitem[Fagotto et al. (1994b)]{fagotto94b} Fagotto, F., Bressan,
A., Bertelli, G., Chiosi, C., 1994b, A\&AS, 105, 29

\bibitem[Franceschini et al. (1999)]{franceschini99} Franceschini, A.,
Hasinger, G., Miyaji, T., \& Malquori, D., 1999, MNRAS, 310, L5

\bibitem[Gallerani et al. (2005)]{gallerani05} Gallerani, S., et
al., 2010, A\&A, 523, 85

\bibitem[Gebhardt et al. (2000)]{gebhardt00} Gebhardt, K., et al. 2000, AJ, 539, L13

\bibitem[Girardi et al. (1996)]{girardi96} Girardi, L., Bressan, A.,
Chiosi, C., Bertelli, G., Nasi, E., 1996, A\&AS, 117, 113

\bibitem[Glikman et al. (2006)]{Glikman06}Glikman, E., Helfand, D. J., \&
White, R. L., 2006, ApJ, 640, 579

\bibitem[Glikman et al. (2012)]{Glikman12}Glikman, E., et al. 2012, ApJ,
757, 51

\bibitem[Gordon et al. (2003)]{gordon03} Gordon, K. et al., 2003, ApJ, 594, 279

\bibitem[Granato et al. (2004)]{granato04} Granato, G. L., De Zotti, G., Silva, L., Bressan, A., Danese,
L., 2004, ApJ, 600, 580

\bibitem[Hao et al. (2010)]{hao10} Hao, H., et al., 2010, ApJ,
724, L59

\bibitem[Hao et al. (2011)]{hao11} Hao, H., et al., 2011, ApJ, 733,
108

\bibitem[Hao et al. (2013a)]{hao13a} Hao, H., et al., 2012a, MNRAS
submitted, arXiv:1210.3033

\bibitem[Hao et al. (2013b)]{hao13b} Hao, H., et al., 2013b, ApJL,
to be submitted

\bibitem[Hasinger et al. (2007)]{hasinger07} Hasinger, G., et al., 2007, ApJS, 172, 29

\bibitem[Ho (2008)]{ho08} Ho, L. C., 2008, ARA\&A, 46, 475

\bibitem[Hopkins et al. (2004)]{hopkins04} Hopkins, P. F., et al.,
2004, AJ, 128, 1112

\bibitem[Hopkins et al. (2006)]{hopkins06} Hopkins, P. F., et al., 2006, ApJS, 163, 1

\bibitem[Hopkins et al. (2007)]{hopkins07} Hopkins, P. F., Richards, G. T., \& Hernquist, L., 2007, ApJ, 654, 731

\bibitem[Ilbert et al. (2009)]{ilbert09} Ilbert, O., et al., 2009,
ApJ, 690, 1236

\bibitem[Ilbert et al. (2010)]{ilbert10} Ilbert, O., et al., 2010,
ApJ, 709, 644

\bibitem[Jiang et al. (2010)]{jiang10} Jiang, L. et al., 2010,
Nature, 464, 380

\bibitem[Kelly et al. (2007)]{kelly07} Kelly, Brandon C. 2007, ApJ,
665, 1489

\bibitem[Kewley et al. (2006)]{kewley06} Kewley, L. J., Groves, B.,
Kauffmann, G., \& Heckman, T., 2006, MNRAS, 372, 961

\bibitem[Komatsu et al. (2009)]{komatsu09} Komatsu, E., et al.,
2009, ApJS, 180, 330

\bibitem[Kormendy \& Richstone (1995)]{kormendy95} Kormendy, J. \& Richstone, D. 1995, ARA\&A, 33,581

\bibitem[Lapi et al. (2006)]{lapi06}Lapi, A., et al., ApJ, 650, 42

\bibitem[Li et al. (2007)]{li07} Li, Y., et al., 2007, ApJ, 665, 187

\bibitem[Lilly et al. (2007)]{lilly07}Lilly, S. J., et al., 2007, ApJS
172, 70

\bibitem[Lilly et al. (2009)]{lilly09}Lilly, S. J., et al., 2009, ApJS
184, 218

\bibitem[Mainieri et al. (2007)]{mainieri07} Mainieri, V., et al., 2007, ApJS, 172,
368

\bibitem[Maiolino et al. (2001)]{maiolino01} Maiolino, R., et al.
2001, A\&A, 365, 28

\bibitem[Malkan \& Sargent (1982)]{malkan82} Malkan, M. A., \& Sargent, W. L. W., 1982, ApJ, 254, 22

\bibitem[Mao et al. (2007)]{mao07} Mao, J., Lapi, A., Granato, G. L., De Zotti, G., Danese,
L., 2007, ApJ, 667, 655

\bibitem[Marconi \& Hunt (2003)]{marconi03} Marconi, A. \& Hunt, L. K. 2003, ApJ, 589,
L21

\bibitem[McCracken et al. (2010)]{mccracken10} McCracken, H. J., et
al., 2010, ApJ, 708, 202

\bibitem[Merloni et al. (2010)]{merloni10} Merloni, A., et al. 2010 ApJ 708, 137

\bibitem[Merloni \& Heinz (2012)]{merloni12} Merloni, A., \& Heinz,
S., 2012, Planets, Stars and Stellar Systems, vol 6, ed W. Keel

\bibitem[Misselt et al. (1999)]{misselt99} Misselt, K. A., Clayton,
G. C., Gordon, K. D., 1999, ApJ, 515, 128

\bibitem[O'Brien et al. (1988)]{obrien88} O'Brien, P. T., Wilson,
R., \& Gondhalekar, P. M. 1988, MNRAS, 233, 801

\bibitem[O'Donnell (1994)]{odonell94} O'Donnell, J. E., 1994, ApJ,
422, 158

\bibitem[Polletta et al. (2007)]{polletta07} Polletta, M., et al. 2007, ApJ, 663, 81

\bibitem[Richards et al. (2003)]{richards03} Richards, G. T., et
al. 2003, AJ, 126, 1131

\bibitem[Richards et al. (2006)]{richards06} Richards, G. T. et al. 2006, ApJS, 166, 470

\bibitem[Richstone et al. (1998)]{richstone98} Richstone, D., et al. 1998, Nature, 395, 14

\bibitem[Salvato et al. (2009)]{salvato09} Salvato, M. et al. 2009
ApJ 690, 1250

\bibitem[Sandage et al. (1971)]{sandage71} Sandage, A. 1971 SWNG
conf, 271

\bibitem[Sanders et al. (1989)]{sanders89} Sanders, D. B., Phinney,
E. S., Neugebauer, G., Soifer, B. T., \& Matthews, K. 1989, ApJ,
347, 29

\bibitem[Schawinski et al. (2009)]{schawinski09} Schawinski, K., et
al., 2009, ApJL, 692, 19

\bibitem[Schmidt \& Green (1983)]{schmidt83} Schmidt, M. \& Green,
R. F. 1983, ApJ, 269, 352

\bibitem[Schneider et al. (2007)]{schneider07} Schneider, D. P. et
al. 2007, AJ, 134, 102

\bibitem[Schramm \& Silverman (2013)]{schramm13} Schramm, M. \&
Silverman, J. D., 2013, ApJ, 767, 13

\bibitem[Scoville et al. (2007)]{scoville07a}Scoville, N. Z., et al.,
2007, ApJS, 172, 1

\bibitem[Shang et al. (2011)]{shang11} Shang, Zhaohui, et al., 2011,
ApJS, 196, 2

\bibitem[Sikora et al. (2007)]{sikora07}Sikora, M., Stawarz, L. \&
Lasota, J. P., 2007, ApJ, 658, 815

\bibitem[Silva et al. (1998)]{Silva98} Silva, L., Granato, G. L.,
Bressan, A., \& Danese, L. 1998, ApJ, 509, 103

\bibitem[Silverman et al. (2005)]{silverman05} Silverman, J. D. et
al. 2005, ApJ, 624, 630

\bibitem[Soltan (1982)]{Soltan82} Soltan, A., 1982, MNRAS, 200, 115

\bibitem[Taniguchi (1999)]{taniguchi99} Taniguchi, Y. 1999, ApJ, 524, 65

\bibitem[Trump et al. (2009)]{trump09}Trump, J. R. et al. 2009 ApJ,
696, 1195

\bibitem[Trump et al. (2011)]{trump11}Trump, J. R. et al. 2011 ApJ,
733, 60

\bibitem[Vasudevan et al. (2009)]{vasudevan09} Vasudevan, R. V., Mushotzky, R. F., Winter,
L. M., \& Fabian, A. C. 2009, MNRAS, 399, 1553

\bibitem[Vestergaard \& Peterson (2006)]{vestergaard06} Vestergaard, M. \& Peterson,
B. M. 2006 ApJ 641, 689

\bibitem[Ueda et al. (2003)]{ueda03} Ueda, Y. et al. 2003 ApJ 598,
886

\bibitem[Ward et al. (1987)]{ward87} Ward, M. et al. 1987 ApJ, 315, 74

\bibitem[Weedman (1973)] {weedman73} Weedman, D. 1973 ApJ, 183, 29

\bibitem[Wild et al. (2010)]{wild10} Wild, V., et al., 2010, MNRAS,
405, 933

\bibitem[Wills, Netzer \& Wills (1985)]{wills85} Wills, B. J., Netzer, H., \& Wills,
D. 1985, ApJ, 288, 94

\bibitem[Xiao et al. (2012)]{xiao12} Xiao, T., et al. 2012, MNRAS,
421, 486

\bibitem[Young et al. (2008)]{young08} Young, M., Elvis, M., \& Risaliti,
G. 2008, ApJ, 688, 128


\end{thebibliography}
\end{document}